\documentclass[iop]{emulateapj}
\usepackage{longtable,subfigure,float}

\usepackage[bookmarks=false]{hyperref}
\usepackage{times}
\usepackage{url}
\usepackage{xspace,booktabs,amsmath} 
\usepackage{booktabs} %

\newcommand {\Lsixum} {$L_{\rm 6\mu m}$\xspace}

\newcommand {\feka} {Fe~K$\alpha$\xspace}

\newcommand{\hbeta}{H$\beta$}

\newcommand {\nh} {$N_{\mathrm{H}}$\xspace}
\newcommand {\loiii} {$L_{\rm [O\ {\scriptscriptstyle III}]}$\xspace}
\newcommand {\lx} {$L_{\mathrm{X}}$\xspace}

\newcommand {\Lsoft} {$L_{\mathrm{2-10\ keV}}$\xspace}
\newcommand {\Lsoftobs} {$L^{\rm obs}_{\mathrm{2-10\ keV}}$\xspace}
\newcommand {\Lbat} {$L^{\rm obs}_{\mathrm{14-195\ keV}}$\xspace}

\newcommand {\xspec}{{\sc Xspec}\xspace}
\newcommand {\torus}{{\sc BNTorus}\xspace}
\newcommand {\mytorus}{{\sc MYTorus}\xspace}

\newcommand {\modelM}{$\mathtt{Model\ M}$\xspace}
\newcommand {\modelSC}{$\mathtt{Model\ SC}$\xspace}

\newcommand {\modelT}{$\mathtt{Model\ T}$\xspace}

\newcommand {\scbat}{SC$_\mathtt{BAT}$\xspace}
\newcommand {\scnustar}{SC$_\mathtt{NuSTAR}$\xspace}

\newcommand {\ctspersec} {cts~s$^{-1}$\xspace}
\newcommand {\ergpersec} {erg~s$^{-1}$\xspace}

\newcommand {\nhunit} {cm$^{-2}$\xspace}
\newcommand {\degrees} {$^{\circ}$\xspace}

\def\micron{{\mbox{$\mu{\rm m}$}}}
\def\arcsec{{\mbox{$^{\prime \prime}$}}}
\def\arcmin{{\mbox{$^{\prime}$}}}
\def\degree{{\mbox{$^{\circ}$}}}

\def \nustar {{\em NuSTAR\ }}
\def \nustarsh {{\em NuSTAR}}
\def \swift {{\em Swift\ }}
\def \swiftxrt {{\em Swift}/XRT\ }
\def \swiftxrtsh {{\em Swift}/XRT}
\def \swiftbat {{\em Swift}/BAT\ }
\def \swiftbatsh {{\em Swift}/BAT}
\def \chandra {{\em Chandra\ }}
\def \XMM{{\em XMM-Newton\ }}

\def \xmm{{\em XMM-Newton\ }}
\def \galex{{\em GALEX\ }}
\def \wise{{\em WISE\ }}

\def \chandrash {{\em Chandra}}

\def \astroh{{\em Astro-H}}
\def \athena{{\em Athena}}

\def \xspec {{\tt Xspec\ }}

\def \rosat{{\em ROSAT\ }}
\def \erosita{{\em eROSITA\ }}

\def \ascash{{\em ASCA}}
\def \bepposax{{\em BeppoSAX\ }}

\newcommand{\oiii}{[O\,{\sc iii}]\xspace }
\def\arcsec{{\mbox{$^{\prime \prime}$}}}

\def\erg{{\rm\thinspace erg}}

\def\Msun{\hbox{$\rm\thinspace M_{\odot}$}}

\def\s{{\rm\thinspace s}}

\def\ergps{\hbox{$\erg\s^{-1}\,$}}

\def\micron{{\mbox{$\mu{\rm m}$}}}
\def\arcsec{{\mbox{$^{\prime \prime}$}}}
\def\arcmin{{\mbox{$^{\prime}$}}}
\def\degree{{\mbox{$^{\circ}$}}}

\begin{document}
\title{A New Population of Compton-Thick AGN Identified Using the Spectral Curvature Above 10 keV}

\author{Michael J. Koss\altaffilmark{1,2,18}, R. Assef\altaffilmark{3}, M. Balokovi{\' c}\altaffilmark{4}, D. Stern\altaffilmark{5}, P. Gandhi\altaffilmark{6}, I. Lamperti\altaffilmark{1}, D. M. Alexander\altaffilmark{7}, D. R. Ballantyne \altaffilmark{8},  F.E. Bauer\altaffilmark{9,10}, S. Berney\altaffilmark{1}, W. N. Brandt\altaffilmark{11,12,13}, A. Comastri\altaffilmark{14}, N. Gehrels\altaffilmark{15}, F. A. Harrison\altaffilmark{4},   G. Lansbury\altaffilmark{7}, C. Markwardt\altaffilmark{15},  C. Ricci\altaffilmark{9}, E. Rivers\altaffilmark{4}, K. Schawinski\altaffilmark{1},  E. Treister\altaffilmark{16}, C. Megan Urry \altaffilmark{17}}
\email{mkoss@phys.ethz.ch}

\altaffiltext{1}{Institute for Astronomy, Department of Physics, ETH Zurich, Wolfgang-Pauli-Strasse 27, CH-8093 Zurich, Switzerland; mkoss@phys.ethz.ch}
\altaffiltext{2}{Institute for Astronomy, University of Hawaii, 2680 Woodlawn Drive, Honolulu, HI 96822, USA}
\altaffiltext{3}{N\'ucleo de Astronom\'ia de la Facultad de Ingenier\'ia, Universidad Diego Portales, Av. Ej\'ercito 441, Santiago, Chile}
\altaffiltext{4}{Cahill Center for Astronomy and Astrophysics, California Institute of Technology, Pasadena, CA 91125, USA}
\altaffiltext{5}{Jet Propulsion Laboratory, California Institute of Technology, Pasadena, CA 91109, USA}
\altaffiltext{6}{School of Physics and Astronomy, University of Southampton, Highfield, Southampton SO17 1BJ, UK}
\altaffiltext{7}{Department of Physics, Durham University, South Road, Durham DH1 3LE, UK}
\altaffiltext{8}{Center for Relativistic Astrophysics, School of Physics, Georgia Institute of Technology, Atlanta, GA 30332, USA}
\altaffiltext{9}{Instituto de Astrof\'{\i}sica, Facultad de F\'{\i}sica, Pontificia Universidad Cat\'olica de Chile, Casilla 306, Santiago 22, Chile}
\altaffiltext{10}{Space Science Institute, 4750 Walnut Street, Suite 205, Boulder, CO 80301, USA}
\altaffiltext{11}{Department of Astronomy \& Astrophysics, 525 Davey Lab, The Pennsylvania State University, University Park, PA 16802, USA}
\altaffiltext{12}{Institute for Gravitation and the Cosmos, The Pennsylvania State University, University Park, PA 16802, USA}
\altaffiltext{13}{Department of Physics, 104 Davey Lab, The Pennsylvania State University, University Park, PA 16802, USA}
\altaffiltext{14}{INAF -- Osservatorio Astronomico di Bologna, Via Ranzani 1, 40127 Bologna, Italy}
\altaffiltext{15}{Astrophysics Science Division, NASA Goddard Space Flight Center, Greenbelt, MD, USA}
\altaffiltext{16}{Departamento de Astronom\'{\i}a, Universidad de Concepci\'{o}n, Casilla 160-C, Concepci\'{o}n, Chile}
\altaffiltext{17}{Yale Center for Astronomy and Astrophysics, Physics Department, Yale University, P.O. Box 208120, New Haven, CT 06520-8120, USA}
\altaffiltext{18}{SNSF Ambizione Fellow}

\begin{abstract}
We present a new metric that uses the spectral curvature (SC) above 10~keV to identify Compton-thick AGN in low-quality \swiftbat X-ray data. Using \nustarsh, we observe nine high SC-selected AGN.  We find that high-sensitivity spectra show the majority are Compton-thick (78\% or 7/9) and the remaining two are nearly Compton-thick (\nh$\simeq5-8\times10^{23}$ \nhunit).  We find the \scbat and \scnustar measurements are consistent, suggesting this technique can be applied to future telescopes.  We tested the SC method on well-known Compton-thick AGN and find it is much more effective than broad band ratios (e.g. 100\% using SC vs. 20\% using 8-24/3-8~keV).   Our results suggest that using the $>10$ keV emission may be the only way to identify this population since only two sources show Compton-thick levels of excess in the Balmer decrement corrected \oiii to observed X-ray emission ratio ($F_{\rm [O\ {\scriptscriptstyle III}]}/F^{\rm obs}_{\mathrm{2-10\ keV}}>1$) and \wise colors do not identify most of them as AGN.    Based on this small sample, we find that a higher fraction of these AGN are in the final merger stage ($<$10 kpc) than typical BAT AGN.  Additionally, these nine obscured AGN have, on average, $\approx 4\times$ higher accretion rates than other BAT-detected AGN ($\langle \lambda_{\mathrm{Edd}}\rangle=0.068\pm0.023$ compared to $\langle \lambda_{\mathrm{Edd}} \rangle=0.016\pm0.004$).  The robustness of SC at identifying Compton-thick AGN implies a higher fraction of nearby AGN may be Compton-thick ($\approx22\%$) and the sum of black hole growth in Compton-thick AGN (Eddington ratio times population percentage), is nearly as large as mildly obscured and unobscured AGN. 
\end{abstract}

\section{Introduction}

While there has been great progress understanding the origin of the cosmic X-ray background (CXB) and the evolution of active galactic nuclei (AGN) with \xmm and \chandra \citep[e.g.,][]{Brandt:2015:1} it is clear that a significant fraction of the $>$8~keV background is not produced by known 2-8~keV sources \citep{Worsley:2005:1281,Luo:2011:37,Xue:2012:129}. This background probably originates from a high column density, low redshift population ($z<1$). However, the source of the bulk of the CXB's surface brightness, peaking at $\approx$30~keV, is still unknown. The measurement of the space density and evolution of this population of highly absorbed AGN as well as the derivation of their column-density distribution function with luminosity and redshift is crucial for understanding the cosmic growth of black holes. Population-synthesis models attempt to explain the CXB by introducing appropriate numbers of absorbed Seyferts \citep[e.g.,][]{Treister:2005:115,Gilli:2007:79}. However, studies suggest that the number of Compton-thick AGN (\nh$>10^{24}$~\nhunit) is a factor of 3-4 smaller than expected in the population synthesis models \citep[e.g.,][]{Treister:2009:110}, at least in the local universe \citep{Ajello:2012:21}.  Additional studies suggest Compton-thick AGN evolve differently than other obscured sources and are more likely associated with rapid black hole growth at higher redshift \citep[e.g.,][]{Draper:2010:L99,Treister:2010:600}.  These problems limit our current knowledge of the origin of the CXB at $>$10~keV. \\

	In many well-studied objects, obscuration significantly attenuates the soft X-ray, optical, and UV signatures of AGN. There are only two spectral bands, the ultra-hard X-ray ($>$10~keV) and the mid-infrared (MIR, 5-50 $\micron$), where this obscuring material is optically thin up to high column densities \nh$<$10$^{24}$ \nhunit (Compton-thin).  Thus, these spectral bands are optimal for less-biased AGN searches \citep[e.g.,][]{Treister:2004:123,Stern:2005:163,Alexander:2008:835}. Radio selection of AGN is also largely obscuration independent, though only $\sim 10$\% of AGN are radio loud \citep[e.g.,][]{Miller:1990:207,Stern:2000:1526} and finding a radio excess in radio quiet AGN can be difficult because of the host galaxy contribution from star formation \citep{DelMoro:2013:A59} and significant free-free absorption absorption from the ionized torus \citep{Roy:2000:173}.   Mid-IR selection is very effective at identifying high-luminosity AGN, where the nuclear emission dominates, but moderate-luminosity AGN, like those common in the local universe, are harder to identify because the host galaxy contribution is relatively larger \citep[e.g., ][]{Cardamone:2008:130,Eckart:2010:584,Donley:2012:142,Stern:2012:30}.  In contrast, X-ray surveys suffer little contamination from non-nuclear emission at typical survey depths, and thus a hard X-ray survey can efficiently find both low and high luminosity AGN in a uniform fashion, including even the heavily obscured, lower luminosity AGNs which we expect to be important contributors to the CXB.  \\

	The Burst Alert Telescope \citep[BAT,][]{Barthelmy:2005:143}, a large field of view (1.4 steradian) coded aperture imaging instrument on the \swift satellite, has surveyed the sky to unprecedented depth. The all-sky BAT survey is a factor of $\approx$20 more sensitive than previous satellites such as \emph{HEAO 1} \citep{Levine:1984:581}. BAT selection is particularly powerful because it uses the 14--195~keV band which can pass through obscuring material of \nh$>10^{24}$ \nhunit,  though it is still biased against the most obscured AGN \citep[e.g., $>$$10^{25}$ \nhunit][]{Lanzuisi:2015:A120}.  It is therefore sensitive to most obscured AGN where even moderately hard X-ray surveys ($\sim$10~keV) are severely reduced in sensitivity.    The 70-month \swiftbat survey has identified 1210 objects of which 823 are AGNs while the rest are overwhelmingly Galactic in origin \citep{Baumgartner:2013:19}.  Higher angular resolution X-ray data for every source was obtained with \swift X-ray Telescope \citep[][]{Burrows:2005:165} because of the large positional uncertainty of \swiftbat ($\approx6\arcmin$) for fainter sources.    \\
	
	Unfortunately, due to the large number of sources spread across the sky and the limited sensitivity of \swiftxrt to obscured sources, X-ray follow-up and identification of the entire BAT catalog of $\approx$800 AGN has been difficult.  Survey programs typically used the first year or two of stacked data (e.g., 9 Month Survey, PI R. Mushotzky; Northern Galactic Cap 22-month Survey, PI N. Brandt).  After 10 years of the mission, one can detect many more obscured AGN which are critical to estimating the fraction of Compton-thick AGNs and the source of the CXB.  Additionally, the majority of sources had \swiftxrt coverage which is insufficient for measuring the column density (\nh) in heavily obscured AGNs \citep{Winter:2009:1322}.  Finally, accurately estimating column densities based on \swiftxrt and BAT data is problematic because of time variability and the low signal-to-noise of typical BAT detections \citep{Vasudevan:2013:111}.  \\

	To make progress in this area requires (1) an ultra-hard X-ray survey of sufficient sensitivity, angular resolution, and solid angle coverage at $\approx$30~keV to identify a large number of sources and (2) high-sensitivity observations to obtain the column density of the sources, their detailed X-ray spectral properties, and confirmations of their identifications.  With the new focusing optics on the {\it Nuclear Spectroscopic Telescope Array} \citep[{\it NuSTAR};][]{Harrison:2013:103}, the entire 3-79~keV energy range can be studied at sensitivities more than 100$\times$ better than those of previous coded aperture mask telescopes such as \swiftbat or {\it INTEGRAL}.  This enables detailed X-ray modeling of heavily obscured AGN \citep[e.g.,][]{Balokovic:2014:111,Brightman:2015:41,Gandhi:2014:117,Puccetti:2014:26,Arevalo:2014:81,Koss:2015:149,Bauer:2014:670}.  	\\
	  
	  In this article, we combine the all-sky nature of \swiftbat with the unprecedented \nustar sensitivity over a wide energy range to develop a new technique to find previously unknown heavily obscured AGNs.  The 10-100~keV spectral becomes increasingly curved with increasing absorption.  This is especially useful in selecting Compton-thick AGN because of its effectiveness up to very high column densities (\nh$\sim10^{25}$ \nhunit).  Additionally, detection based solely on spectral curvature (SC) offers an important test of AGN torus models and is less biased against Compton-thick AGNs.  In Section 2, we detail the \nustar sample and the spectral curvature selection. Section 3 describes the data reduction and analysis procedures for the \nustar observations.  Section 4 focuses on the results of SC on the full BAT sample of 241 nearby AGN ($z<0.03$), a subset of 84/241 \nustarsh-observed AGN, and the X-ray spectral and multiwavelength analyses for the nine \nustarsh-observed SC-selected AGN.  Finally, Section 5 gives a summary of our results and discusses the implications of the full survey in terms of the black hole growth.  Throughout this work, we adopt $\Omega_m$= 0.27, $\Omega_\Lambda$= 0.73, and $H_0$ = 71 km s$^{-1}$ Mpc$^{-1}$.  Errors are quoted at the 90\% confidence level unless otherwise specified.   

\section{Sample Selection}
Here we describe the simulations used to derive the SC measurement in \swiftbat and \nustar data (Section \ref{SC_sim}).  We then discuss the results of applying this SC measurement to nearby \swiftbat AGN (Section \ref{SC_bat}).  Finally, we select nine high SC BAT AGN for  \nustar follow-up.

	
\subsection{Spectral Curvature Measurement}	\label{SC_sim}
\begin{figure*}
\includegraphics[width=6cm]{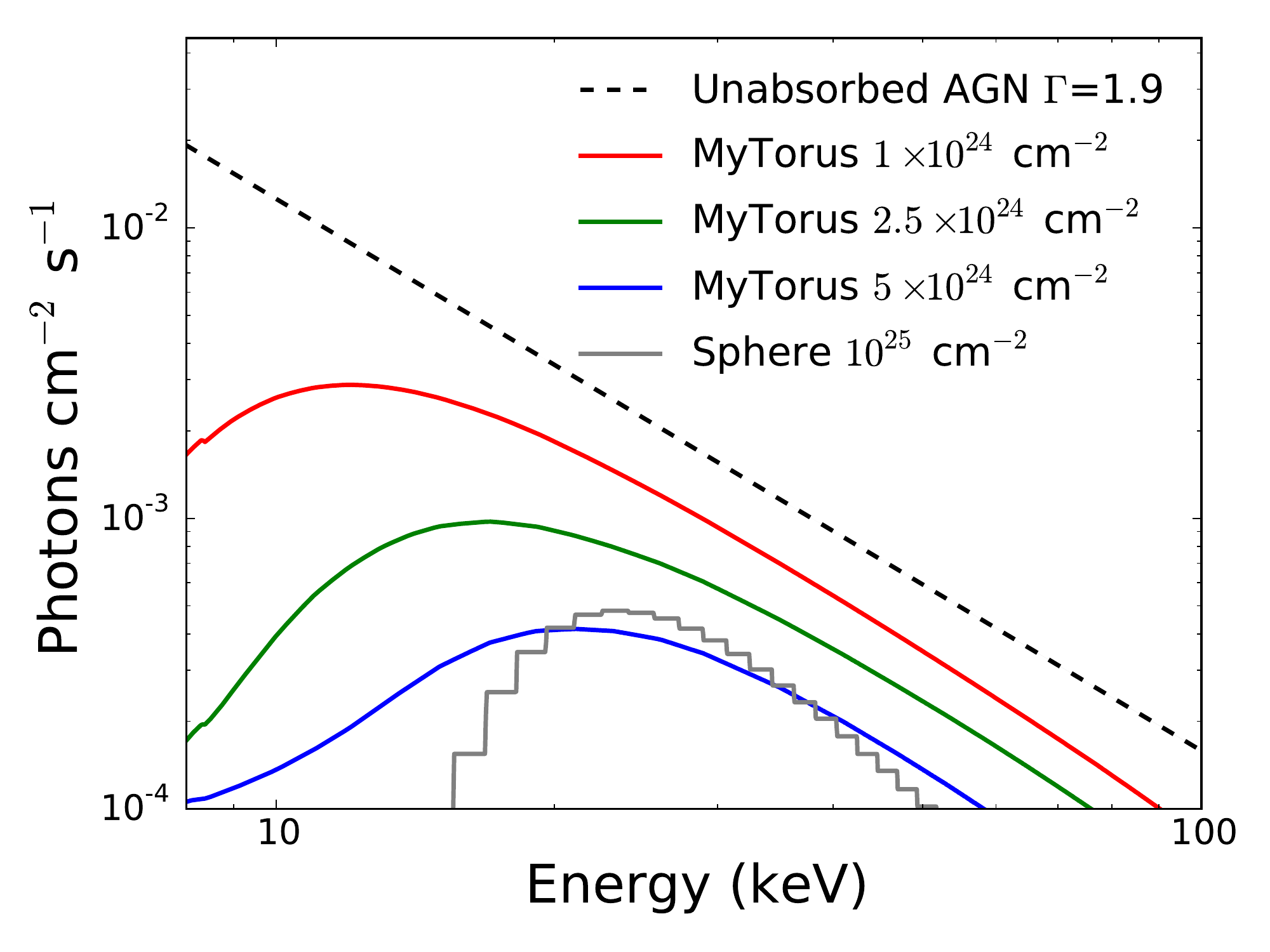}
\includegraphics[width=6cm]{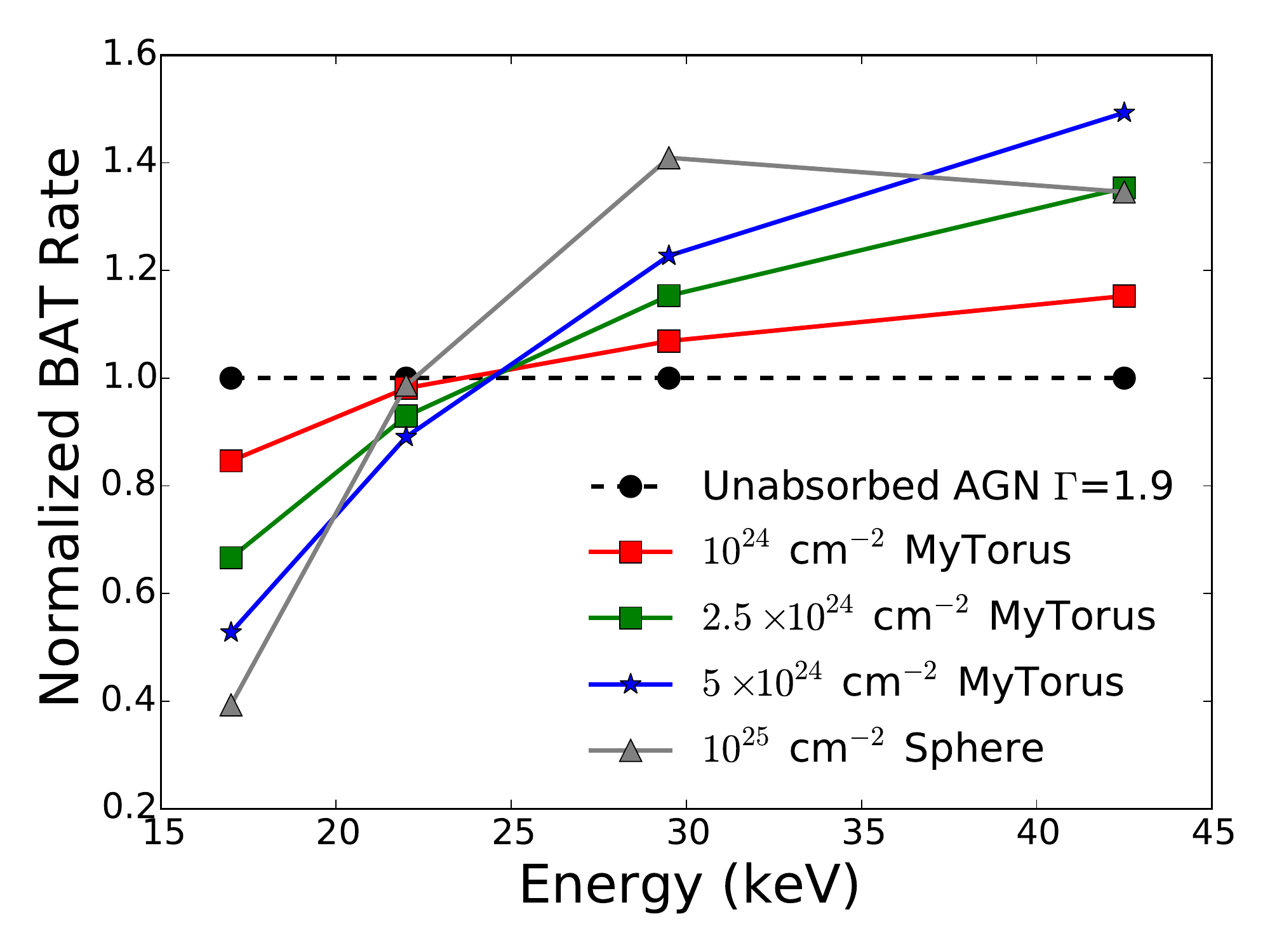}
\includegraphics[width=6cm]{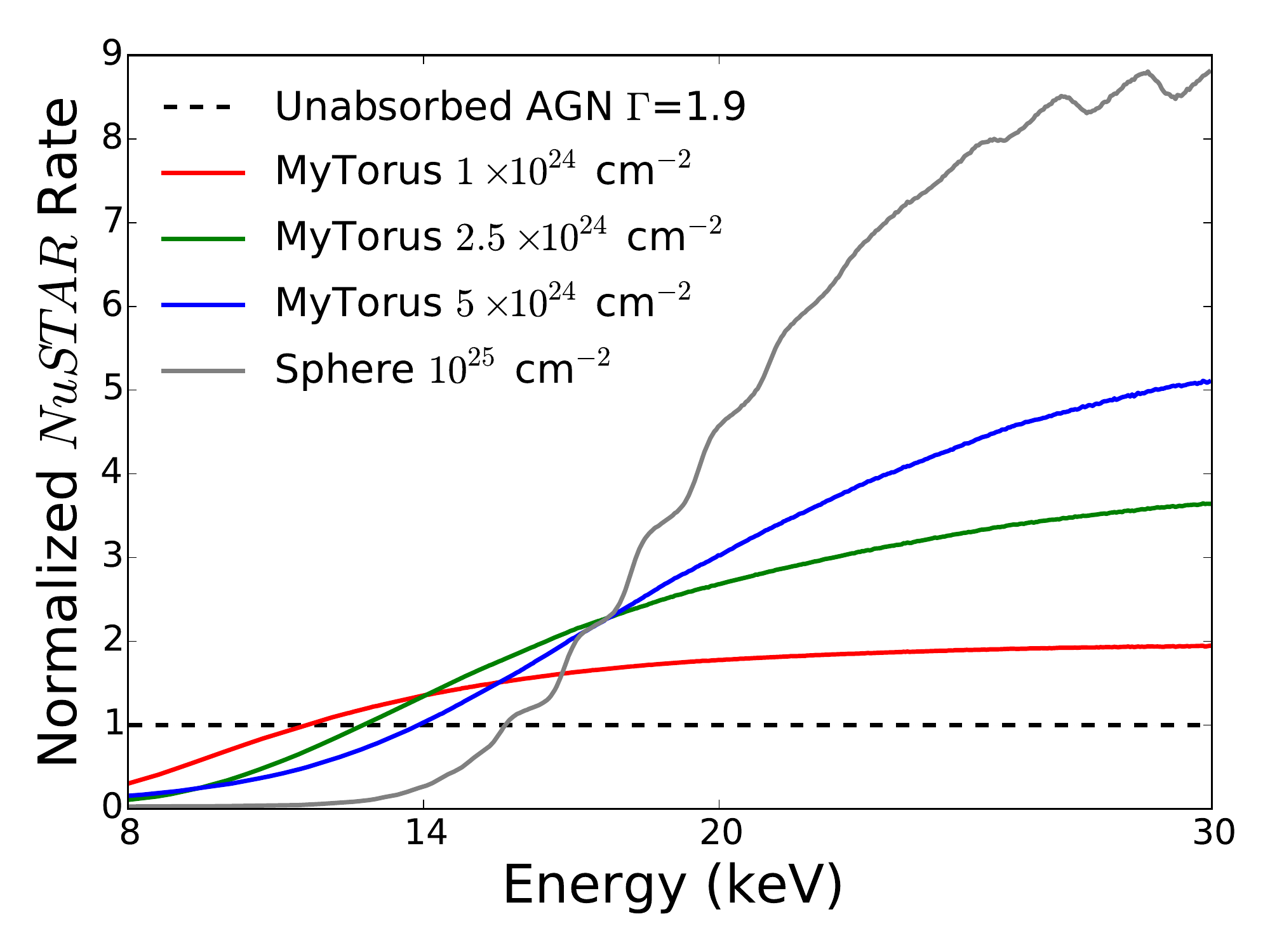}
\caption{\emph{Left}:  Simulated Compton-thick AGN compared to an unabsorbed power-law source with $\Gamma$=1.9 showing the increasing spectral curvature with column density.  \emph{Center}:  \swiftbat count rates for these same sources normalized by the rate of an unobscured source for the four BAT channels between 14-50~keV.  The energy bands above 24~keV show an excess while the bands between 14-24~keV show a decrement.   \emph{Right}: \nustar count rates for these same sources normalized by the rate of an unobscured source in the range 8-30~keV. At energies between 14-30~keV, a Compton thick source has an excess while the 8-14~keV energy band shows a decrement compared to the count rates of an unobscured source.  The weighted average of BAT and \nustar energy bands can be used to find Compton-thick sources.    }
\label{counts_torusfig}
\end{figure*}

	We define a curvature parameter to estimate Compton thickness using the distinctive spectral shape created by Compton reflection and scattering (Figure \ref{counts_torusfig}). We hereafter call it the spectral curvature (SC).   We generated the simulated data (using the {\sc Xspec} $\mathtt{fakeit}$ feature) from Compton-thick obscuration using a \mytorus model \citep{Murphy:2009:1549}.  \\

The \mytorus-based model used for calculating the SC (\modelSC hereafter) has the following form: 
\begin{flushright}
$\mathtt{Model\ SC} =\mathtt{MYTZ}\times \mathtt{POW} + \mathtt{MYTS} + \mathtt{MYTL}$
\end{flushright}

\noindent Here, $\mathtt{MYTZ}$ represents the zeroth-order transmitted
continuum ($\mathtt{POW}$) through photoelectric absorption and the
Compton scattering of X-ray photons out of the line-of-sight, $\mathtt{MYTS}$ is
the scattered/reflected continuum produced by scattering X-ray photons into
the line of sight, and $\mathtt{MYTL}$ is the fluorescent emission
line spectrum.  The torus model we use for measuring SC is viewed nearly edge-on ($\theta_{inc}=80$\degrees) with a cutoff power law ($\Gamma=1.9$, $E_c>200$~keV).  The SC model assumes a half opening angle of 60\degrees, which is equivalent to a covering factor of 0.5.    The SC is calculated so that a heavily Compton-thick source in an edge-on torus model has a value of one (e.g. SC=1 for \nh$=5 \times10^{24}$ \nhunit) and an unabsorbed AGN has a value of zero.

\begin{figure}	
\includegraphics[width=8.5cm]{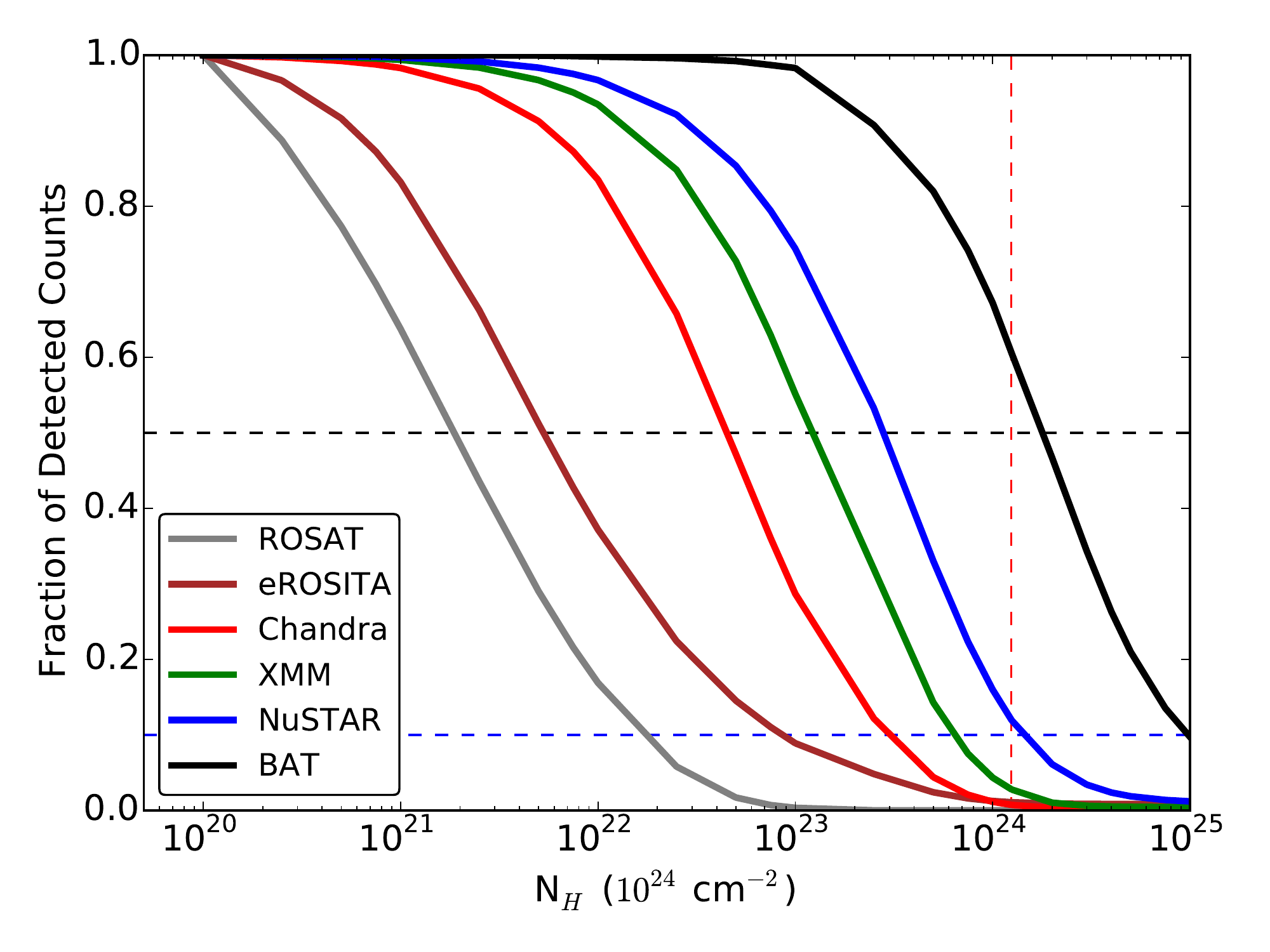}
\caption{Simulations using an edge-on MYTORUS model showing the reduction in count rates for instruments observing AGN of different column densities.  The black and blue horizontal dashed lines indicate a 50\% and 90\% reduction, respectively.   Surveys below 10~keV like \rosat and \erosita are strongly biased against detecting heavily obscured AGNs while \chandra and \XMM are heavily biased against detecting Compton-thick AGN.  While \nustar and \swiftbat are less biased against Compton-thick AGN, gainst Compton-thick AGN with very high column densities (\nh$>5 \times10^{24}$ \nhunit) because of the large reduction in count rates. }
\label{CTbiasfig}
\end{figure}

	The SC can be applied to X-ray observations from any satellite with energy coverage of the ``Compton hump" ($\approx$10-30~keV).  This ``hump" occurs because of the energy dependence of photoelectric absorption, whereby soft X-rays are mostly absorbed, and higher energy photons are rarely absorbed and tend to Compton scatter \citep[see e.g.,][]{Reynolds:1999:178}.  The SC measurement uses weighted averages of different energy ranges as compared to the count rate in an unobscured AGN.    To estimate the SC for BAT data, we focus on data below 50~keV because this shows the strongest difference in curvature compared to an unobscured source.  Additionally, the BAT sensitivity is significantly reduced in the 50-195~keV energy ranges.   For \nustarsh, we use the 8-14, 14-20, and 20-30~keV energy ranges, because of the reduced sensitivity of \nustar above 30~keV compared to the preceding energy ranges.   We note that the SC method will be biased against Compton-thick AGN with very high column densities (\nh$>5 \times10^{24}$ \nhunit) because of the large reduction in count rates (Figure \ref{CTbiasfig}).  
	
\begin{figure*}
\includegraphics[width=8.5cm]{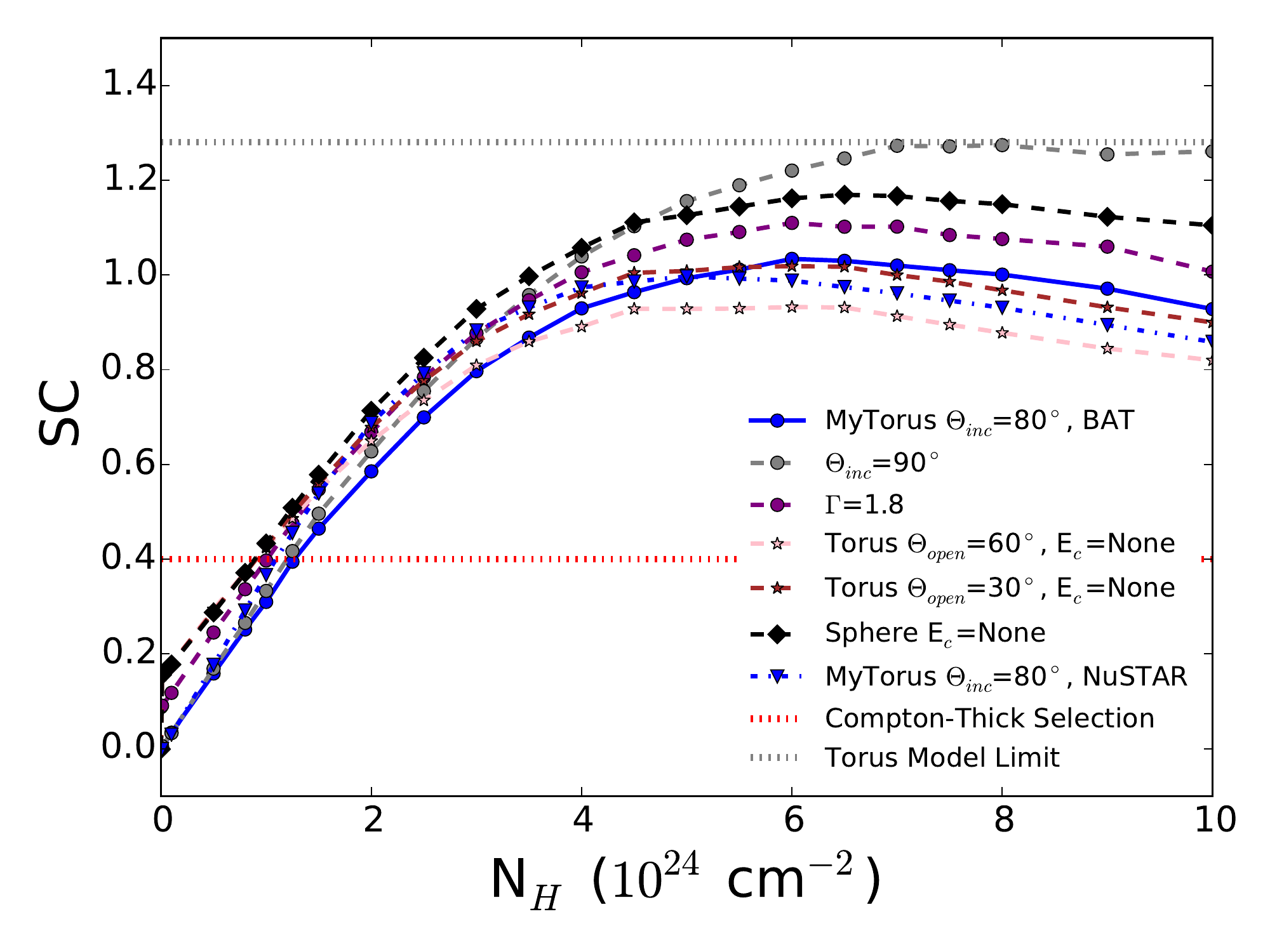}
\includegraphics[width=8.5cm]{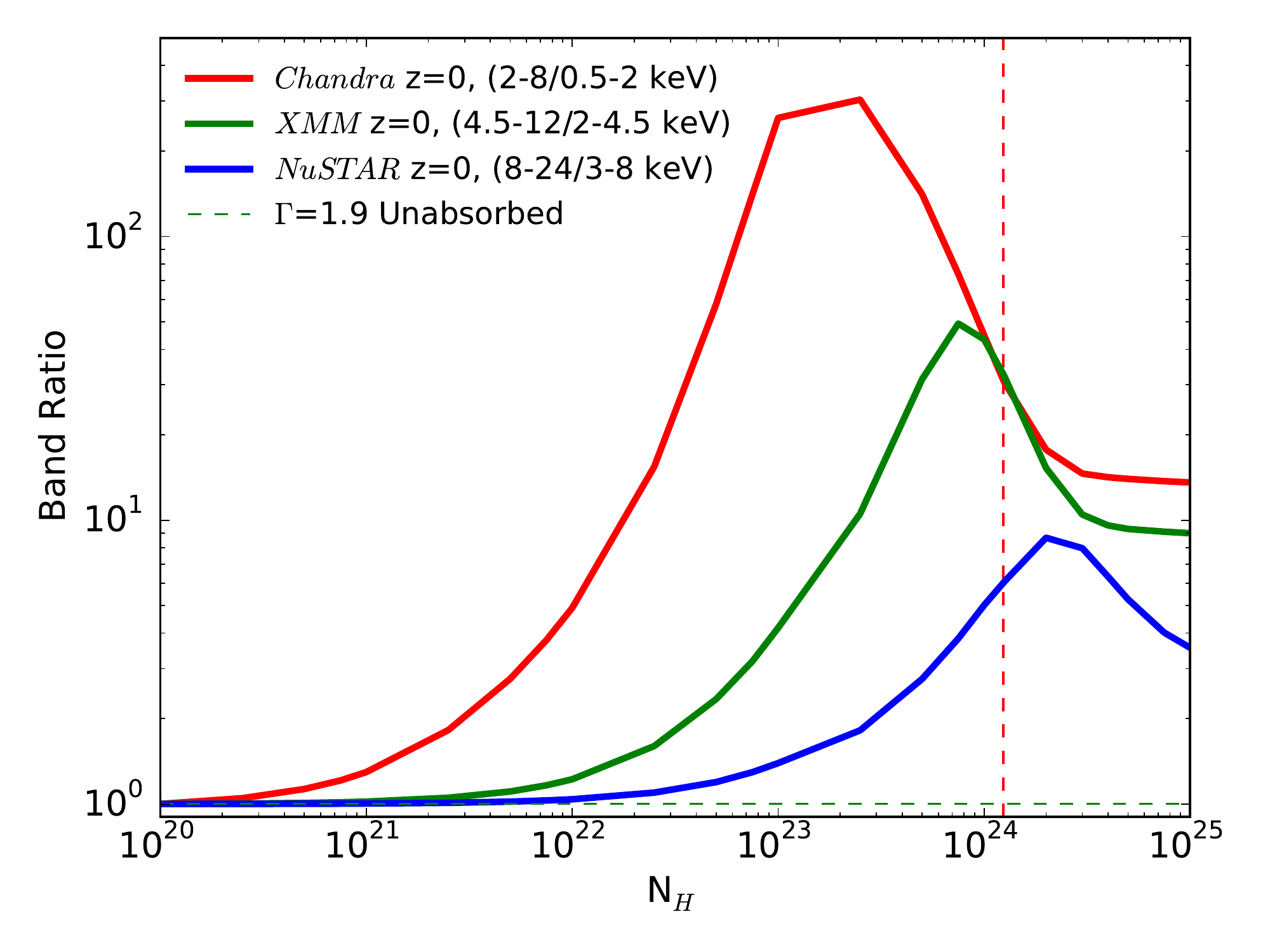}
\caption{{\em Top:} Variations in SC with column density based on \xspec simulations.  The blue solid line shows the \mytorus model used for the \scbat definition with $\Gamma$=1.9, $E_c=$200~keV, and $\theta_{inc}$=80$\degree$.  \scnustar (blue, dashed line) only shows small differences from \scbat ($\Delta$SC$<0.07$) at all column densities.  We also show the dependency of \scbat on inclination angle ($\theta_{inc}$, grey circle), intrinsic power law ($\Gamma$, purple circle), opening angle with \torus model \citep[$\theta_{open}$, pink and brown stars,][]{Brightman:2011:1206}, or a simple sphere model \citep[black diamonds,][]{Brightman:2011:1206}.  The differences between models are small ($\Delta$\scbat$<0.1$) except at very high column densities (\nh $>4\times10^{24}$ \nhunit).  At very high column densities (\nh $>4\times10^{24}$ \nhunit) the SC measure does not increase at larger column densities.   At low column densities (\nh $<5\times10^{23}$ \nhunit), the lack of the high energy cutoff in the \torus and sphere models raises the SC because of the additional flux in the high energy emission. A red dotted line shows the Compton-thick lower limit (\scbat=0.4) used for this study and a grey dotted line shows the upper limit from the torus model (\scbat=1.28).  {\em Right:}  Simulations of obscured sources using band ratios. The band ratio, unlike the SC method, is ineffective at selecting Compton-thick AGNs because of a degeneracy with sources at lower columns.  }
\label{ctcurve_modeldep}
\end{figure*}

The two SC equations take the following form:
{\small
\begin{eqnarray}
\footnotesize
\mathtt{SC_{BAT}}=\dfrac{-3.42\times A-0.82\times B+ 1.65\times C+ 3.58\times D}{ Total  \, Rate}
\end{eqnarray}
}
\noindent where A, B, C, and D refer to the 14-20, 20-24, 24-35, and 35-50~keV channel \swiftbat count rates, respectively and the total rate refers to the 14-50~keV total rate; and  
{\small
\begin{eqnarray}
\mathtt{SC_{NuSTAR}}=  \dfrac{-0.46\times E+0.64\times F+2.33\times G}{ Total \, Rate}
\end{eqnarray}
}
\noindent here E, F, and G refer to the \nustar 8-14, 14-20, and 20-30~keV on-axis count rates, respectively and the total rate refers to the 8-30~keV  total rate in the A and B telescope.  Simulations of the SC measurement with \swiftbat and \nustar as well as different model parameters and column densities are shown in Figure \ref{ctcurve_modeldep}.  The differences between SC measurements at a specific column density using a variety of different torus model parameters or using \nustar or \swiftbat are small except at very high column densities (\nh $>4\times10^{24}$ \nhunit).  The SC measure is more sensitive to sources that are mildly Compton-thick (\nh$\approx3\times10^{24}$ \nhunit) as can be seen by the flattening of the slope at very high column densities.  As \scbat only uses emission above 14~keV it is insensitive to differences between unobscured sources and mildly obscured sources (\nh$< 7\times10^{23}$ \nhunit) that fail to obscure any of the softest 14-20~keV emission.  Finally, we show simulations of broad band ratios which, unlike the SC measurement, are ineffective at selecting Compton-thick AGNs because of a degeneracy with sources at lower columns. \\





\subsection{SC of Nearby BAT AGN and NuSTAR Targets} \label{SC_bat}

We applied the SC to study Compton-thickness in a sample of nearby ($z<0.03$) BAT-detected AGN.  We use the 70-month catalog \citep[][]{Baumgartner:2013:19} with a low-luminosity cut ($L_{14-195\; \mathrm{keV}}>10^{42}$ \ergpersec) to avoid detecting purely star forming galaxies such as M82.  This luminosity limit corresponds to the 90\% BAT all-sky sensitivity limit at 100 Mpc ($z$=0.023), so that our survey sensitivity is more uniform in the volume explored.  This does exclude one lower luminosity ($L_{14-195\; \mathrm{keV}}<10^{42}$ \ergpersec) BAT-detected Compton-thick AGN (NGC 4102).  We also exclude four sources in the Galactic plane with heavy obscuration ($E[B-V]>1.2$, galactic latitude $\lvert b \rvert<10\degree$) because of the difficulty of multi-wavelength study, as well as eight BAT detections that are heavily contaminated by a secondary BAT source.   The final sample includes 241 BAT-detected nearby AGN.  Of these,  35\% (84/241) have been observed by \nustarsh. \\

\begin{figure*}	
\plotone{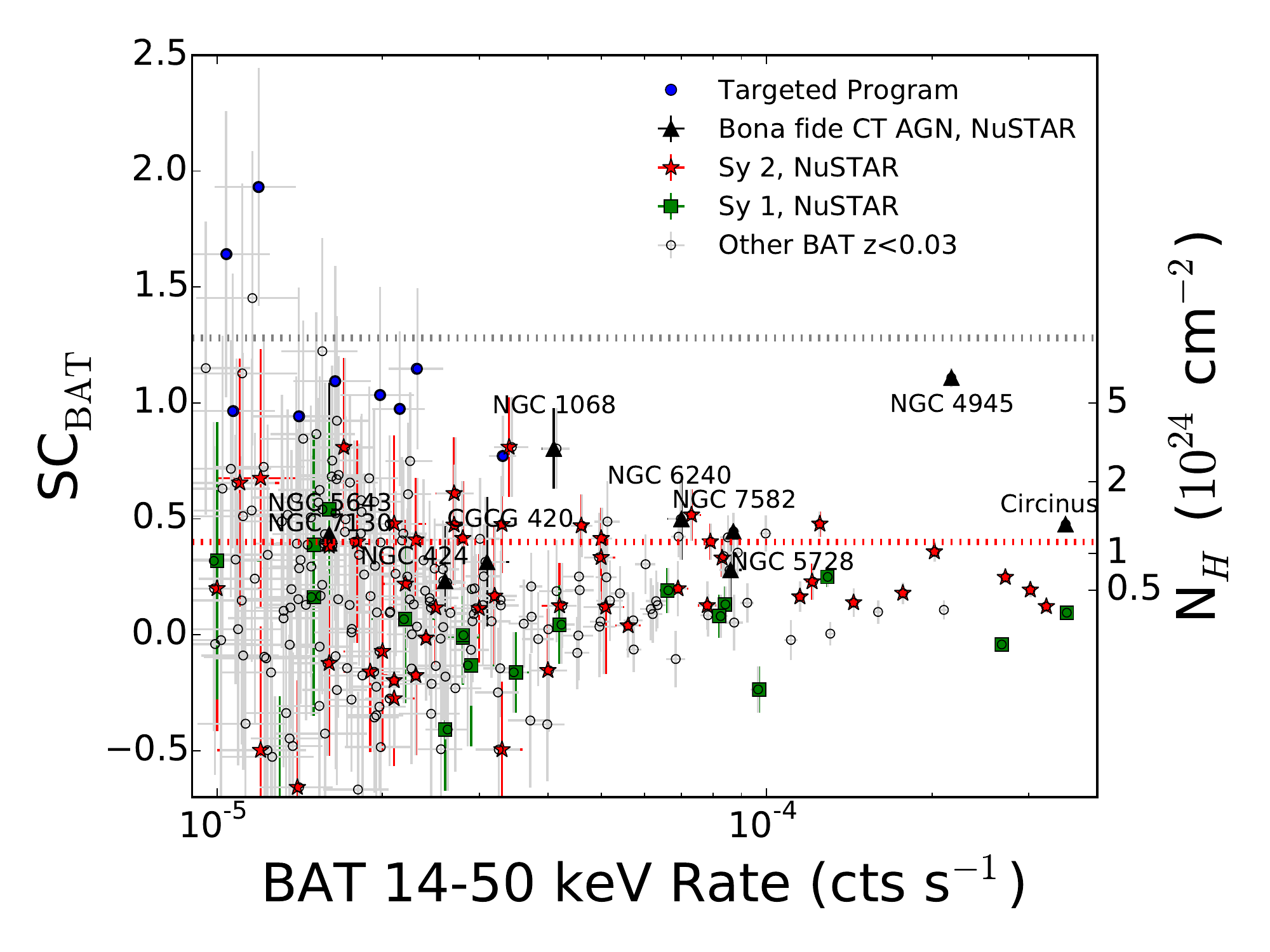}
\caption{Spectral curvature measurement for our sample of 241 BAT AGN at $z<0.03$. Compton-thick AGN confirmed with \nustar are shown in black, mildly obscured Seyfert 2s observed with \nustar are shown in red, and \nustarsh-observed Seyfert 1s are shown in green.  The simulation upper limits of any torus model are shown as a horizontal grey dotted line while a Compton-thick column is shown by a horizontal red dotted line.  The SC-selected \nustar targets in our program (blue) were selected to have the highest SC measure along with archival optical imaging and spectroscopy.  The majority of the \nustar program is at much fainter fluxes than previously known Compton-thick AGN.   }
\label{scbatall}
\end{figure*}

	We measure the SC in these 241 BAT AGNs (Figure \ref{scbatall}).   We have highlighted 10 ``bona fide" Compton-thick AGN in our sample that have been observed with \nustar and have been confirmed to be Compton-thick based on spectral fitting \citep{Gandhi:2014:117}.  We targeted nine northern hemisphere objects with archival optical imaging and spectroscopy from past studies \citep[e.g.,][]{Koss:2011:57,Koss:2012:L22} and very high SC values for \nustar follow-up.  Two of the nine targeted \nustar sources have measured curvatures above simulation upper limits. The majority of our sample is at much fainter fluxes than previously known Compton-thick AGN observed by \nustar \citep[e.g.,][]{Arevalo:2014:81,Balokovic:2014:111,Bauer:2014:670,Puccetti:2014:26,Brightman:2015:41}.  A list of likely Compton-thick AGN based on SC is found in Table \ref{scmeas}.\\

The brightest three targets from the \nustar SC program (NGC 3079, NGC 3393, and NGC 7212) have already been claimed to be Compton-thick, but we observed them for the first time with \nustar to confirm this.   NGC 3079 was suggested to be Compton-thick based on a large \feka equivalent width ($>$1~keV) and prominent emission above 10~keV with \bepposax \citep{Iyomoto:2001:L69}.   For NGC 3393, \bepposax observations in 1997 suggested columns of \nh$=3\times10^{23}$ \nhunit, but the large Fe K$\alpha$ line equivalent width ($>$1~keV), high ratio of \oiii to soft X-ray  flux, and a $>20$~keV excess suggested a Compton-thick AGN \citep{Salvati:1997:L1}.  \citet{Risaliti:2000:13} suggested that NGC 7212 is Compton-thick from analysis of a low signal-to-noise \ascash, spectrum, based on a flat continuum and a prominent \feka line.  \\

The remaining six targets (2MFGC 02280, CGCG 164-019, MCG +06-16-028, NGC 3588 NED0201, NGC 6232, and UGC 3157) have never been suggested to be Compton-thick in past literature. They were first observed with \swiftxrt for BAT counterpart identification.  2MFGC 02280 (SWIFT J0251.3+5441) was observed for 10.9 ks, with no counterpart detected above 3$\sigma$ in the BAT error circle \citep[][]{Baumgartner:2013:19}.  The remaining five sources were observed for 8-13 ks and confirmed to be the brightest X-ray source within the BAT error circle.   Most of these AGN are just above the BAT detection limit of 4.8$\sigma$ (NGC 3588 S/N=5.0, CGCG 164-019 S/N=5.1, MCG +06-16-028 S/N=6.1, NGC 6232 S/N=5.1, UGC 3157 S/N=5.4), with a more significant detection of 2MFGC 02280 (S/N=8.9).  Of these, only NGC 3588 NED02 was observed using \chandrash, in a study searching for dual AGN in close mergers \citep{Koss:2012:L22}.  

\begin{table*}[]
\centering
\caption{Likely Compton-Thick AGN using Spectral Curvature}
\scriptsize
\begin{tabular}{lccrrrr} 
\toprule \toprule
Object & Type\tablenotemark{a}&Sey.\tablenotemark{b}  &$z$ \tablenotemark{c} & \scbat  & \scnustar\tablenotemark{d}& BAT 14-50~keV\tablenotemark{e}\\

\bottomrule
NGC 6232&P&2&0.015&1.93$\pm$0.51&0.27$\pm$0.14&1.19E-05\\
CGCG 164-019&P&1.9&0.030&1.64$\pm$0.62&0.39$\pm$0.06&1.04E-05\\
ESO 406- G 004&\ldots&2&0.029&1.45$\pm$0.63&\ldots&1.16E-05\\
MCG +00-09-042&\ldots&2&0.024&1.22$\pm$0.49&\ldots&1.55E-05\\
NGC 3393&P&2&0.013&1.15$\pm$0.35&0.77$\pm$0.04&2.31E-05\\
ESO 323-32&\ldots&2&0.016&1.13$\pm$0.82&\ldots&1.11E-05\\
NGC 4945&CTB&2&0.002&1.11$\pm$0.04&1.01$\pm$0.01&2.17E-04\\
MCG +06-16-028&P&1.9&0.016&1.09$\pm$0.50&0.56$\pm$0.04&1.64E-05\\
UGC 3157&P&2&0.015&1.03$\pm$0.47&0.28$\pm$0.04&1.98E-05\\
2MFGC 02280&P&2&0.015&0.97$\pm$0.33&0.73$\pm$0.07&2.15E-05\\
NGC 3588 NED01&P&2&0.026&0.96$\pm$0.59&0.26$\pm$0.04&1.07E-05\\
NGC 7212 NED02&P&2&0.027&0.94$\pm$0.55&0.35$\pm$0.04&1.41E-05\\
NGC 1106&\ldots&2&0.015&0.92$\pm$0.45&\ldots&1.65E-05\\
ESO 565- G 019&\ldots&2&0.016&0.87$\pm$0.53&\ldots&1.52E-05\\
CGCG 229-015&\ldots&1.5&0.028&0.85$\pm$0.51&\ldots&1.43E-05\\
NGC 1194&\ldots&2&0.014&0.81$\pm$0.22&0.56$\pm$0.03&3.44E-05\\
2MASX J07262635-3554214&\ldots&2&0.029&0.81$\pm$0.39&0.37$\pm$0.03&1.71E-05\\
NGC 1068&CTB&2&0.004&0.80$\pm$0.17&0.66$\pm$0.02&4.15E-05\\
NGC 3079&P&1.9&0.004&0.77$\pm$0.17&0.87$\pm$0.04&3.31E-05\\
UGC 12282&\ldots&2&0.017&0.75$\pm$0.41&\ldots&1.62E-05\\
ESO 426- G002&\ldots&2&0.022&0.75$\pm$0.30&\ldots&2.25E-05\\
ESO 005- G 004&\ldots&2&0.006&0.61$\pm$0.24&\ldots&2.72E-05\\
UGC 12741&\ldots&2&0.017&0.61$\pm$0.31&\ldots&2.23E-05\\
MCG +04-48-002&\ldots&2&0.014&0.52$\pm$0.11&0.45$\pm$0.04&7.29E-05\\
NGC 6240&CTB&1.9&0.025&0.50$\pm$0.18&0.50$\pm$0.02&6.98E-05\\
Fairall 51&\ldots&1.5&0.014&0.49$\pm$0.18&\ldots&5.13E-05\\
Mrk 3&CTB&1.9&0.014&0.48$\pm$0.05&0.35$\pm$0.01&1.25E-04\\
Circinus Galaxy&CTB&2&0.001&0.48$\pm$0.03&0.69$\pm$0.01&3.50E-04\\
NGC 612&\ldots&2&0.030&0.47$\pm$0.13&0.39$\pm$0.03&4.61E-05\\
NGC 7582&CTB&2&0.005&0.45$\pm$0.08&0.39$\pm$0.02&8.71E-05\\
NGC 3281&\ldots&2&0.011&0.44$\pm$0.08&\ldots&9.96E-05\\
ESO 297-018&\ldots&2&0.025&0.42$\pm$0.09&\ldots&6.92E-05\\
NGC 3081&\ldots&2&0.008&0.42$\pm$0.09&\ldots&8.50E-05\\
NGC 1365&\ldots&2&0.006&0.40$\pm$0.08&0.26$\pm$0.01&7.92E-05\\
\bottomrule
ARP 318&\ldots&2&0.013&0.72$\pm$0.57&\ldots&1.22E-05\\
UGC 07064&\ldots&1.9&0.025&0.72$\pm$0.55&\ldots&1.06E-05\\
HE 1136-2304&\ldots&1.9&0.027&0.69$\pm$0.51&\ldots&1.66E-05\\
NGC 452&\ldots&2&0.017&0.68$\pm$0.43&\ldots&1.62E-05\\
NGC 7479&\ldots&1.9&0.008&0.68$\pm$0.40&\ldots&1.89E-05\\
MCG -01-30-041&\ldots&1.8&0.019&0.67$\pm$0.46&\ldots&1.65E-05\\
NGC 2788A&\ldots&\ldots&0.013&0.66$\pm$0.38&\ldots&1.75E-05\\
CGCG 122-055&\ldots&1.5&0.021&0.63$\pm$0.66&\ldots&1.02E-05\\
Mrk 1310&\ldots&1.5&0.019&0.62$\pm$0.49&\ldots&1.54E-05\\
NGC 7465&\ldots&2&0.007&0.59$\pm$0.52&\ldots&1.39E-05\\
NGC 1125&\ldots&2&0.011&0.58$\pm$0.36&\ldots&1.83E-05\\
NGC 3035&\ldots&1.5&0.015&0.57$\pm$0.40&\ldots&1.93E-05\\
ESO 553- G 043&\ldots&2&0.028&0.57$\pm$0.45&\ldots&1.53E-05\\
NGC 3786&\ldots&1.9&0.009&0.54$\pm$0.38&0.09$\pm$0.05&1.55E-05\\
ESO 317- G 038&\ldots&2&0.015&0.54$\pm$0.65&\ldots&1.16E-05\\
UGC 11397&\ldots&2&0.015&0.53$\pm$0.42&\ldots&1.83E-05\\
ESO 374- G 044&\ldots&2&0.028&0.52$\pm$0.46&\ldots&1.64E-05\\
ESO 533- G 050&\ldots&2&0.026&0.51$\pm$0.70&\ldots&1.11E-05\\
PKS 2153-69&\ldots&2&0.028&0.51$\pm$0.48&\ldots&1.48E-05\\
NGC 7679&\ldots&1.9&0.017&0.50$\pm$0.44&\ldots&1.71E-05\\
Mrk 622&\ldots&1.9&0.023&0.49$\pm$0.55&\ldots&1.31E-05\\
UGC 03995A&\ldots&2&0.016&0.48$\pm$0.38&0.36$\pm$0.02&2.06E-05\\
NGC 7319&\ldots&1.9&0.023&0.48$\pm$0.21&0.40$\pm$0.07&3.29E-05\\
MCG -01-05-047&\ldots&2&0.017&0.47$\pm$0.26&0.28$\pm$0.03&2.71E-05\\
Mrk 590&\ldots&1.5&0.026&0.44$\pm$0.41&\ldots&1.71E-05\\
NGC 6552&\ldots&2&0.027&0.44$\pm$0.34&\ldots&1.83E-05\\
ESO 549- G 049&\ldots&1.9&0.026&0.44$\pm$0.31&\ldots&2.18E-05\\
NGC 5643&CTB&2&0.004&0.43$\pm$0.65&0.47$\pm$0.05&2.18E-05\\
2MASX J07394469-3143024&\ldots&2&0.026&0.42$\pm$0.25&0.25$\pm$0.02&2.80E-05\\
NGC 4992&\ldots&2&0.025&0.42$\pm$0.13&0.30$\pm$0.02&5.02E-05\\
CGCG 367-009&\ldots&2&0.027&0.41$\pm$0.27&0.10$\pm$0.03&2.32E-05\\
2MASX J18263239+3251300&\ldots&2&0.022&0.40$\pm$0.44&0.07$\pm$0.01&1.78E-05\\
NGC 7130&CTB&1.9&0.016&0.40$\pm$0.55&0.86$\pm$0.08&1.56E-05\\
\bottomrule
CGCG 420-015&CTB&1.9&0.029&0.31$\pm$0.28&0.44$\pm$0.03&3.08E-05\\
NGC 5728&CTB&2&0.009&0.28$\pm$0.14&0.61$\pm$0.02&8.56E-05\\
NGC 424&CTB&1.9&0.012&0.23$\pm$0.25&0.58$\pm$0.05&2.58E-05\\
2MASX J00253292+6821442&\ldots&2&0.012&-0.1$\pm$0.40&0.56$\pm$0.05&1.59E-05
\end{tabular}
\begin{minipage}[l]{1\textwidth}
NOTE. -- List of likely Compton-thick AGN based on spectral curvature from our BAT sample of 241 $z<$0.03 AGN.  The errors indicated here are 1$\sigma$.  The upper section lists any likely Compton-thick BAT AGN (34/241; \scbat$>$0.4 and an error lower bound of \scbat$>0.3$).  The middle section lists the remaining sources with \scbat$>$0.4, which are less likely to be Compton-thick (33/241) because of the large error bars. Finally,  the bottom section shows four sources with SC$>$0.4 in \nustarsh, but \scbat$<$0.4.   \end{minipage}
\footnotetext[1]{Type of AGN, here P=SC-selected BAT AGN observed in our \nustar program, CTB=``bona fide" Compton-thick AGN confirmed to be Compton-thick based on spectral fitting \citep{Gandhi:2014:117}, U=above upper limit in \scbat for torus simulations.}
\footnotetext[2]{AGN type based on optical spectra from Koss et al. (2015, in prep) or NED.}
\footnotetext[3]{Measured redshift from NED.}
\footnotetext[4]{Where ellipses are shown, no public \nustar observations exist.}
\footnotetext[5]{BAT 14-50~keV total count rate in cts s$^{-1}$.  Total count rate errors are small ($<3\times10^{-6}$ cts s$^{-1}$).}
\label{scmeas}
\end{table*}

\section{Data and Reduction}
\subsection{NuSTAR}
  Table \ref{obs_table} provides details, including dates and exposure times, for the nine \nustar observations of SC-selected BAT AGN.  We analyzed these sources, as well as 75 other low redshift ($z<0.03$) AGNs in the \nustar public archive, for a total of 84 \nustar observations.  We have chosen not to include the \swiftbat data in the model fits because our selection method will bias our fits to Compton-thick obscuration and the \swiftbat data were taken over a period of six years (2004-2010).\\
  
  The raw data were reduced using the {\tt NuSTARDAS} software package (version 1.3.1) jointly developed by the ASI Science Data Center (ASDC) and the California Institute of Technology. {\tt NuSTARDAS} is distributed with the HEAsoft package by the NASA High Energy Astrophysics Archive Research Center (HEASARC).  We extracted the \nustar source and background spectra using the {\tt nuproducts} task included in the {\tt NuSTARDAS} package using the appropriate response and ancillary files.  Spectra were extracted from circular regions 40$\arcsec$ in radius, centered on the peak of the centroid of the point-source images. The background spectra were extracted from a circular region lying on the same detector as the source.   We also applied the same reduction procedure to other 75 low redshift ($z<0.03$) \nustar observed AGNs in the public archive for \scnustar measurements of a total of 84 \nustar observed nearby BAT-detected AGNs.

\begin{table*}[]
\centering
\caption{X-ray Observation Log}
\scriptsize
\begin{tabular}{lcccccccc} 

\toprule\toprule 
\multicolumn{3}{l}{} & \multicolumn{3}{c}{\nustar Observations} &
\multicolumn{3}{c}{XRT Observations} \\
\noalign{\smallskip}
\cmidrule(rl){4-6} \cmidrule(rl){7-9} 
\noalign{\smallskip}
\multicolumn{1}{c}{Object Name} & BAT ID  & $z$  & Observation ID & UT Date &
 $t_{\rm eff}$ & Observation ID & UT Date & $t$ \\
\multicolumn{1}{c}{(1)} & (2) & (3) & (4) & (5) & (6) & (7) & (8) &
(9)\\
\toprule
2MFGC 02280 & SWIFTJ0251.3+5441 & 0.0152 & 60061030002 & 2013-02-16 & 15 & 00080255001 & 2013-02-16 & 6 \\
CGCG 164-019 & SWIFTJ1445.6+2702 & 0.0299 & 60061327002 & 2013-09-13 & 24 & 00080536001 & 2013-09-13 & 6 \\
MCG +06-16-028 & SWIFTJ0714.2+3518 & 0.0157 & 60061072002 & 2013-12-03 & 23 & 00080381001 & 2013-12-03 & 7 \\
NGC 3079 & SWIFTJ1001.7+5543 & 0.0037 & 60061097002 & 2013-11-12 & 21 & 00080030001 & 2013-11-12 & 6 \\
NGC 3393 & SWIFTJ1048.4-2511 & 0.0125 & 60061205002 & 2013-01-28 & 15 & 00080042001 & 2013-01-28 & 7 \\
NGC 3588 NED02 & SWIFTJ1114.3+2020 & 0.0262 & 60061324002 & 2014-01-17 & 23 & 00080533001 & 2014-01-17 & 5 \\
NGC 6232 & SWIFTJ1643.2+7036 & 0.0148 & 60061328002 & 2013-08-17 & 18 & 00080537001 & 2013-08-17 & 6 \\
NGC 7212 NED02 & SWIFTJ2207.3+1013 & 0.0267 & 60061310002 & 2013-09-01 & 24 & 00080283001 & 2013-09-02 & 3 \\
UGC 3157 & SWIFTJ0446.4+1828 & 0.0154 & 60061051002 & 2014-03-18 & 20 & 00041747001 & 2010-10-22 & 10 \\
\toprule
\end{tabular}
\begin{minipage}[l]{1\textwidth}
NOTE. -- (1): Full NED object name for BAT counterpart. 
(2): \swiftbat name.
(3): Redshift. 
(4) and (5): \nustar observation ID and start date (YYYY-MM-DD), respectively.
(6): Effective exposure time (ks). This is the net value after data cleaning and correction for vignetting. 
(7) and (8): \swiftxrt observation ID and start date (YYYY-MM-DD), respectively. 
(9): Net on-axis, flaring-corrected exposure time (ks).
\end{minipage}
\label{obs_table}
\end{table*}

\subsection{Soft X-ray Observations of SC-selected AGN}
Most \nustar observations were accompanied by a short observation (3-7 ks) with \swiftxrt within 24 hr, although for one source, UGC 3157, the only available observation was from 4 years earlier.  These observations provided mostly simultaneous coverage in the soft X-rays ($<3$~keV) where \nustar is not sensitive.   All the \swiftxrt data were collected in Photon Counting (PC) mode.  We built \swiftxrt spectra using the standard point source processing scripts from the UK \swift Science Data Centre in Leicester \citep{Evans:2009:1177}.  Table \ref{obs_table} provides the complete list of observations. The \swiftxrt observations of 2MFGC 02280 and NGC 3588 NED02 did not yield a detection below 3~keV, so we use them here only to place an upper limit on the soft X-ray emission. \\

	In addition to the \nustar and \swiftxrt data, there are archival \xmm (for NGC 3079 and NGC 7212 NED02) and \chandra data available (for NGC 3393 and NGC 3588 NED02).  Studies of NGC 3079, NGC 7212, and NGC 3393 found no signs of variability \citep{Hernandez-Garcia:2015:A90, Koss:2015:149}, so we use these spectra because of their much higher sensitivity than \swiftxrtsh.  For NGC 3588 NED02, the source is very faint with a total of 9 counts, all above 3~keV, in the \swiftxrt observation,  compared to 72 counts in the \chandra observation.  We fit the \chandrash, \nustarsh, and \swiftxrt spectra between 3 and 8~keV with a power law and a cross-normalization factor for the \chandra data.  We find \chandra is consistent with no variability (1.2$\pm$0.5).  \\
		
	We processed the \xmm observations using the Science Analysis Software (SAS), version 13.5.0, with the default parameters of $\mathtt{xmmextractor}$.  NGC 3079 had two \xmm observations, which we combined using $\mathtt{epicscombine}$ for a total exposure time of 9.4 ks after filtering.  After filtering, NGC 7212 had an exposure time of 9.3 ks.  We also used $\mathtt{epicscombine}$ to combine the MOS1 and MOS2 instruments into a single spectrum before fitting.  We reduced and combined the two \chandra observations of NGC 3393, with a total exposure of 99.9 ks, and NGC 3588 NED02, with a total exposure of 9.9 ks, following \citet{Koss:2015:149}.

\subsection{X-ray Spectral Fitting}

For the three brightest sources with \xmm and \chandra data, NGC 3079, NGC 7212 NED02, and NGC 3393, we binned to a minimum of 20 photons per bin using the $\mathtt{HEAsoft}$ task $\mathtt{grppha}$.  We use $\mathtt{statistic\ cstat}$ \citep{Wachter:1979:274} in {\sc Xspec} for the remaining six sources, which is more appropriate than  $\chi^2$ in the case of Poisson distributed data \citep{Nousek:1989:1207}. In the case of unmodeled background spectra, $\mathtt{cstat}$ applies the $W$~statistic\footnote{See also \url{http://heasarc.gsfc.nasa.gov/docs/xanadu/xspec/wstat.ps}}.  While the $W$~statistic is intended for unbinned data, bins containing zero counts can lead to erroneous results\footnote{See \url{https://heasarc.gsfc.nasa.gov/xanadu/xspec/manual/XSappendixStatistics.html}}, so we group the \swiftxrt and \nustar data by a minimum of $3$ counts per bin, respectively \citep[e.g.,][]{Wik:2014:48} using the $\mathtt{HEAsoft}$ task $\mathtt{grppha}$. \\

We let the \swiftxrt and \nustar FPMA and FPMB cross-normalizations vary independently within 5\% based on the most recent calibrations \citep{Madsen:2015:1672}.  We also use varying cross-normalizations of 0.93$\pm$0.05 for \chandrash, 1.05$\pm$0.05 for \xmm pn, and 1.02$\pm$0.05 for the combined MOS observation based on recent \nustar calibrations. \\

Conventional spectral fitting and error estimation can sometimes underestimate the likely range of model parameters.  Additionally, several past studies have found degeneracies between photon index ($\Gamma$) and column density (\nh) for \nustar observed Compton-thick AGN \citep[e.g.,][]{Gandhi:2014:117}.  We therefore use Markov chain Monte Carlo (MCMC) methods built into XSPEC for error estimation.   We use the default Goodman-Weare algorithm to sample parameter space, constructing a chain of parameter values.  The algorithm works by holding multiple sets of parameters, called walkers, for each step in the chain and generating walkers for the next step using those from the current step.

\subsection{X-ray Spectral Modeling}\label{specmod}

For spectral modeling, we follow the strategies of past studies of single AGN observed with \nustar \citep[e.g, ][]{Gandhi:2014:117} and use the \mytorus and \torus models.   Since the soft X-ray data ($<$3~keV) from \swiftxrt typically have a small number of counts ($\approx0-30$) and therefore lack the statistics to model this emission in detail, we use two components corresponding to scattered AGN emission on larger scales in the host galaxy and a thermal plasma.  In the absence of high signal-to-noise and spectral resolution soft X-ray data, these are only meant as a simple prescription to describe the spectral shape in this regime.  The scattered emission was simulated using a single power-law (of photon-index $\Gamma$ and normalization tied to that of the AGN) and with a scattering fraction, $f_{scatt}$, relative to the intrinsic power-law.  We also include a low-energy component with a thermal plasma component \citep[$\mathtt{APEC}$;][]{Smith:2001:L91}, fixing the abundance to solar, similar to past studies \citep[e.g.,][]{Guainazzi:1999:10}.\\

The \mytorus-based model (\modelM hereafter) has the following form: 
\begin{flushright}
$\mathtt{Model\ M} = \mathtt{PHABS}\times (\mathtt{MYTZ}\times \mathtt{POW} + \mathtt{f_{refl}}*(\mathtt{MYTS} + \mathtt{MYTL}) + \mathtt{f_{scatt}}*\mathtt{POW}+\mathtt{APEC})$
\end{flushright}
The common parameters of
$\mathtt{MYTZ}$, $\mathtt{MYTS}$, and $\mathtt{MYTL}$ (\nh and $\theta_{\rm inc}$) are
tied together.  \nh is defined along the equatorial plane of the torus.   There is a fitted constant ($\mathtt{f_{refl}}$) between the zeroth-order continuum and the scattered/reflected and fluorescent emission-line spectrum.   The intrinsic (unprocessed) photon indices and normalizations are tied to those of the
zeroth-order continuum ($\mathtt{POW}$).   The torus opening angle ($\theta_{\rm tor}$) is fixed at $60\degree$ in the current version of \mytorus. \\
	

Our second choice of physically motivated model (\modelT hereafter) uses the \torus model, and has the following form:
\begin{flushright}
$\mathtt{Model\ T} = \mathtt{PHABS}\times (\mathtt{BNTORUS} + \mathtt{f_{scatt}}*\mathtt{POW}+ \mathtt{APEC})$.
\end{flushright}
\torus self-consistently includes photoelectric absorption, Compton-scattering, and fluorescent line emission due to the obscuration of an intrinsic power-law continuum by a biconical torus \citep{Brightman:2011:1206}, so no $\mathtt{f_{refl}}$ factor is used in this model. \nh is defined along the line of sight, and is independent of $\theta_{\rm inc}$.

\section{Results}
Here we discuss our results measuring the spectral curvature for the full BAT sample of 241 nearby AGN as well as a subset of 84/241 \nustarsh-observed nearby BAT AGN (Section \ref{SC_nustar}).  We then discuss results for the nine \nustarsh-observed SC-selected AGN focusing on simple and complex X-ray spectral fits (Section \ref{xrayspec}).  Next, we examine other measurements of Compton-thickness using the mid-IR and \oiii emission (Section \ref{sedchapt}).  We conclude with a comparison of the host galaxy properties and accretions rates compared to all BAT-detected nearby AGN (Section \ref{hostgal}).

\subsection{Spectral Curvature} \label{SC_nustar}
	The SC measure may be used to measure the overall fraction of Compton-thick sources and whether the curvature of these sources is consistent with torus models.  Using \scbat about $28\pm5\%$ (67/241) of AGN fall within the Compton-thick region (Figure \ref{scbatall}).  However, the large error bars of \scbat at low fluxes ($<3\times10^{-5}$ \ctspersec) may overestimate this fraction because of the larger number of sources below the limit.  A more conservative estimate using only bright sources ($>3\times10^{-5}$ \ctspersec) finds a Compton-thick fraction of $22\pm8\%$ (18/79), which is lower, but not statistically different than the fraction for the whole sample.  The \scbat identifies most well-known Compton-thick AGN in the ``bona fide" sample (7/10), with the remaining three identified within their 1$\sigma$ error.    \\
	
	We can also estimate the local Compton-thick space density.  Assuming a luminosity threshold of \Lsoft$>10^{43}$ \ergps and a conversion factor between 14-195~keV and 2-10~keV of 2.67 \citep{Rigby:2009:1878}, we find 74 BAT AGN. We estimate 16 are Compton-thick (\scbat$>0.4$). Given the sample is within $z<0.03$ excluding the 10 degrees within the Galactic plane and confused sources, and the 91\% completeness at this luminosity threshold  \citep{Baumgartner:2013:19}, this implies a volume $7.0\times10^6$ Mpc$^{3}$.   The Compton-thick number density is therefore $2.3\pm0.3\times10^{-6}$~Mpc$^{-3}$ above \Lsoft$>10^{43}$~\ergps.\\ 
	
\begin{figure}
\includegraphics[width=8.5cm]{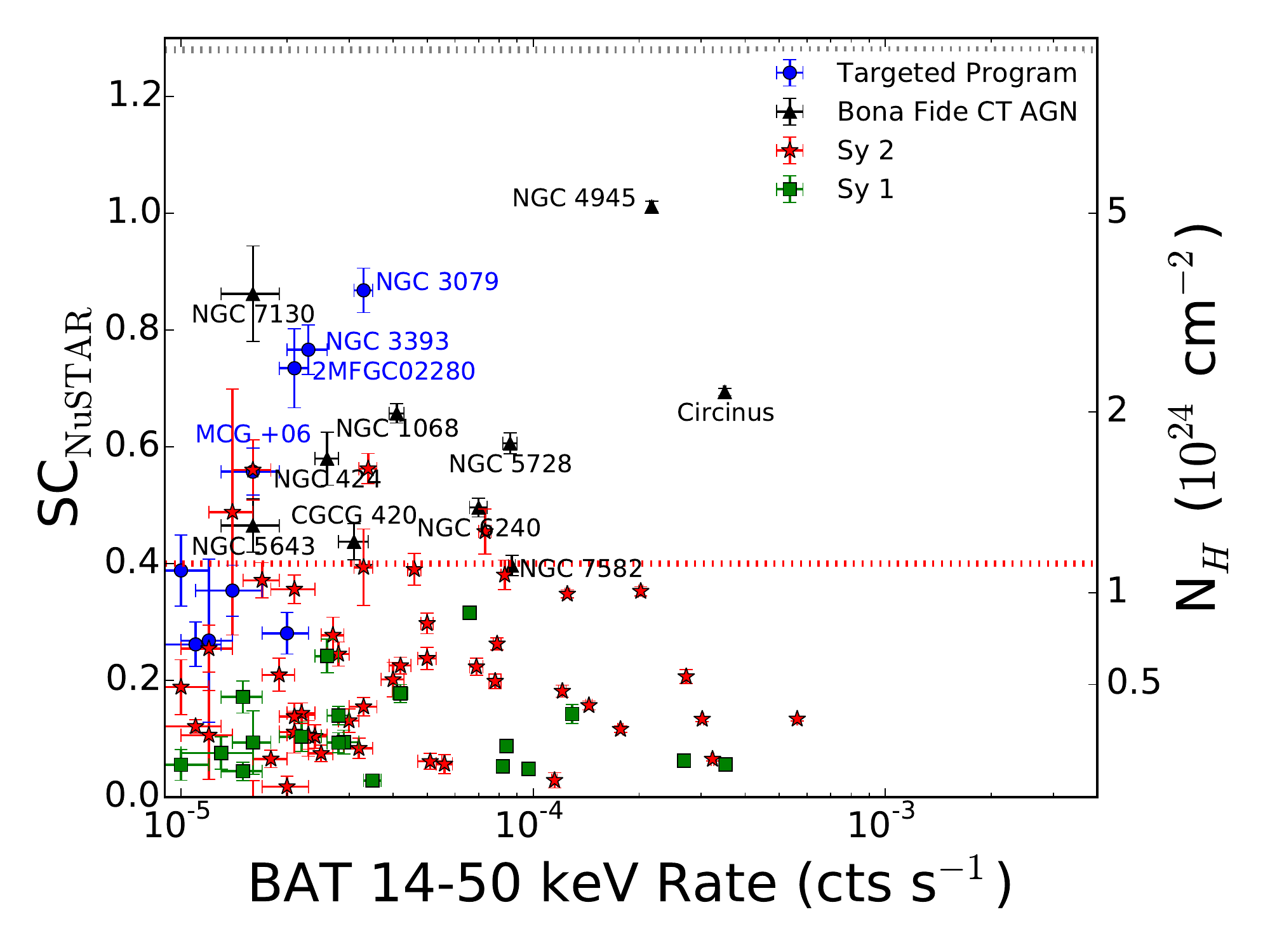}
\includegraphics[width=8.5cm]{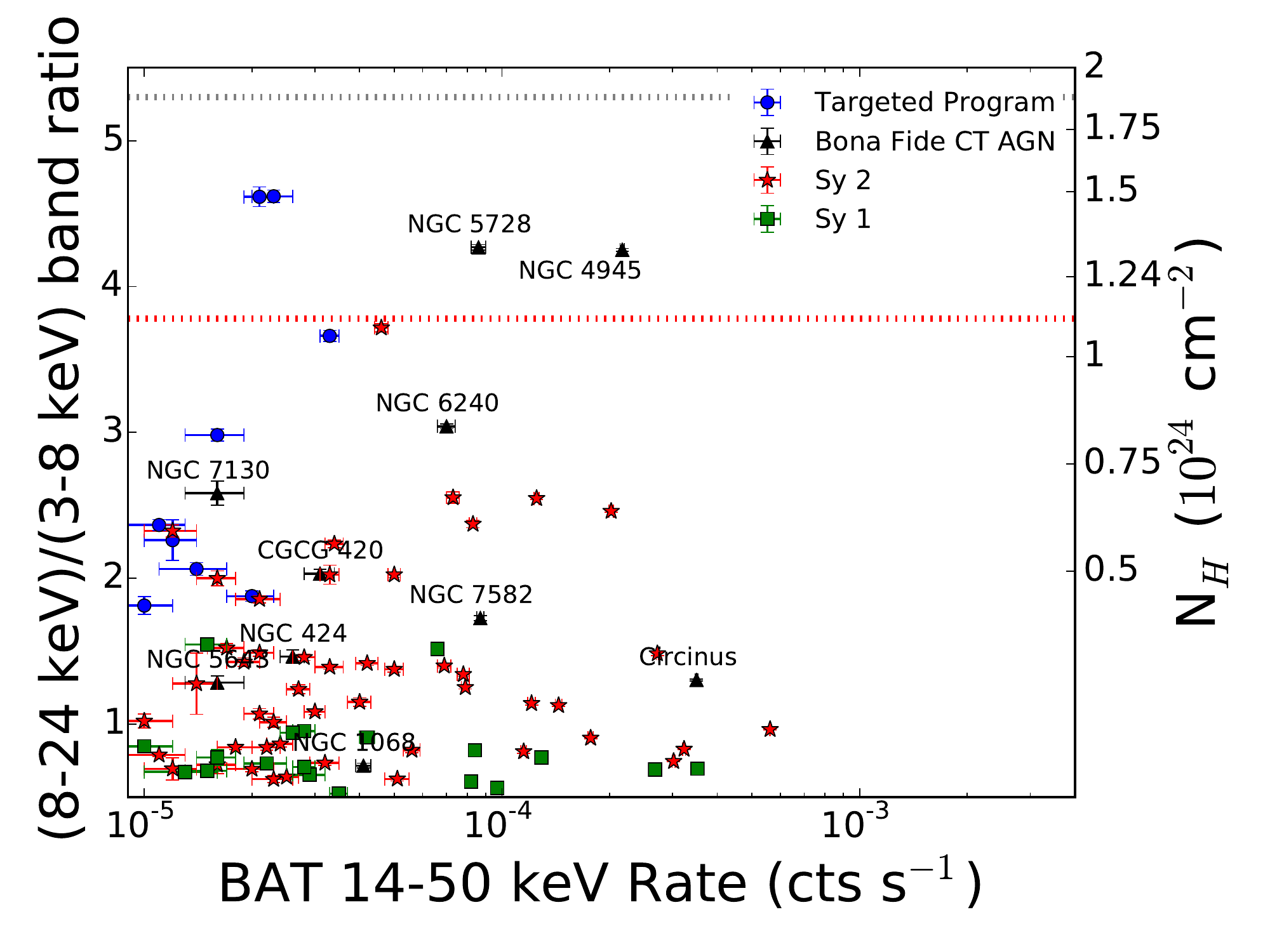}
\caption{\emph{Top:} Plot of the \scnustar measurement for nearby ($z<0.03$) BAT AGN.  The nine BAT AGN targeted based on 14-50~keV curvature are shown in blue. Well known ``bona fide" Compton-thick AGN observed with \nustar are shown in black.  The simulation limits for an edge-on torus are shown as a grey dotted line while lower limit of a Compton-thick column is shown by a red dotted line.  \emph{Bottom:}  Band ratios of BAT AGN observed with \nustar (8-24~keV/3-8~keV).     }
\label{nustar_curv}
\end{figure}

The higher sensitivity of \nustar allows a more precise SC measurement than \swiftbat because of the higher sensitivity.  We measure \scnustar for those nearby AGN with \nustar observations (35\%, 84/241, Figure \ref{nustar_curv}).   While the SC measurements are designed to be independent of the telescope, \nustar is studying a somewhat softer energy range because of its reduced sensitivity above 30~keV (14-50~keV vs. 8-30~keV).  Thus, the two SC measurements may have systematically different average values.  However,  for the sample of 84 overlapping sources \scbat$=0.29\pm0.07$ and \scnustar$=0.27\pm0.03$ showing no evidence of a significant difference, at least on average.  We can also compare our SC measurements for ``bona fide" \nustarsh-observed Compton-thick AGN that were confirmed based on spectral fitting \citep{Gandhi:2014:117}.  We find that the Compton-thick fraction estimated using \scnustar is also similar ($21\pm7\%$, 16/84) to the \scbat measurement.   \\

\begin{figure*}	
\plotone{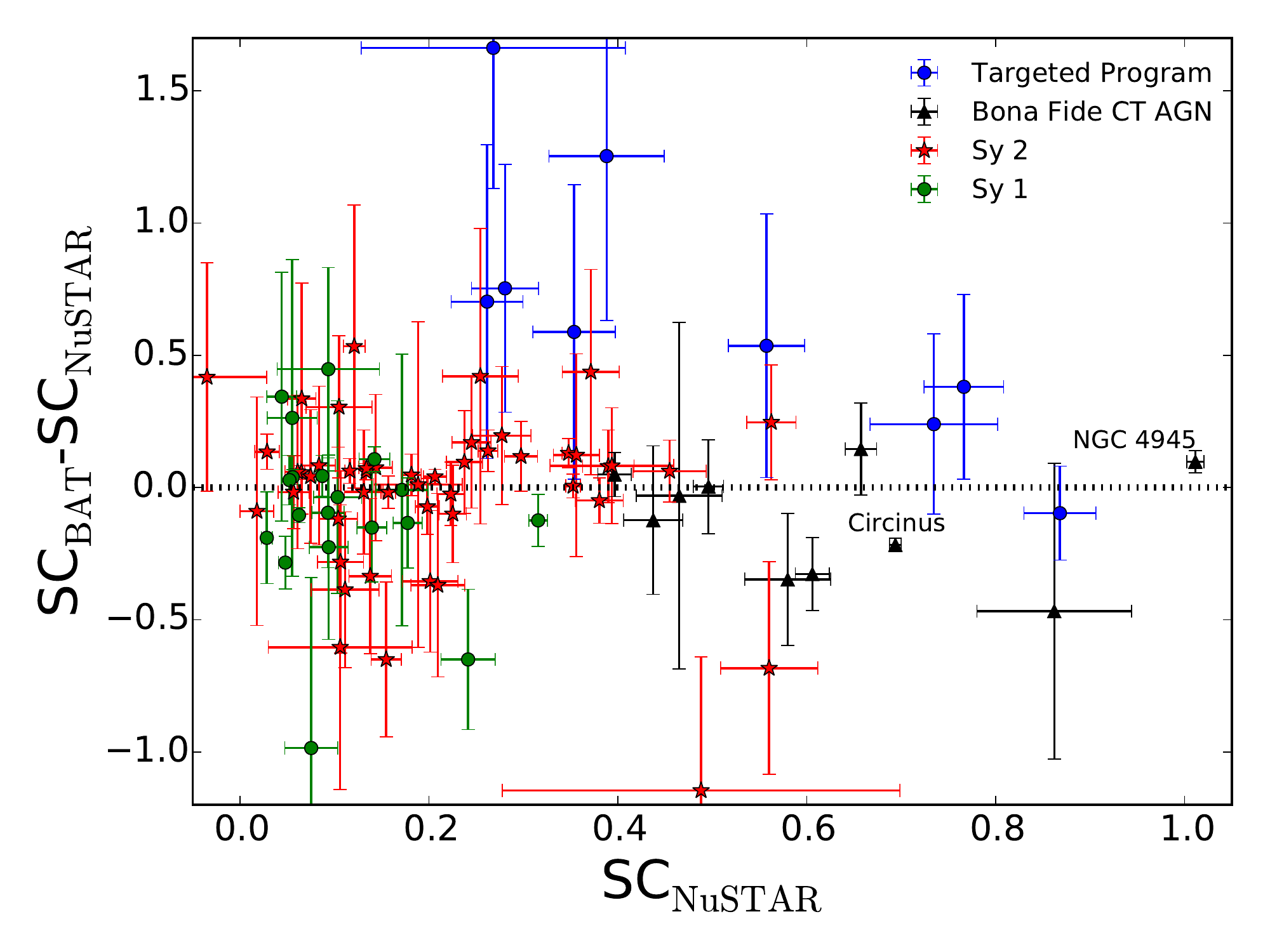}
\caption{Comparison of Spectral Curvature measurement for all 84 BAT AGN with \nustar observations. Compton-thick AGN confirmed with \nustar are shown in black, mildly obscured Seyfert 2s observed with \nustar are shown in red, and \nustarsh-observed Seyfert 1s are shown in green.  A horizontal black dotted line is plotted through zero.  The two measurements show no significant difference on average (\scbat$=0.29\pm0.05$ and \scnustar$=0.27\pm0.02$).      }
\label{scbatvsnustar}
\end{figure*}

We also look for individual sources whose \scbat and \scnustar measurements differ significantly (Figure \ref{scbatvsnustar}).  The \scbat is based on a time average spectra over 6 years between November, 2004 and August, 2010 which may differ from the \nustar observations which occurred after 2012.  We look for sources above 2.6$\sigma$ where we expect only one source from statistical noise with a sample size of 84.  We find five sources whose change (\scbat-\scnustar) was greater than 2.6$\sigma$.  The majority of these are Seyfert 1s which are already known to be variable above 10~keV based on \nustar \citep[MCG-06-30-015, MCG-05-23-016, NGC4151;][]{Parker:2014:721, Keck:2015:149,Balokovic:2015:62}.  Another source with a significant difference is Circinus.  However, analysis of the \nustar and \swiftbat spectra found that the 14-20~keV \swiftbat energy band was significantly contaminated by a nearby ULX which would explain the lower value of \scbat compared to \scnustar \citep[][]{Arevalo:2014:81}.  The final source is NGC 6232, one of the program sources, whose \scnustar was 3.1$\sigma$ below \scbat.  Further \nustar observations would be necessary to confirm whether this source is variable.  \\

We can also test whether the \nustar SC or a broad band ratio (8-24~keV/3-8~keV) is more efficient at finding Compton-thick AGN.  For the well known Compton-thick AGN observed with \nustarsh, all (10/10) are found in the Compton-thick region of the 8-30~keV SC measurement.  This is an improvement over \scbat because of the greater \nustar sensitivity.  By contrast, only 2/10 well known Compton-thick AGN are found in the Compton-thick region based on the 8-24/3-8~keV ratio (Figure \ref{nustar_curv}).   \\

We compare the SC measurement from \nustar to the one we used with \swiftbat for selection of the nine sources.  The five sources with a \swiftbat SC above the simulation upper limits for an edge-on torus (NGC 6232, CGCG 164-019, NGC 3393, MCG +06-16-028, UGC 3157) are now within the simulation results based on the \nustar SC.   We find that 4/9 of the selected sources in our sample are firmly in the Compton-thick region at the 3$\sigma$ level (NGC 3079, NGC 3393, MCG +06-16-028, 2MFGC 02280).  Three fainter sources are just below the Compton-thick cutoff, but are consistent within errors of being Compton-thick (NGC 6232, NGC 7212 NED02, CGCG 164-019).  Finally, NGC 3588 NED02 and UGC 3157 are below the cutoff at the 3$\sigma$ level. \\


\subsection{X-ray Spectral Fitting} \label{xrayspec}
\begin{figure*}
\centering
\includegraphics[width=17cm]{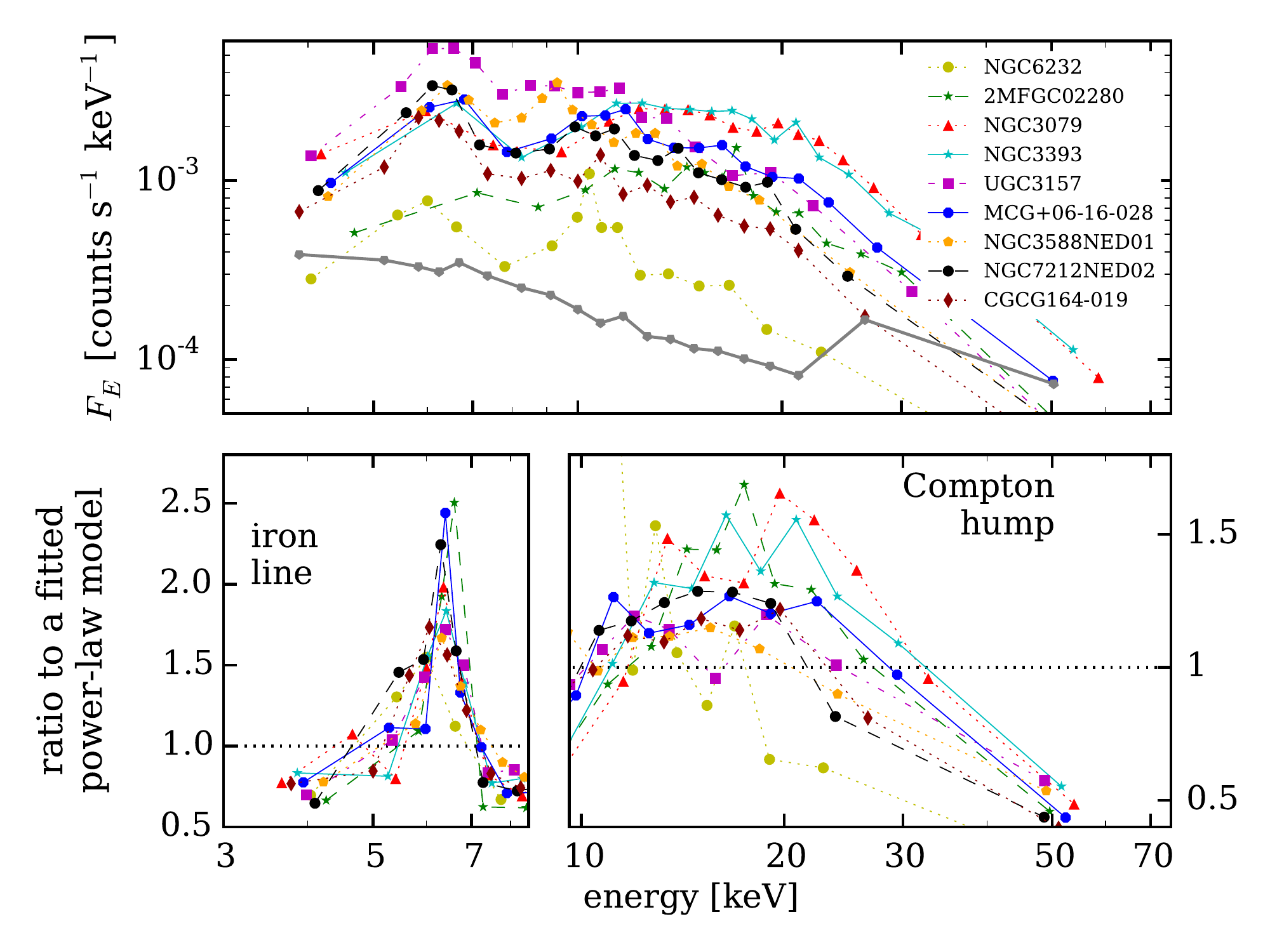}
\caption{\emph{Top Panel:} Observed \nustar spectra of nine galaxies selected based on their SC from \swiftbatsh. Spectra for the two focal plane modules have been co-added.  We have rebinned the spectra for each galaxy to have similar numbers of points in each panel at levels between 2.5$\sigma$ (NGC 6232) and 15$\sigma$ (NGC 3079). The typical background is shown by the filled gray symbols.  Almost all sources are above the background level between 3 and 30~keV, except for NGC 6232, which is only above the background between 5 and 20~keV.     \emph{Lower left}: Ratio of the spectra and a simple power-law model fitted to the 3-10~keV spectra (symbols as in the panel above).  We find that all sources show spectra consistent with a prominent \feka line (a decrement at 3-5~keV, an excess at 6-7~keV, and a decrement at 7-10~keV).   \emph{Lower right}:  Ratio of the spectra and a simple power-law model fitted to the 10-70~keV spectra (symbols as in the panel above).  We find that all sources show an excess at 10-25~keV, with the one exception, NGC 6232 showing an excess between 10 and 20~keV where it is significantly detected.    }
\label{nustar_allxray}
\end{figure*}

A plot of all the \nustar spectra before model fitting can be found in the top panel of Figure \ref{nustar_allxray}.  We highlight the spectral features by showing the best-fit power law model in the bottom two panels of Figure \ref{nustar_allxray}.  Fitting the 3-10~keV \nustar spectra with a power law model indicates $\Gamma<1$ for all sources.  This suggests that complex models are required. Additionally, a prominent excess is found at 6.4~keV, matching the \feka emission-line.  In order to measure the \feka equivalent width, we add a Gaussian component.  We find a large equivalent width ($>$ 1~keV) for all nine sources.  The high value of the equivalent width of the \feka lines is consistent with Compton-thick AGN \citep[e.g.][]{Krolik:1987:L5,Levenson:2002:L81}.  At higher energies ($>$10~keV) an excess is seen between 10-25~keV for 8/9 AGN; the remaining source, NGC 6232, shows an excess between 10 and 20~keV.  The hard photon index ($\Gamma<$1), high equivalent width fluorescent \feka lines, and Compton hump suggest a strong reflection component \citep[e.g.,][]{Matt:2000:173} which require more complex models to accurately measure the column density.


\subsubsection{Individual Torus Model Fits} 
\begin{figure*}	
\includegraphics[width=6.0cm]{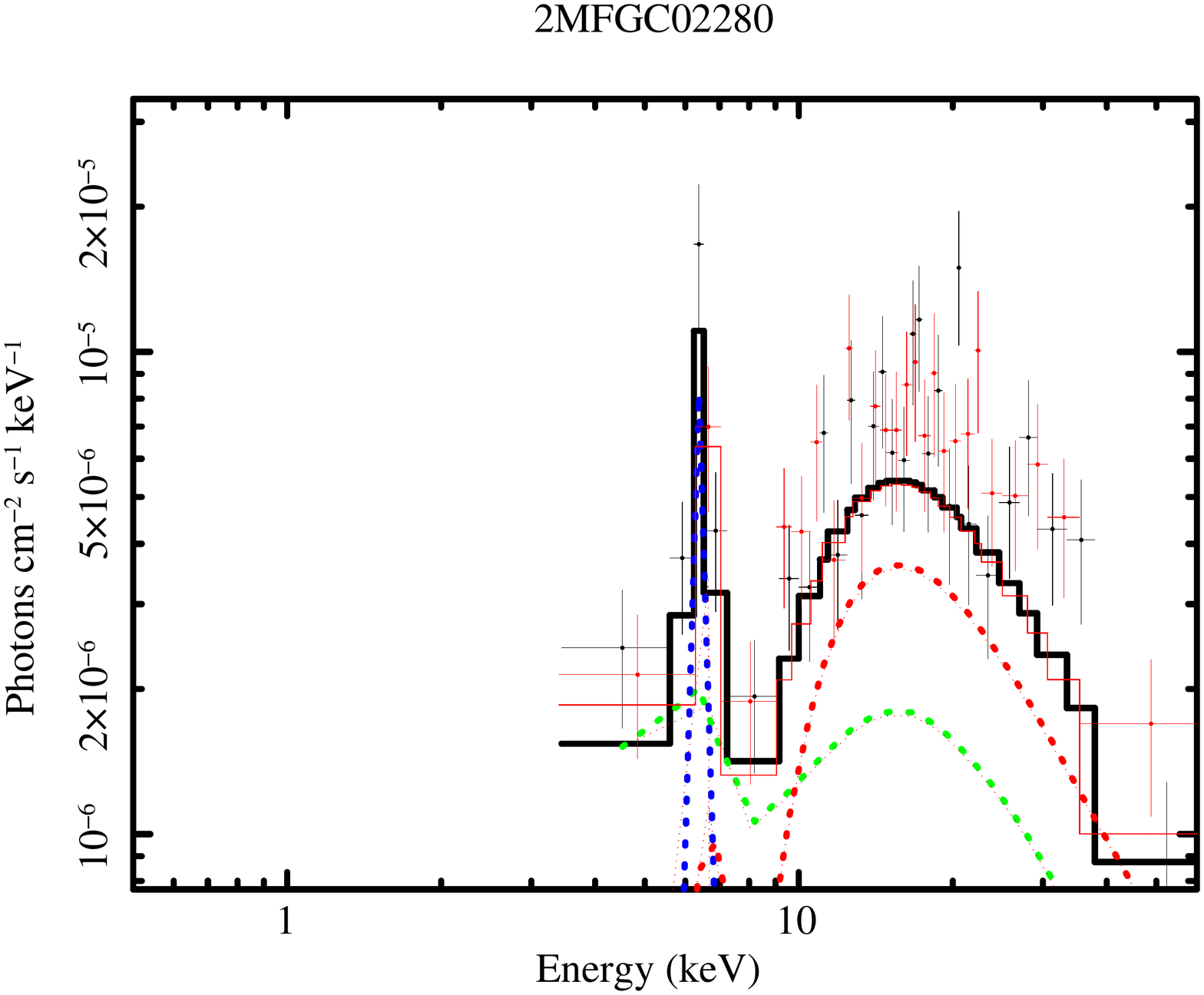}
\includegraphics[width=6.0cm]{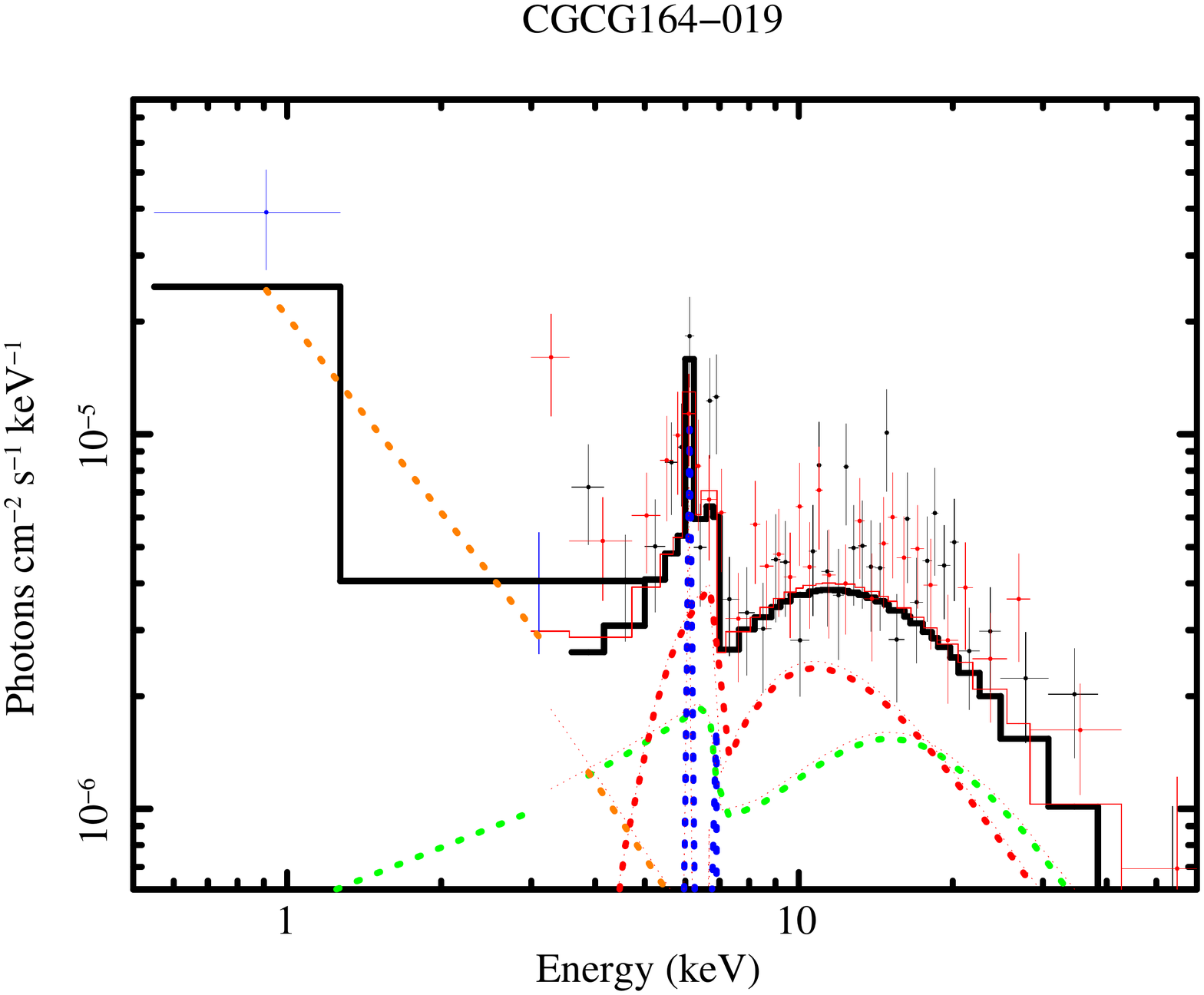}
\includegraphics[width=6.0cm]{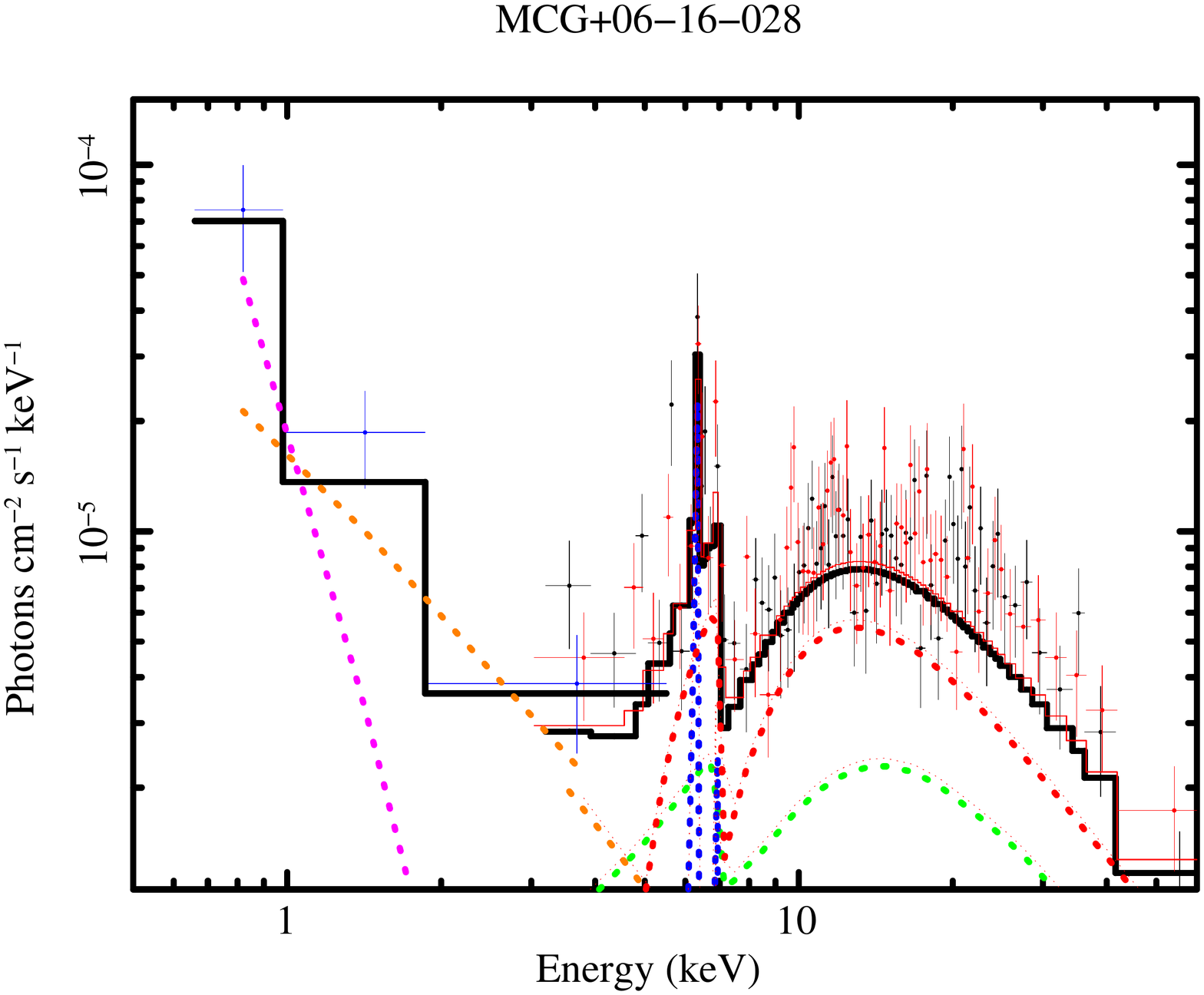}
\includegraphics[width=6.0cm]{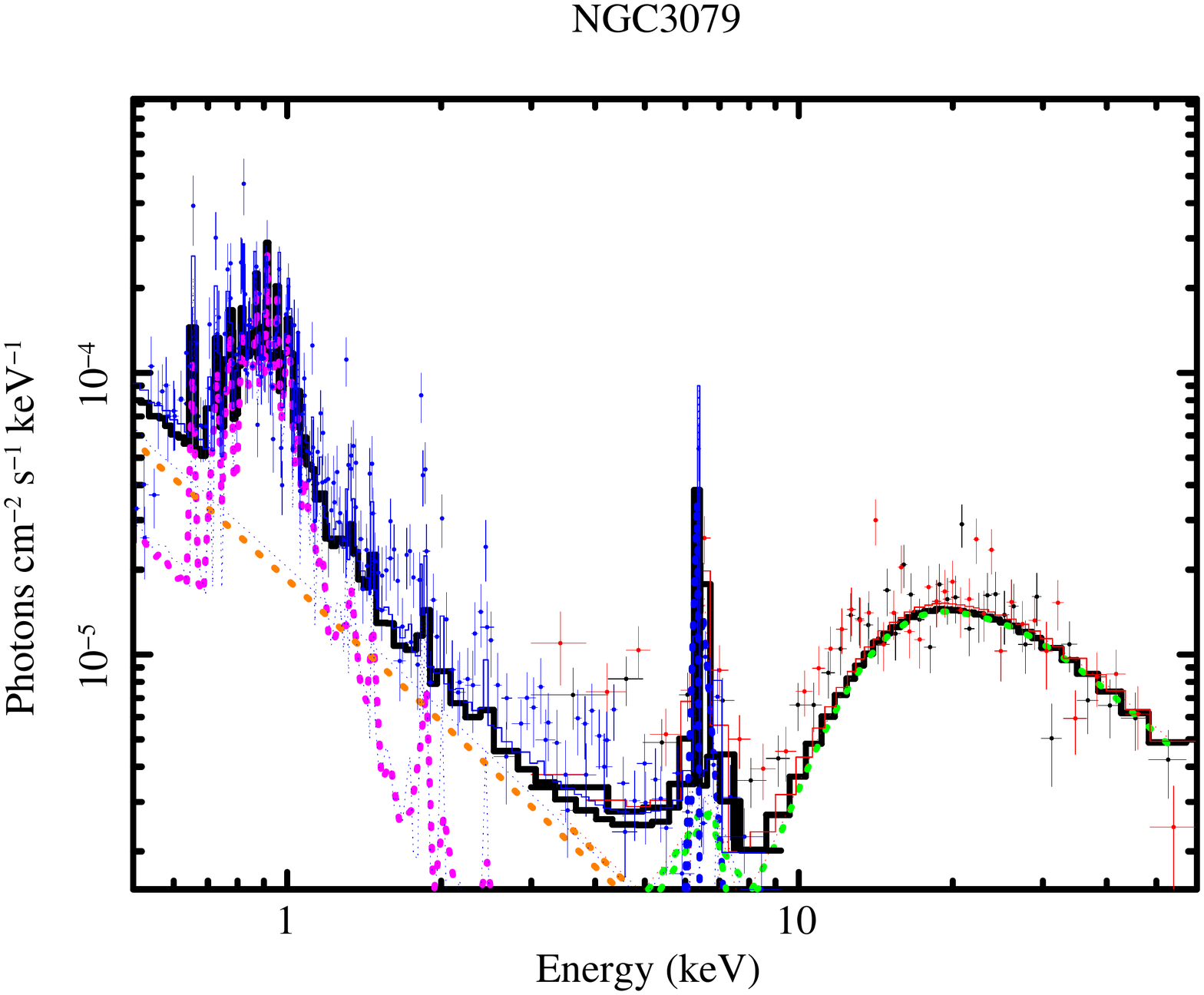}
\includegraphics[width=6.0cm]{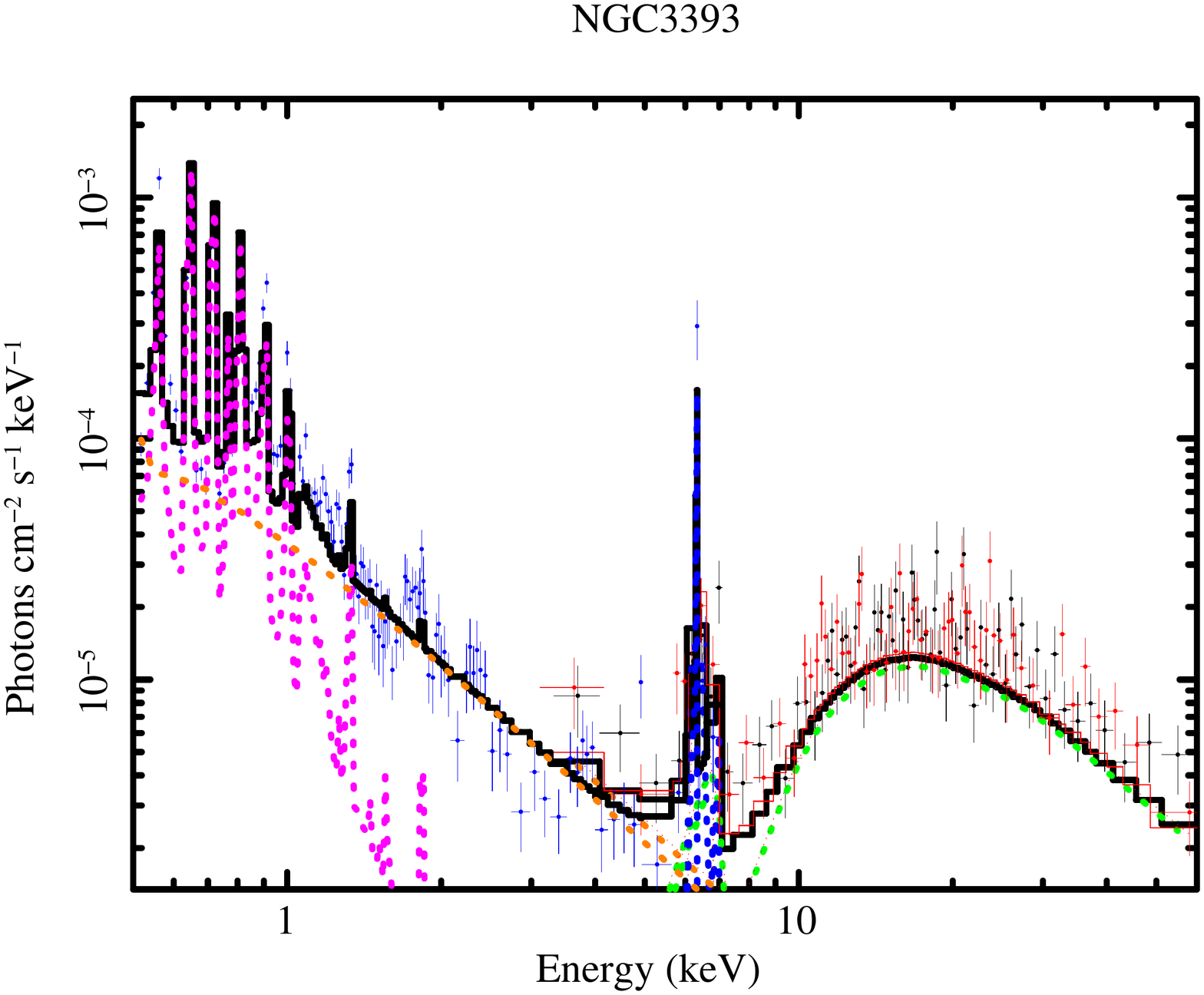}
\includegraphics[width=6.0cm]{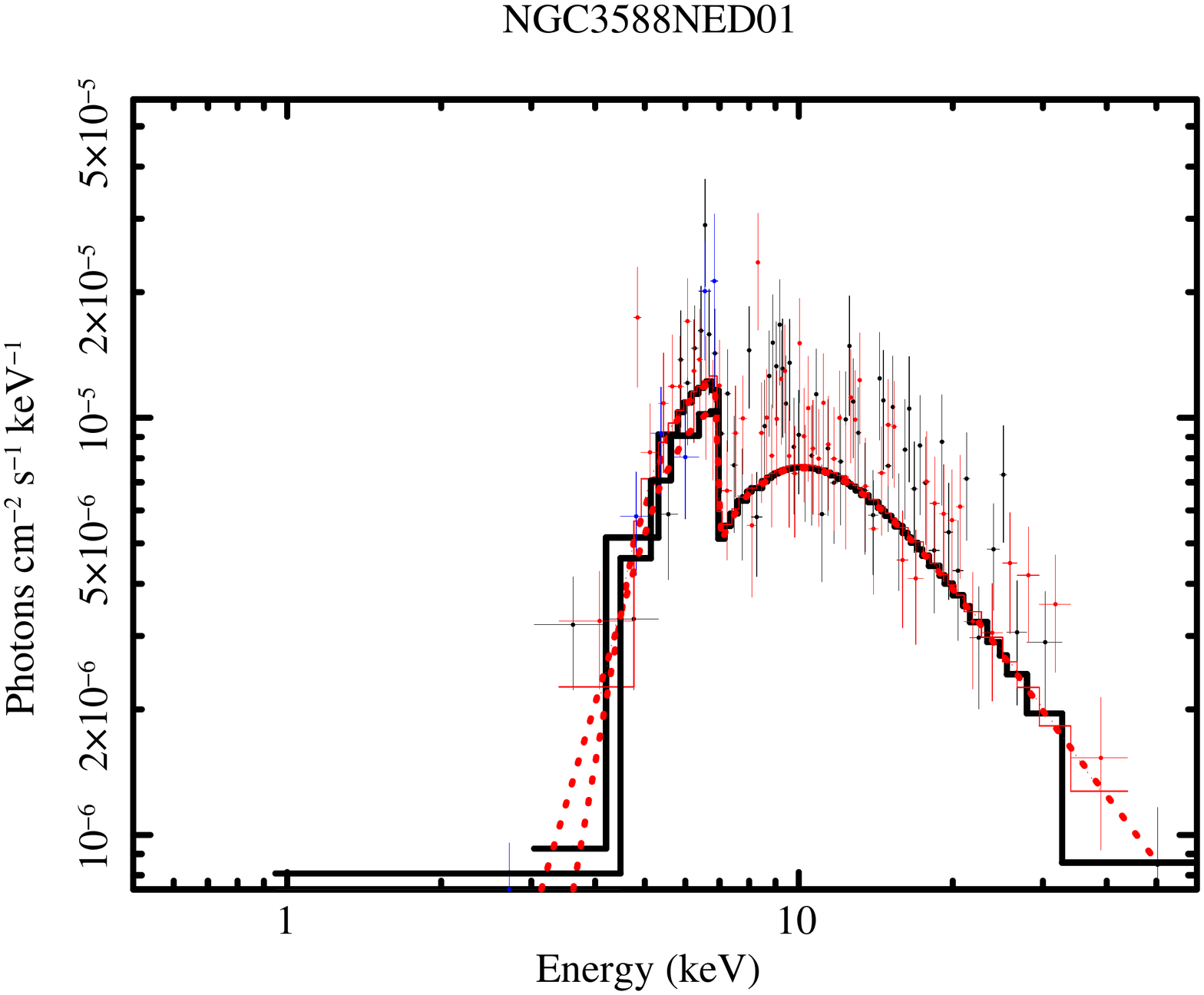}
\includegraphics[width=6.0cm]{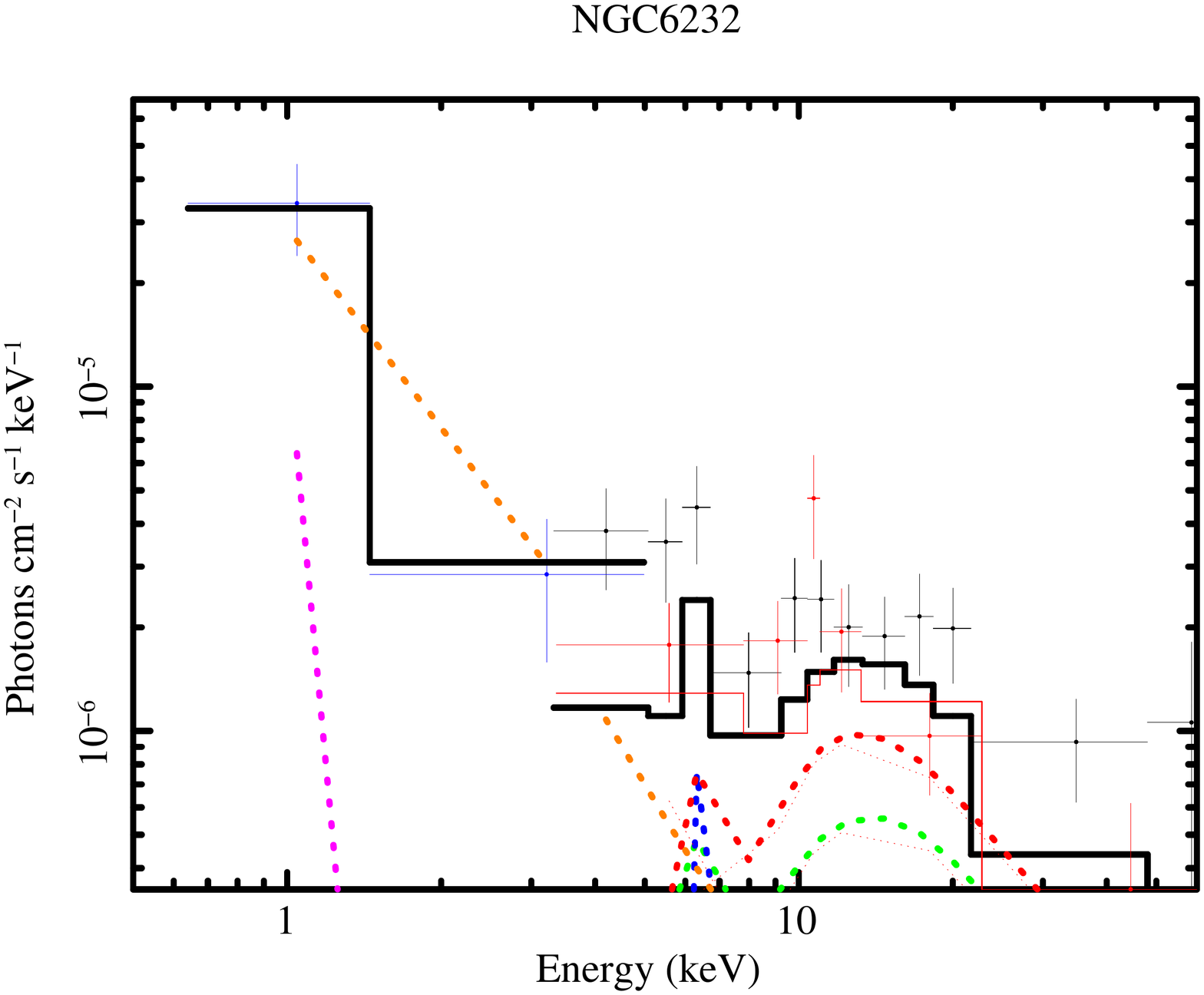}
\includegraphics[width=6.0cm]{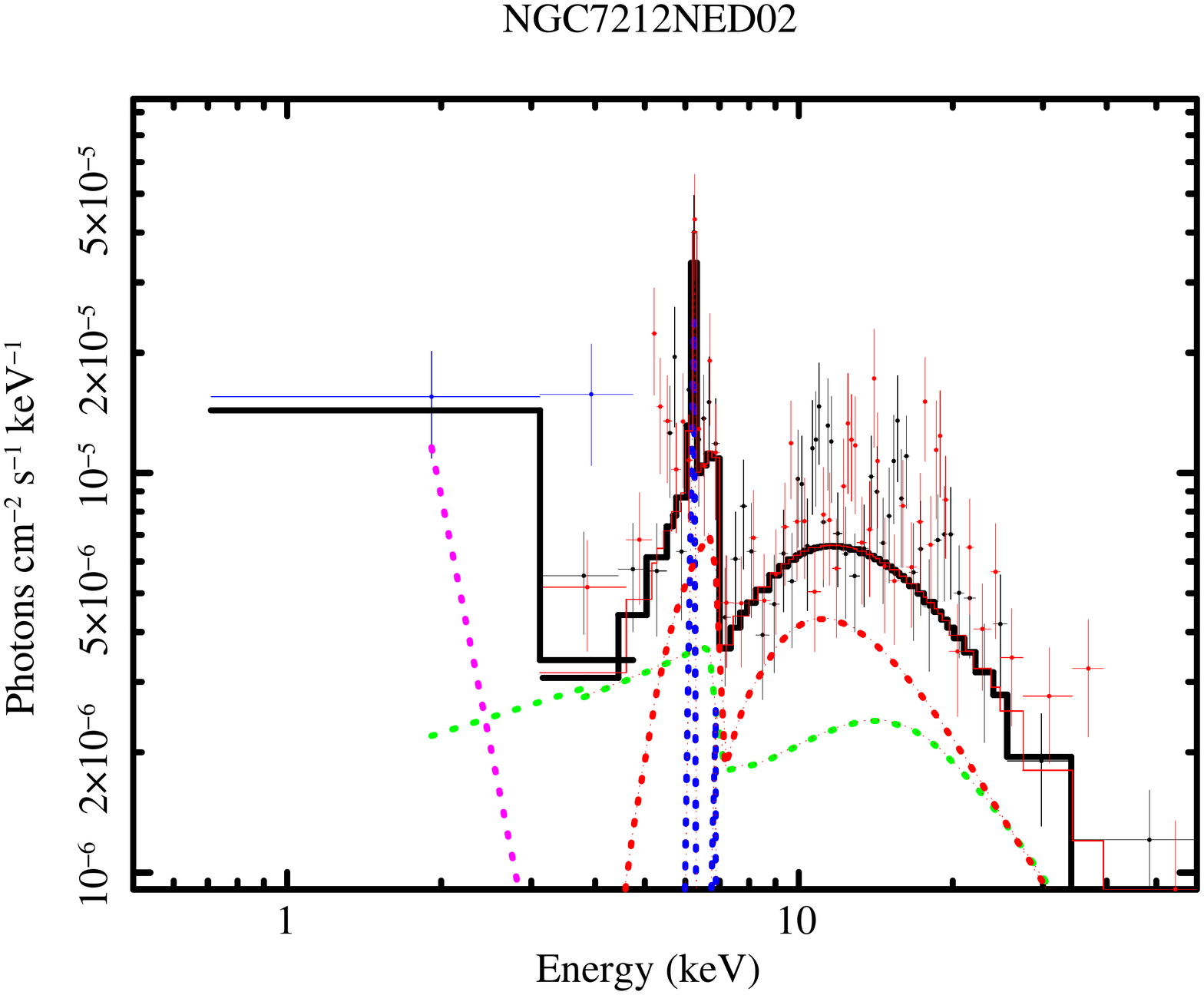}
\includegraphics[width=6.0cm]{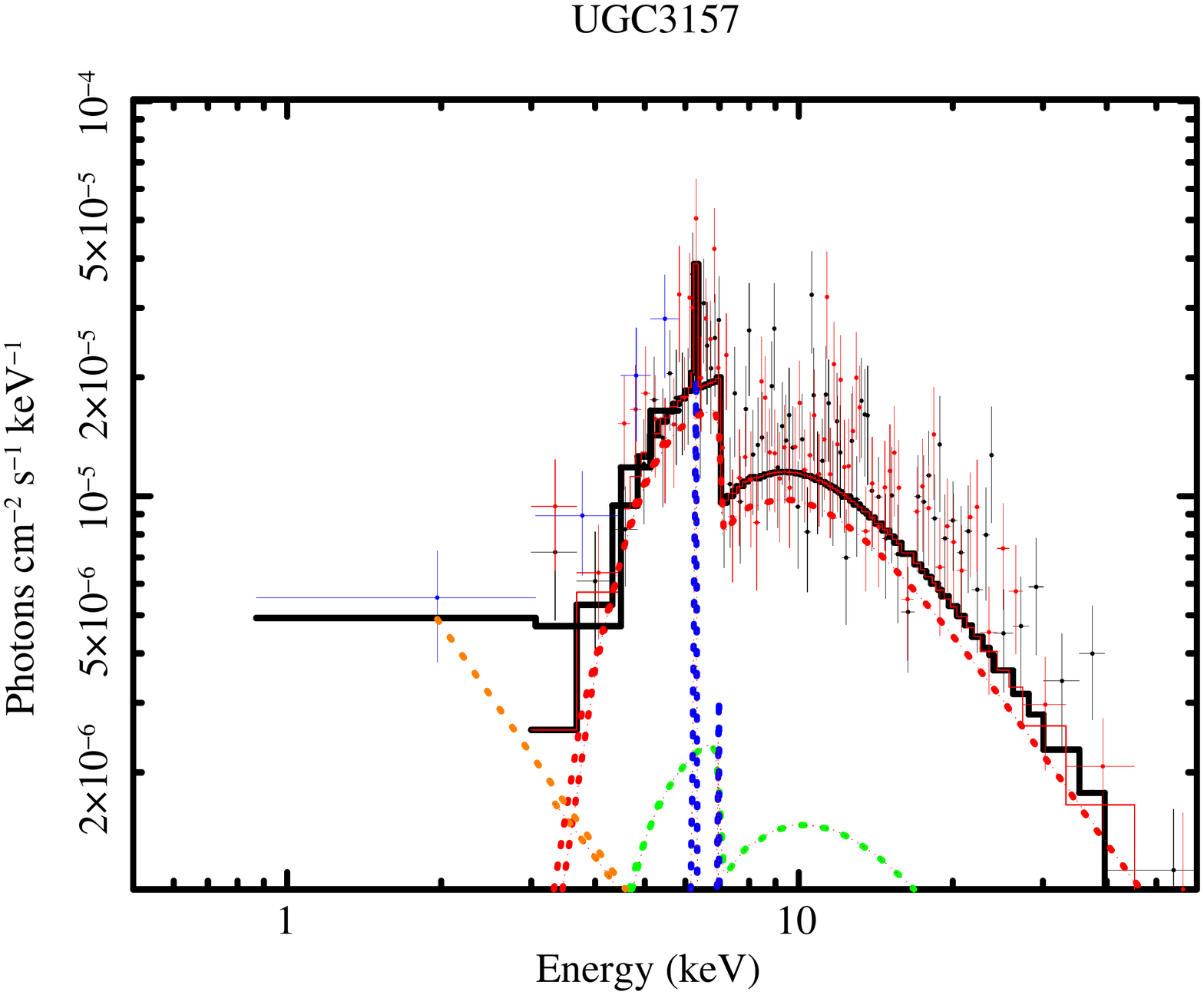}
\caption{X-ray spectra of the nine \nustarsh-observed \swiftbat AGN selected based on SC.   The best-fit \mytorus-based model (\modelM; described in Section \ref{specmod}) is shown binned to match the unfolded data.  \nustar is shown in black (FPMA) and red (FPMB) crosses while blue crosses represent the soft X-ray data.  The soft X-ray data are from \xmm for NGC 3079 and NGC 7212 NED02, \chandra for NGC 3393 and NGC3588 NED02, and \swiftxrt for the remaining sources.  The data are shown grouped to a minimum significance of 3.5$\sigma$ per bin for visual purposes.  The sum of the model is represented by a solid black line.  The model components are represented by dashed lines indicating the zeroth-order transmitted continuum through photoelectric absorption ($\mathtt{MYTZ}$, red), the scattered/reflected component ($\mathtt{MYTS}$, green), and fluorescent emission-line spectrum ( $\mathtt{MYTL}$, dark blue).  At softer energies ($<$3~keV), there is a model component for scattered AGN emission on larger scales in the host galaxy ($f_{scatt}$, orange) and in some models a thermal plasma component (APEC, pink).   }
\label{modelM_fits}
\end{figure*}

	We next self consistently fit the X-ray absorption and scattering adopting a toroidal structure with fluorescent lines to derive the covering factor and torus inclination.     A summary of the best fits to the intrinsic absorbing column density (\nh), the intrinsic photon index ($\Gamma$), APEC model ($kT$), diffuse scattering component ($f_{scatt}$), and intrinsic X-ray luminosity (\lx) can be found in Table \ref{xrayfitsumm}.   Figure \ref{modelM_fits} shows the X-ray spectral fits with the best-fit torus-based model (\modelM; described in Section 3.4) for the sample of nine \nustarsh-observed BAT AGN.  For \modelT, none of our fits constrained the torus inclination angle, so we fix it here to edge-on ($\theta_{\rm inc}=85\degree$). For the faintest source in our program, NGC 6232, \modelM and \modelT fits to the intrinsic absorbing column density (\nh) are poorly constrained, so we fix $\Gamma$=1.9, $\theta_{\rm inc}=85\degree$, and $\theta_{\rm tor}=60\degree$, consistent with typical AGN observed edge-on.\\
	
	Our results confirm that the two brightest sources above 10~keV, NGC 3079 and NGC 3393, are Compton-thick in both \modelM and \modelT. This is in agreement with past studies with other X-ray telescopes \citep[e.g.,][]{Iyomoto:2001:L69,Koss:2015:149}.   \modelM also suggests the reflection/scattering component is dominant over the transmitted component.  For both NGC 3079 and NGC 3393, the large-scale scattered component contributes the majority of the emission between 1-5~keV, with the APEC model dominant below 1~keV.	For NGC 3079, the large scattering fraction ($f_{scatt}<$0.36) may indicate additional soft components that are poorly fit. For NGC 7212 NED02, \modelM suggests a Compton-thick source whereas \modelT suggests a source that is nearly Compton-thick ($N_{\mathrm{H}}=0.90^{+0.13}_{-0.09}\times10^{24}$~$\mathrm{cm}^{-2}$). For NGC 7212 NED02,  \modelM yields a better fit in terms of $\chi^{2}/dof$ and the difference in column with \modelT is likely associated with the strength of the large-scale diffuse scattered component, which is dominant below 3~keV and better fit with \modelM.  Finally, 2MFGC 02280 was also found to be Compton-thick by  both \modelM and \modelT, consistent with an earlier \nustar study \citep{Brightman:2015:41}. 2MFGC 02280 does not have a detection below 3~keV in \swiftxrt and lies in a high Galactic column region ($N_{\mathrm{H}}=4\times10^{21}$~$\mathrm{cm}^{-2}$), so there are almost no constraints on the APEC model or diffuse scattering.\\
	
	 CGCG 164-019, MCG +06-16-028 and NGC 6232 also have model fits suggestive of being Compton-thick.  Both \modelM and \modelT suggest Compton-thick levels of obscuration for MCG +06-16-028 and NGC 6232, with relatively large uncertainties in the column density and weak constraints on the APEC model and diffuse component.  We note that for NGC 6232, the spectral index was fixed at $\Gamma$=1.9 and the source may not be Compton-thick if $\Gamma<1.7$.   CGCG 164-019, like NGC 7212 NED02, has \modelM suggesting a Compton-thick source whereas \modelT suggests a source that is only heavily obscured ($N_{\mathrm{H}}=0.55^{+0.24}_{-0.43}\times10^{24}$~$\mathrm{cm}^{-2}$).  Some of the differences in column can be attributed to the higher $\Gamma$ in \modelM ($\Gamma=2.0^{+0.3}_{-0.0}$) than \modelT ($\Gamma=1.6^{+0.3}_{-0.2}$).  Both \modelM and \modelT yield similar $\chi^2$, so it is difficult to say more.  Again the five detected counts below 3~keV in the \swiftxrt observations limit how well we can constrain the models.\\

	 The sources NGC 3588 NED02 and UGC 3157 have fits suggesting they are Compton-thin, but heavily obscured ($N_{\mathrm{H}}\approx5\times10^{23}$~$\mathrm{cm}^{-2}$) despite large \feka equivalent widths ($>1$~keV).  Past studies have noted, however, that dusty Compton thin gas can boost the \feka equivalent widths \citep{Gohil:2015:1449}.  \swiftxrt detects no counts below 3~keV in NGC 3588 NED02 and only 4 counts in UGC 3157 limiting the model constraints.  Finally, we note that a recently accepted compilation paper submitted on Compton-thick AGN in the BAT sample found broad agreement with our analysis (NGC 6232, NGC 7212 NED02, 2MFGC 02280, NGC 3079, NGC 3393, MCG +06-16-028 were Compton-thick).
	 
\subsubsection{Summary of X-ray Fits}
In general, the  \modelM and \modelT fits provide similar quality of fit $\chi^{2}/dof$ to the data with no systematic trend toward higher column density or power law index for either model.  We also report some overall properties from the SC-selected sample using the results from  \modelM and \modelT.        The mean power law index is $\Gamma=1.88\pm0.07$.  The mean column is $N_{\mathrm{H}}=(1.93\pm0.38)\times10^{24}$~$\mathrm{cm}^{-2}$.  We find that the mean observed-to-intrinsic luminosity at 2-10~keV and 14-50~keV is $L_{\rm \ 2-10 \ keV}^{\rm obs/int}=0.05\pm0.02$ and $L_{\rm 14-50 \ keV}^{\rm obs/int}$=$0.59\pm0.08$, respectively.  For this calculation, we used the \modelT observed-to-intrinsic luminosity estimates for the reflection-dominated sources as determined from \modelM.  The torus inclination is found to have values between 60 and 90\degrees (edge-on), while the torus half opening angle spans 26-78\degrees.	\\
	
	Interestingly, our individual spectral fits broadly agree with our Compton-thick selection based on SC.  Using SC we found that NGC 3079, NGC 3393, MCG +06-16-028,  2MFGC 02280 were all Compton-thick.   The three fainter sources that lie just below the Compton-thick cutoff but within 1$\sigma$ error of being Compton-thick are all sources which might be Compton-thin using \modelT.  NGC 3588 NED02 and UGC 3157 are the only two sources significantly below the SC Compton-thick cutoff and are also Compton-thin when fitting the spectra.\\
	
	Finally, we tried fitting a so-called \mytorus ``decoupled" model, where the \nh of the $\mathtt{MYTS}$ and $\mathtt{MYTL}$ components are allowed to vary compared to the zeroth-order transmitted continuum.  In this ``decoupled" model, the geometry can be thought of as a patchy torus where the global column density experienced by the scattered/reflection and fluorescent emission is different from the line of sight column density.  However, we found that the quality of fit was not significantly better for any of the sources and the two component \nh was poorly constrained given the quality of the data. 	 \\
	      
	    In summary, we find that torus models suggest that Compton-thick column densities are preferred for most  (78\% or 7/9) of the sources selected based on their SC values with the remaining two sources being heavily obscured ($N_{\mathrm{H}}>5\times10^{23}$~$\mathrm{cm}^{-2}$).  We note that our study is limited in that 6/9 sources have low-quality \swiftxrt or \chandra data with only a handful of counts below 3~keV.  Higher quality spectroscopy would be required to place stronger constraints on the large scattered component which may affect the column density measurements.    While much brighter Compton-thick AGN such as NGC 1068 exhibit additional complexities and components which are important to model \citep[e.g.,][]{Bauer:2014:670}, testing for these is currently not possible because of the faintness of these sources and the comparatively low photon statistics from the short, $\approx20$ ks observations.
	   

\begin{table*}[]
\centering
\caption{Best-fit Models for the \nustar + \swiftxrt
  Phenomenological vs. Physically
  Motivated Torus Models}
\scriptsize
\begin{tabular}{lcccccccccccc} 
\toprule\toprule \noalign{\smallskip}
\multicolumn{1}{c}{Object} & Mod. \tablenotemark{a}&$\chi^{2}/dof$\tablenotemark{b}  & $\Gamma \tablenotemark{c}$  & $N_{\mathrm{H}}$\tablenotemark{d}& $\theta_{\rm inc}$ \tablenotemark{e}& $\theta_{\rm tor}\tablenotemark{f}$ & $kT$\tablenotemark{g}& $L_{\rm \ 2-10 \ keV}^{\rm obs/int}$\tablenotemark{h}& $L_{\rm 14-50 \ keV}^{\rm obs/int}$\tablenotemark{h}&$f_{scatt}$\tablenotemark{i}\\
& & & & ($10^{24}$~$\mathrm{cm}^{-2}$) & (\degrees)  &  (\degrees) & (keV)  &  ($10^{41}$~\ergpersec) &  ($10^{41}$~\ergpersec)\\
\noalign{\smallskip} \toprule \noalign{\smallskip}
2MFGC02280 & M & 226/263 & $2.1^{+0.1}_{-0.1}$ & $3.16^{+0.35}_{-0.30}$ & $70.8^{+4.94}_{-4.63}$ & 60 & ND & 0.8/204.6 & 20.5/113.7& ND\\
 & T & 228/264 & $1.9^{+0.0}_{-0.1}$ & $1.97^{+0.71}_{-0.19}$ & 85 & $48.6^{+10.5}_{-5.95}$ & ND & 0.8/46.7 & 21.0/38.1& ND\\
CGCG164-019 & M & 295/377 & $2.0^{+0.3}_{-0.1}$ & $3.02^{+2.36}_{-0.86}$ & $61.3^{+0.51}_{-0.95}$ & 60 & $0.5^{+1.4}_{-0.5}$ & 5.9/57.7 & 43.7/43.0& $0.01^{+0.00}_{-0.01}$\\
 & T & 303/378 & $1.6^{+0.3}_{-0.2}$ & $0.55^{+0.24}_{-0.43}$ & 85 & $51.3^{+31.1}_{-25.3}$ & $0.1^{+0.4}_{-0.1}$ & 6.1/30.2 & 41.7/46.8& $0.05^{+0.00}_{-0.05}$\\
MCG+06-16-028 & M & 465/566 & $1.7^{+0.3}_{-0.0}$ & $1.71^{+4.59}_{-1.11}$ & $67.9^{+17.6}_{-7.94}$ & 60 & $0.3^{+0.4}_{-0.3}$ & 2.2/45.3 & 29.2/55.0& ND\\
 & T & 469/567 & $1.8^{+0.3}_{-0.2}$ & $1.21^{+0.26}_{-0.25}$ & 85 & $62.3^{+19.3}_{-16.5}$ & $0.1^{+0.1}_{-0.0}$ & 2.2/49.3 & 28.6/46.6& ND\\

NGC3079 & M & 543/287 & $1.7^{+0.0}_{-0.0}$ & $2.45^{+0.23}_{-0.11}$ & $88.3^{+1.61}_{-28.3}$ & 60 & $0.8^{+0.0}_{-0.0}$ & 0.1/R\tablenotemark{j} & 4.6/R& $0.36^{+0.00}_{-0.36}$\\
 & T & 534/288 & $1.4^{+0.1}_{-0.0}$ & $1.84^{+0.28}_{-0.35}$ & 85 & $77.9^{+3.72}_{-7.16}$ & $0.8^{+0.0}_{-0.0}$ & 0.1/5.8 & 4.2/12.2& ND\\

NGC3393 & M & 1175/852 & $1.8^{+0.1}_{-0.1}$ & $2.06^{+0.24}_{-0.33}$ & $87.9^{+2.01}_{-27.9}$ & 60 & $0.3^{+0.0}_{-0.0}$ & 1.3/R & 38.0/R& $0.04^{+0.00}_{-0.04}$\\
 & T & 1197/853 & $2.1^{+0.1}_{-0.0}$ & $2.16^{+0.32}_{-0.01}$ & 85 & $26.0^{+7.30}_{-0.0}$ & $0.2^{+0.0}_{-0.0}$ & 1.4/80.6 & 32.1/47.2& ND\\

NGC3588NED01 & M & 386/483 & $1.7^{+0.0}_{-0.0}$ & $0.60^{+0.07}_{-0.03}$ & $89.3^{+0.61}_{-10.4}$ & 60 & ND & 7.3/75.1 & 51.7/94.5& ND\\
 & T & 385/484 & $1.8^{+0.0}_{-0.1}$ & $0.57^{+0.05}_{-0.03}$ & 85 & $77.9^{+6.02}_{-4.99}$ & ND & 7.4/67.2 & 51.9/75.2& ND\\
NGC6232 & M & 139/176 & 1.9 & $1.23^{+0.97}_{-0.23}$ & 85 & 60 & $0.4^{+0.6}_{-0.4}$ & 0.4/10.6 & 4.9/10.0& $0.01^{+0.00}_{-0.01}$\\
 & T & 144/176 & 1.9 & $3.31^{+5.39}_{-1.31}$ & 85 & 60 & $0.1^{+1.8}_{-0.1}$ & 0.5/22.6 & 6.8/21.3& $0.01^{+0.00}_{-0.01}$\\
NGC7212 NED02 & M & 174/135 & $1.9^{+0.1}_{-0.0}$ & $2.64^{+0.46}_{-0.45}$ & $61.1^{+0.15}_{-0.15}$ & 60 & $0.8^{+0.0}_{-0.0}$ & 9.0/R & 76.2/R& $0.02^{+0.00}_{-0.02}$\\
 & T & 227/136 & $2.0^{+0.2}_{-0.1}$ & $0.90^{+0.13}_{-0.09}$ & 85 & $45.6^{+31.3}_{-6.65}$ & $0.8^{+0.0}_{-0.1}$ & 8.9/119.0 & 70.2/89.6& $0.01^{+0.00}_{-0.01}$\\

UGC3157 & M & 452/564 & $1.7^{+0.3}_{-0.0}$ & $0.57^{+1.21}_{-0.18}$ & $87.0^{+3.0}_{-27.0}$ & 60 & $0.0^{+1.9}_{-0.0}$ & 4.6/26.8 & 23.8/33.2& $0.01^{+0.00}_{-0.01}$\\
 & T & 454/565 & $1.8^{+0.0}_{-0.2}$ & $0.55^{+0.09}_{-0.13}$ & 85 & $59.9^{+24.0}_{-22.9}$ & $0.0^{+1.9}_{-0.0}$ & 4.6/25.9 & 23.9/28.7& ND\\

\noalign{\smallskip} \toprule \noalign{\smallskip}
\end{tabular}
\begin{minipage}[l]{1\textwidth}
NOTE. -- Best-fitting model parameters for the $0.5$--$70$~keV spectrum.  Parameters without errors have been fixed. The models are detailed in Section \ref{specmod}.  The errors correspond to 90\% confidence level for a single parameter.   ND refers to model components that are not detected. \end{minipage}
\footnotetext[1]{M=\mytorus; T=\torus model.}
\footnotetext[2]{The sources with \xmm data, NGC7212 NED02 and NGC3079, were binned to 20 counts per bin using $\chi^2$ statistics and the remaining sources were binned to 3 counts per bin using Poisson statistics.}
\footnotetext[3]{The power-law photon index of the direct and scattered component.}
\footnotetext[4]{Column density for the direct component.   In the \mytorus models the column density is equatorial rather than line of sight; however, for torus $\theta_{\rm inc}>60\degree$ (which all the \mytorus models converge to here) the difference between the equatorial and line-of-sight column density is less than 3\%.}
\footnotetext[5]{Best-fitting torus inclination angle to the observer.}
\footnotetext[6]{Best-fitting torus opening angle.  The \mytorus model assumes a 60$\degree$ opening angle.}
\footnotetext[7]{Temperature of the best fitting APEC component.  A non-detection is listed when the 90\% confidence upper limit is less than 0.01~keV.  }
\footnotetext[8]{Observed compared to intrinsic emission. }
\footnotetext[9]{Scattered fraction normalized to the intrinsic direct component.  A non-detection is listed when the 90\% confidence upper limit is less than a fraction of 0.01.      }
\footnotetext[10]{The intrinsic luminosity cannot be estimated because the source is reflection-dominated (reflected/transmitted$>$5) and the transmitted component is not detected.  }
\label{xrayfitsumm}
\end{table*}

\subsection{Other Measures of Compton-Thickness} \label{sedchapt}
Another way to identify Compton-thick AGN is to use additional intrinsic luminosity indicators (e.g. \Lsixum, \loiii) to compare to the observed X-ray luminosity.  A high mid-IR/X-ray ratio and/or high \oiii/X-ray ratio may indicate a Compton-thick AGN.  A summary of all the indicators including the results from the X-rays is provided in Table \ref{ctagnsumm}.

\begin{deluxetable*}{lcccccccc}
 \tabletypesize{\scriptsize}
 \tablewidth{0pt}
 \tablecaption{Compton-thick AGN Indicators}
 \tablehead{ \colhead{Object} & \colhead{\feka EW} & \colhead{$\Gamma$} &\colhead{\nustar}& \colhead{\mytorus} &
 \colhead{$F^{\rm obs}_{\mathrm{2-10\ keV}}$/$F^{\rm pred\  6\mu m}_{\rm 2-10\ keV}$} &
  \colhead{IR SED } &
 \colhead{$F^{\rm obs}_{\mathrm{2-10\ keV}}$/$F_{\rm [O\ {\scriptscriptstyle III}]}$}  \\
 \colhead{ } &
 \colhead{$>1$~keV} &
 \colhead{$<1$} &
  \colhead{SC} &
 \colhead{Model} &
 \colhead{$<20$} &
 \colhead{E(B-V)} &
  \colhead{$<1$} &
}
 \startdata 
2MFGC 02280	&T	&T	&T	&T	&T	&T	&\bf{N}	\\
CGCG 164-019	&T	&T	&T	&T	&T	&\bf{N}	&\bf{N}	\\
MCG +06-16-028	&T	&T	&T	&T	&T	&\bf{N}	&\bf{N}	\\
NGC 3079	&T	&T	&T	&T	&T	&\bf{N}	&T	\\
NGC 3393	&T	&T	&T	&T	&T	&T	&T	\\
NGC 3588 NED02	&T	&T	&\bf{N}	&\bf{N}	&\bf{N}	&T	&\bf{N}	\\
NGC 6232	&T	&T	&\bf{N}	&T	&T	&\bf{N}	&\bf{N}	\\
NGC 7212 NED02	&T	&T	&\bf{N}	&T	&T	&T	&\ldots	\\
UGC 3157	&T	&T	&\bf{N}	&\bf{N}	&T	&T	&\bf{N}	
 \enddata
 
\tablecomments{Results of various tests of Compton-thickness. {\bf N} indicates the object is Compton-thin, T indicates that the object was classified as Compton-thick, and a ellipse indicates that the test could not be performed on the object because of a lack of data.}
 \label{ctagnsumm}
 \end{deluxetable*}
 
\subsubsection{IR emission}  \label{midirfit}
\begin{figure*}	
\plotone{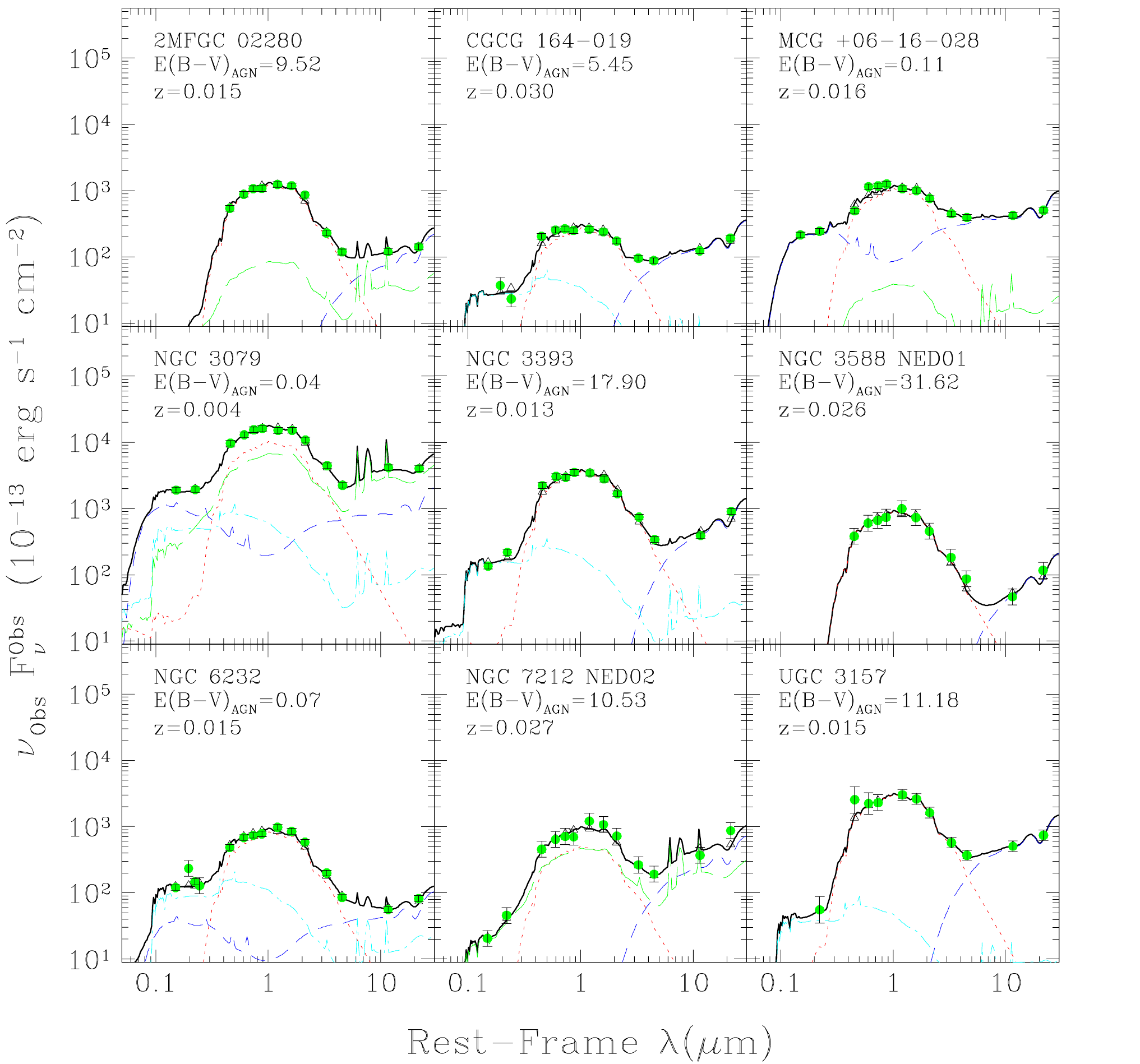}
\caption{Spectral energy distributions (SEDs) of the nine SC-selected galaxies. The observed photometry includes optical ({\em griz}), NIR ({\em JHK}), and mid-IR (3.4-22 $\micron$) as well as FUV and NUV photometry when available.  We used the \citet{Assef:2010:970} 0.03-30 $\micron$ algorithm to model the strength of the AGN emission in the IR using empirical AGN and galaxy templates.    The plot points represent observed data (green circles) and predicted SED model flux (open triangles).  The total best-fit SED template line (solid black) was made by combining the AGN (dashed blue) and old, (E, dashed red), intermediate (Sbc, dashed green), and young (Im, dashed cyan) galaxy stellar populations.}
\label{sed}
\end{figure*}

	The 6 $\mu$m AGN emission provides an additional assessment of the intrinsic AGN luminosity \citep[e.g.,][]{Gandhi:2009:457}.  Moderate luminosity AGN, however, can have the majority of their mid-IR emission from host galaxy contributions \citep[e.g., ][]{Stern:2012:30} rather than the AGN.  We therefore first measure the \wise colors to test whether the sources show colors indicative of AGN and are dominated by AGN emission in the mid-IR.  We find that only three sources have $W1-W2>0.8$, indicating the AGN likely dominates the mid-IR emission (MCG +06-16-028, CGCG 164-019 and NGC 7212 NED02).     It is therefore important to fit the spectral energy distributions (SEDs) to measure the intrinsic 6~$\mu$m AGN emission.   The observed photometry includes optical \citep[{\em griz} from][]{Koss:2011:57}, near-infrared (NIR; 2MASS-{\em JHK}), mid-IR ({\em WISE}, 3.4-22 $\micron$), as well as \galex or \swift far-ultraviolet (FUV) and near-ultraviolet (NUV) photometry when available.  We follow the photometry procedure of \citet{Koss:2011:57} using Kitt Peak or 2MASS data to measure the optical and NIR photometry.  We use the \citet{Assef:2010:970} 0.03-30 $\micron$ algorithm to model the strength of the AGN emission in the mid-IR using empirical AGN and galaxy templates.  The template SEDs (Figure \ref{sed}) suggest a Compton-thick level of obscuration for most of the sample (5/9,55\%) based on the  \nh$/E(B-V)=1.5\times 10^{23}$ \nhunit conversion from reddening to column density \citep[][]{Maiolino:2001:28}.  \\
	
	For NGC 3079, MCG +06-16-028, and NGC 6232, almost no obscuration is detected ($E[B-V]<0.11$) in the best fit SEDS.  This is inconsistent with the X-ray spectral fitting and the lack of broad lines in the optical spectra. To understand the fitting better, we produced 1000 resampled SEDs based on the measured photometry but resampled by the photometric errors (assuming Gaussian noise) and find that the SED fits do include a small percentage of solutions with Compton-thick obscuration ($<$5\%).    The high AGN obscuration (\nh$>10^{24}$ \nhunit) combined with significant host galaxy star formation as shown by the bright UV and IR emission as seen in Herschel \citep{Melendez:2014:152} make SED techniques unable to accurately model the AGN SED in some cases due to the lack of photometric bands where the AGN is dominant. Further mid-IR studies using high spatial resolution imaging of nuclear emission \citep[e.g., Asmus:2014:1648][]{Asmus:2014:1648} are required to resolve these degeneracies.\\

\begin{figure}	
\includegraphics[width=8.5cm]{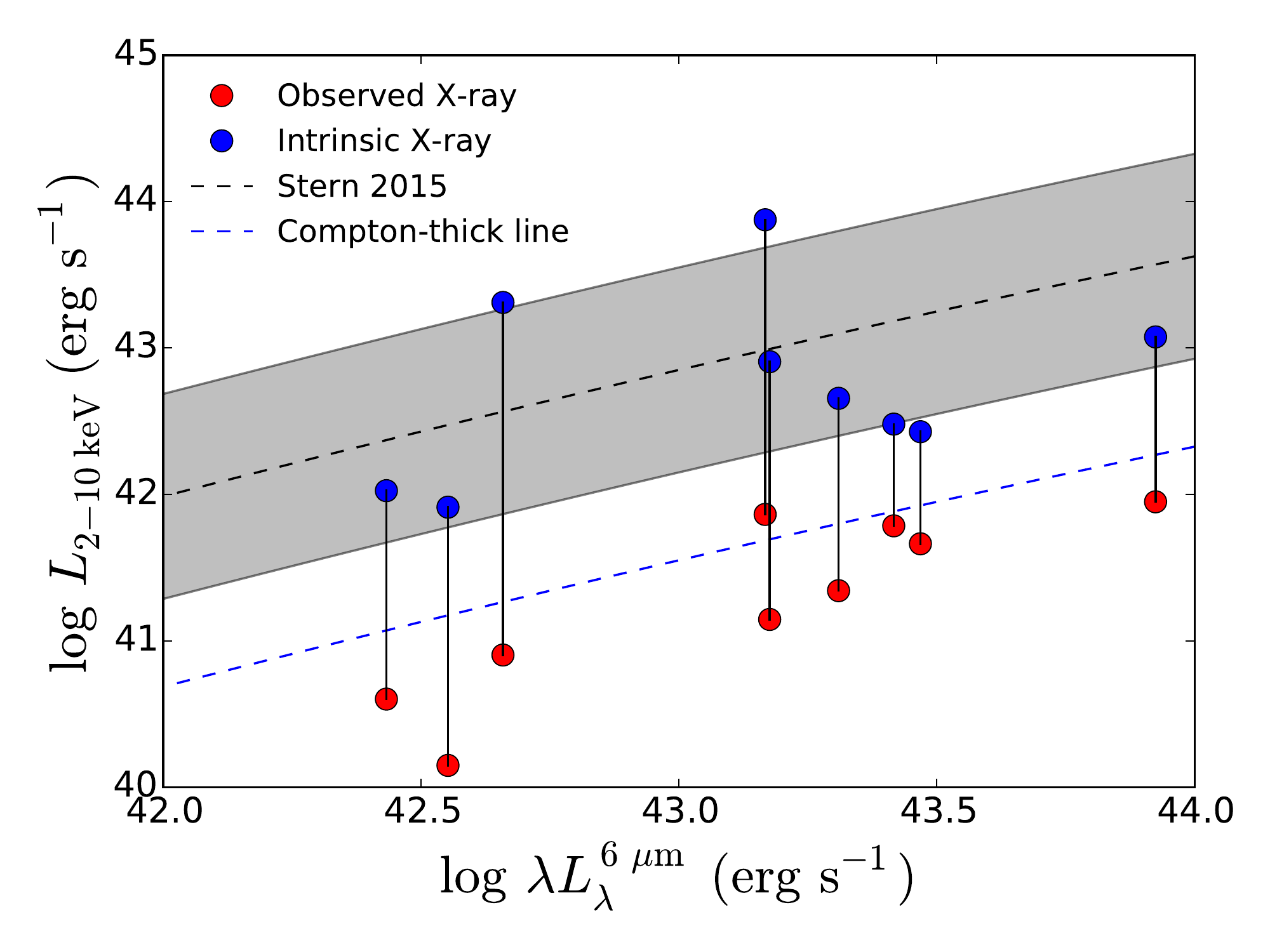}
\caption{ \Lsixum from SED fitting compared to \Lsoftobs (red dot) and \Lsoft (blue dot) based on X-ray model fitting.  A black dashed line indicates the relation from \citet{Stern:2015:129} and a grey region has been shaded within a factor of five of the mean ratio ($\pm$0.7 dex).  Sources below the blue dashed line are likely to be Compton-thick.  All the sources except for NGC 3588 NED02 are below the Compton-thick line based on their mid-IR to observed X-ray emission.  Finally, we find that even with the X-ray spectral fitting the majority of the intrinsic estimates of the source luminosity are below the values expected from the \Lsixum suggesting they may still be underestimated.}
\label{irtoxray}
\end{figure}

	We estimate the intrinsic 6 $\micron$ AGN emission from the template fitting and compare it to the observed 2-10~keV X-ray emission.  We compare it to the ratio obtained from large-sample studies of AGN \citep{Stern:2015:129}.  This ratio is luminosity dependent, with a predicted X-ray to mid-IR ratio of 0.87 for our least luminous mid-IR source (NGC 6232) to 0.44 for the most luminous mid-IR source (NGC 7212 NED02).  Assuming \modelM and $\Gamma$=1.9, the observed 2-10~keV X-ray emission is diminished by a factor of $\approx$20 at Compton-thick obscuration.  We find that the observed X-ray emission is fainter than expected from the mid-IR (Figure \ref{irtoxray}) in all our sources by an average factor of 72$\pm$29. The smallest difference is NGC 3588 NED02 (factor of 13 lower X-ray emission) and the largest difference is NGC 3079 (factor of 298 lower).  NGC 3588 NED02 is also the only AGN whose mid-IR to X-ray ratio (\Lsoftobs/\Lsixum) is consistent with a Compton-thin AGN.    Thus, the intrinsic mid-IR AGN emission confirms that the observed X-ray emission is consistent with Compton-thick AGN for most of the sample.

\subsubsection{Optical Spectroscopy}  \label{oiii}
We compare the optical spectroscopy of these sources with other BAT-detected AGN (Koss et al., in prep). We have optical spectra for all sources except NGC 7212 NED02.  We first apply AGN emission-line diagnostics \citep[e.g.,][]{Kewley:2006:961,Veilleux:1987:295} using the [N~II]/H$\alpha$ diagnostic.    All the AGN are in the Seyfert or LINER region,  except NGC 3588 NED02 which falls in the composite region, but this has been found with many other AGN in close mergers \citep[e.g.,][]{Koss:2011:L42}.  The \hbeta~line for 2MFGC 02280 is not detected, but the limit places it most likely in the Seyfert or LINER region.  The ratio of the observed 2-10~keV X-ray to Balmer decrement corrected \oiii line strength provides a measure of Compton-thickness \citep{Bassani:1999:473}.  We find that only NGC 3393 and NGC 3079 show an excess in the Balmer decrement corrected \oiii vs. X-ray luminosity ratio consistent with a Compton-thick AGN  ($F^{\rm obs}_{\mathrm{2-10\ keV}}$/$F_{\rm [O\ {\scriptscriptstyle III}]}<1$).  \\

	AGN are known to have a wide range in \oiii to X-ray ratios.  One possibility for the low values of \oiii is a low scattering fraction from a ``buried" AGN \citep{Ueda:2007:L79b} with a small opening angle and/or have unusually small amount of gas responsible for scattering.  Additionally, if these AGN have high Eddington ratios they should have relatively weak \oiii as found by the ``Eigenvector 1" relationships \citep[e.g.,][]{Boroson:1992:109}.  Finally, AGN ``flickering" on shorter timescales than the light travel time to the ionized regions can cause some AGN to have much stronger X-ray emission since it has just begun to ionize the narrow line region \citep{Schawinski:2015:2517}.  \citet{Noguchi:2010:144} found that optical emission-line studies are biased against  ``buried" AGN that have a small scattering fraction or a small amount of narrow line region gas.  AGN with a low ratio of \oiii to X-ray luminosity (\loiii/\Lsoft) tend to be ``buried" AGN.  We use the estimated ratio obtained from large studies of AGN \citep{Berney:2015:3622}.  In our sample, we find that only 2MFGC 02280 is consistent with a ``buried" AGN in that the X-ray emission is significantly outside the scatter of the mean \oiii to X-ray ratio  (\loiii/\Lbat). NGC 3588 NED02 does have a higher X-ray to \oiii ratio, but this ratio has been found to be elevated in many merging AGN galaxies \citep{Koss:2010:L125}.  In summary, the majority of the sample does not show evidence of having uniquely high intrinsic X-ray to \oiii values. 
	
	

\subsection{Host Morphology and Accretion Rates} \label{hostgal}
\begin{figure*}	
\includegraphics[width=6.0cm]{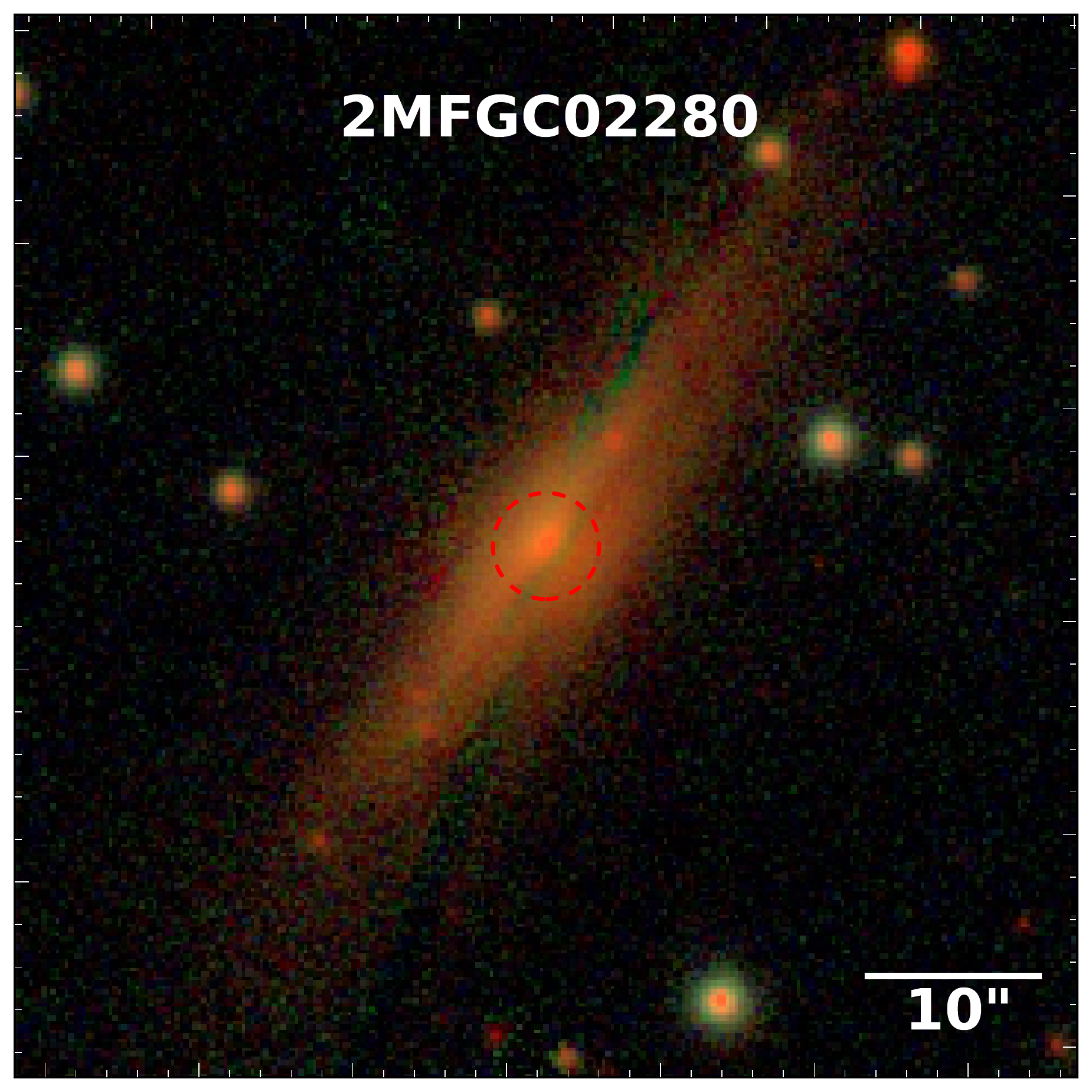}	
\includegraphics[width=6.0cm]{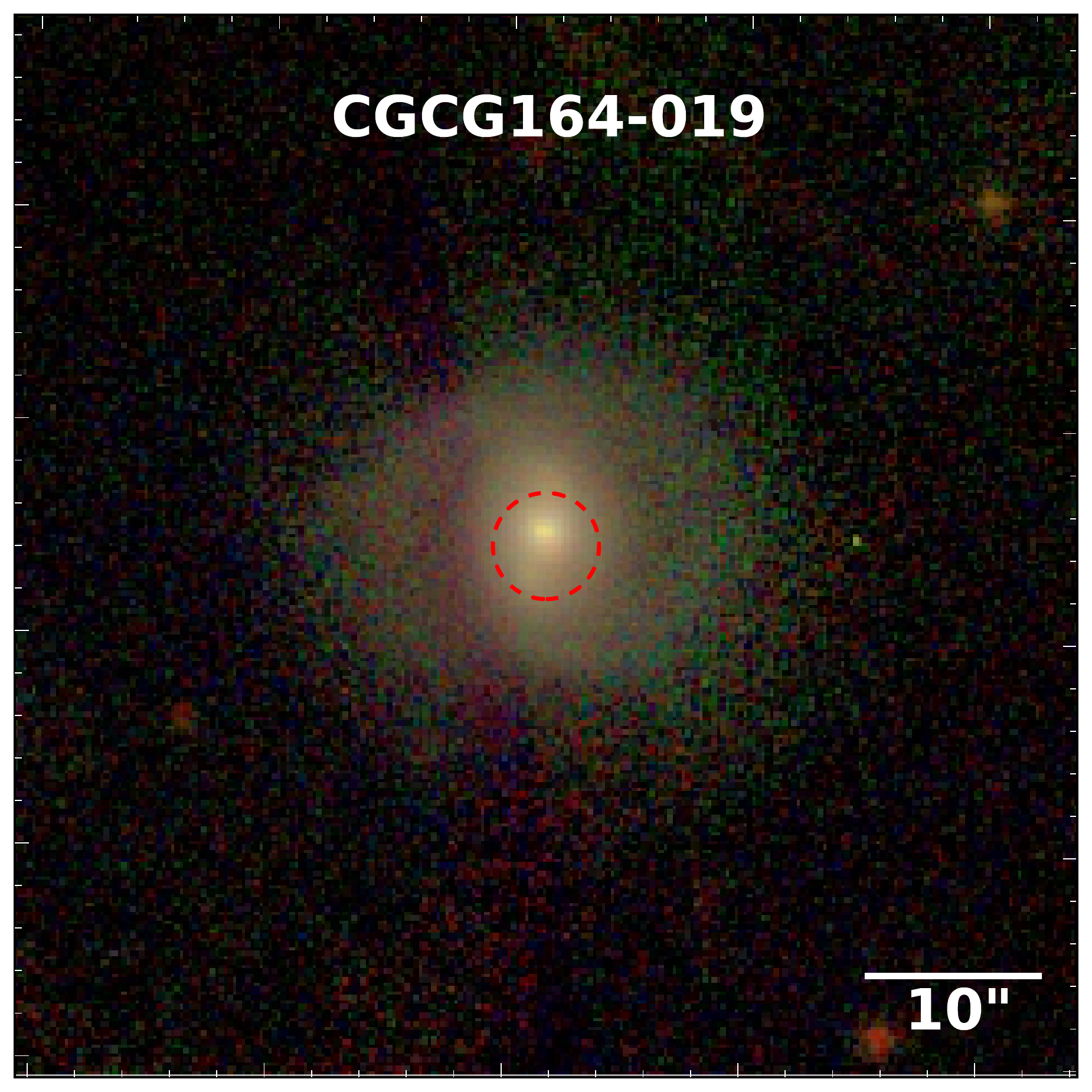}
\includegraphics[width=6.0cm]{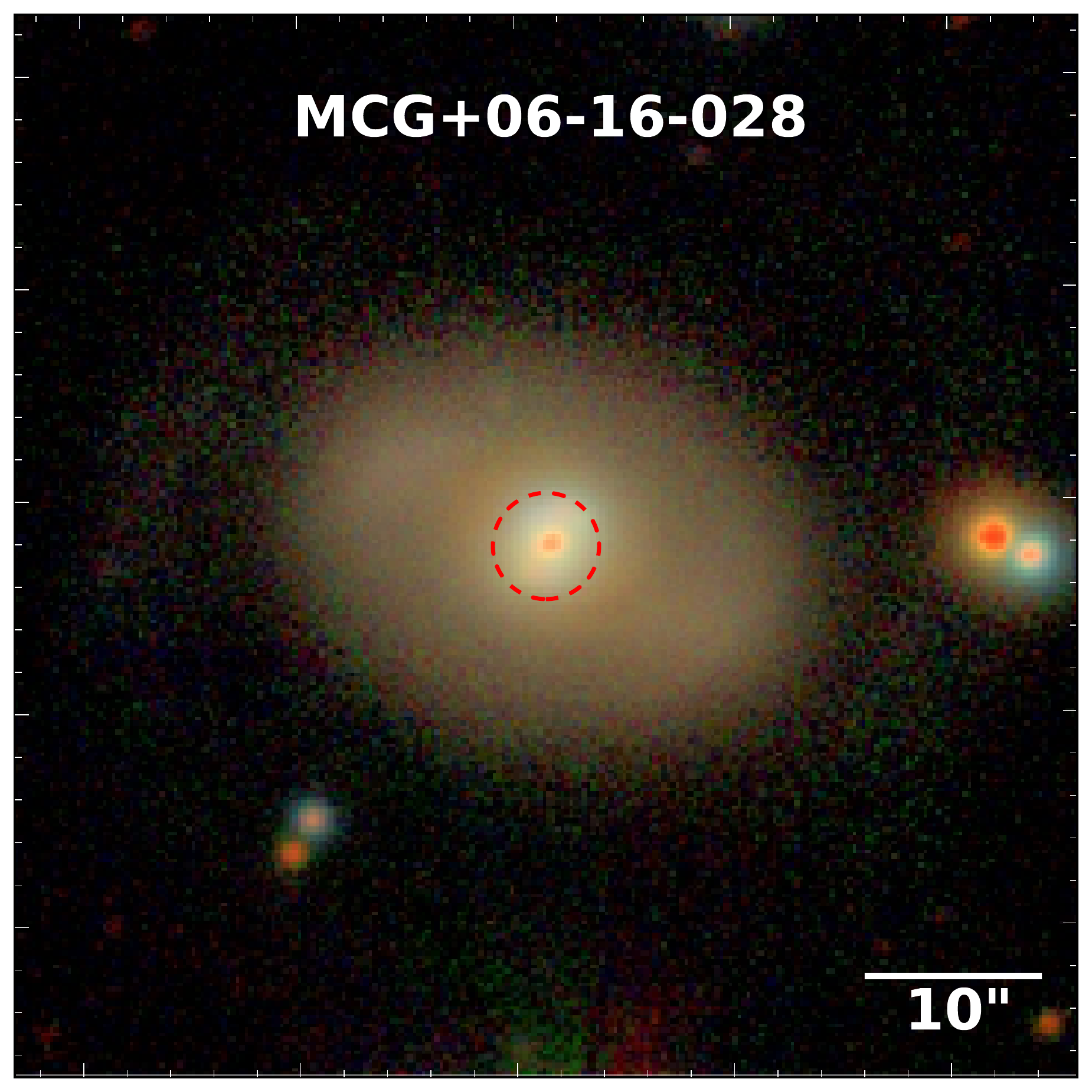}
\includegraphics[width=6.0cm]{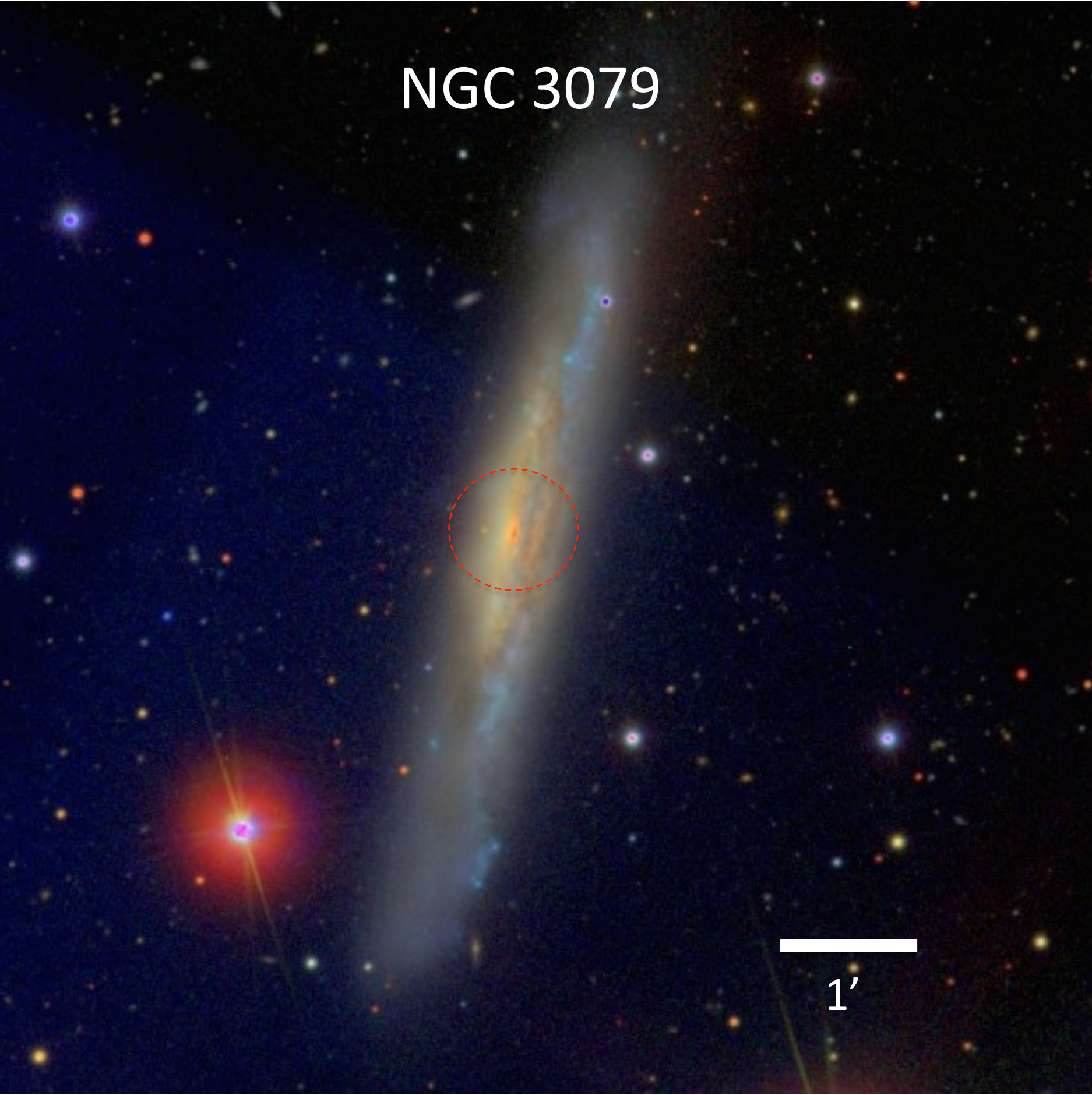}
\includegraphics[width=6.0cm]{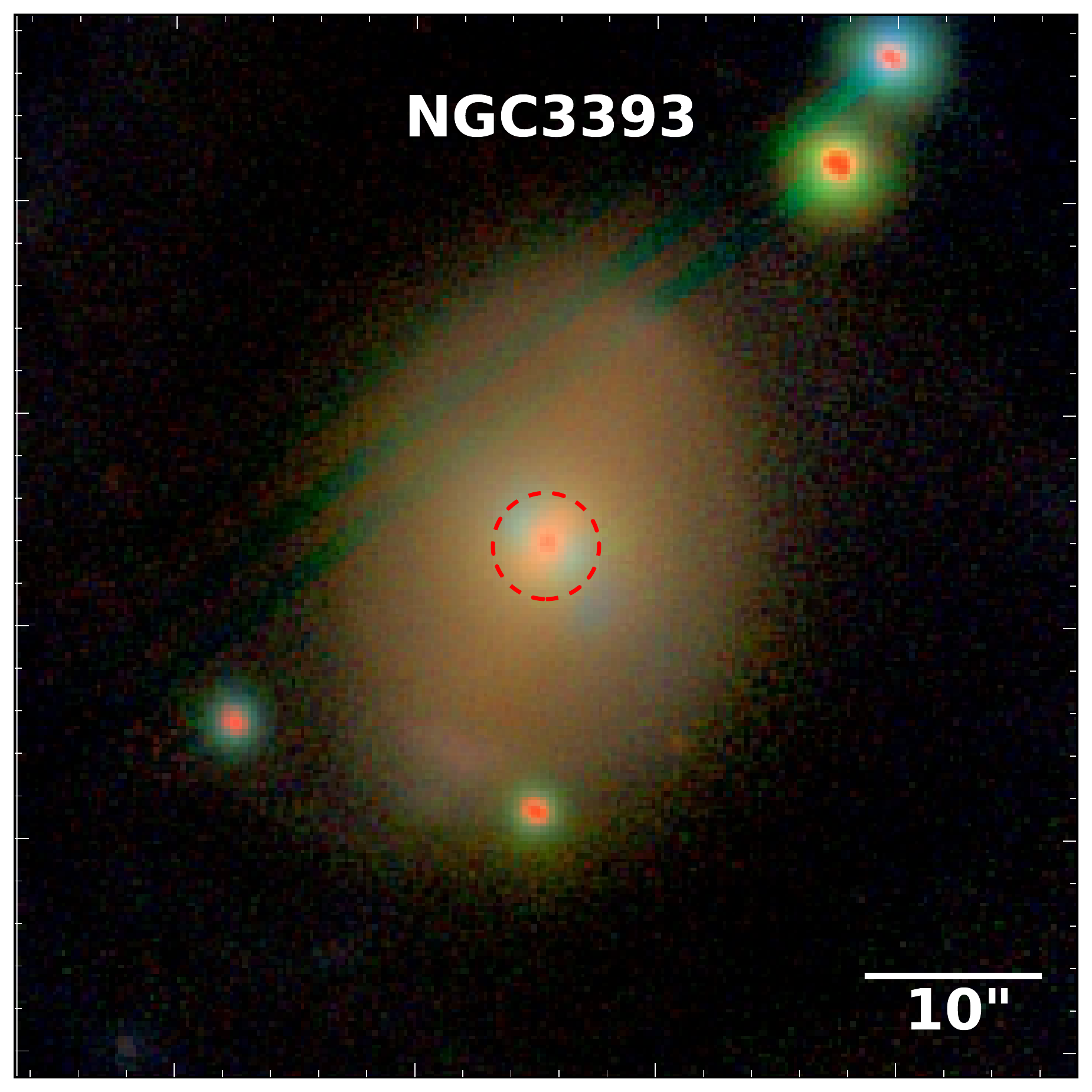}
\includegraphics[width=6.0cm]{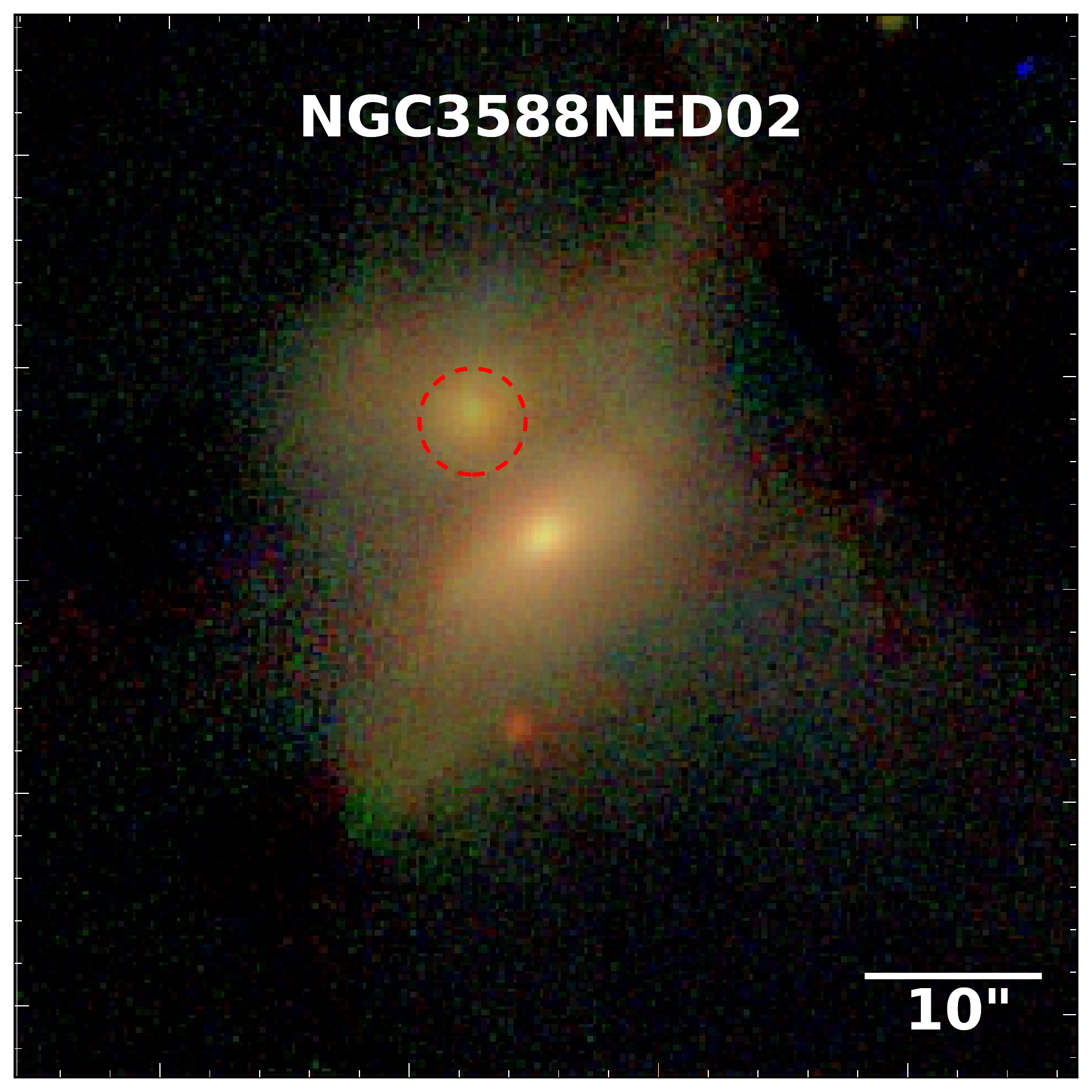}
\includegraphics[width=6.0cm]{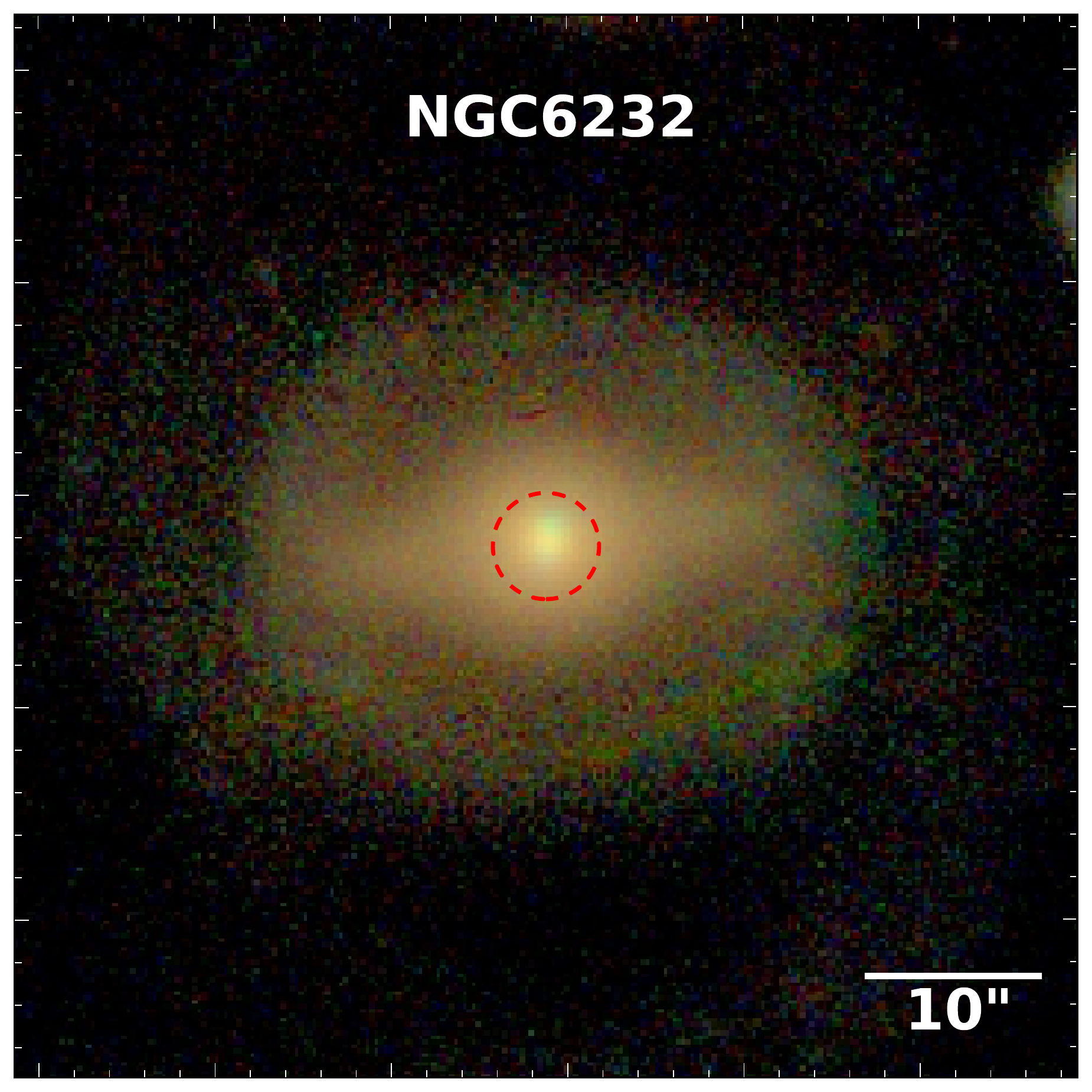}
\includegraphics[width=6.0cm]{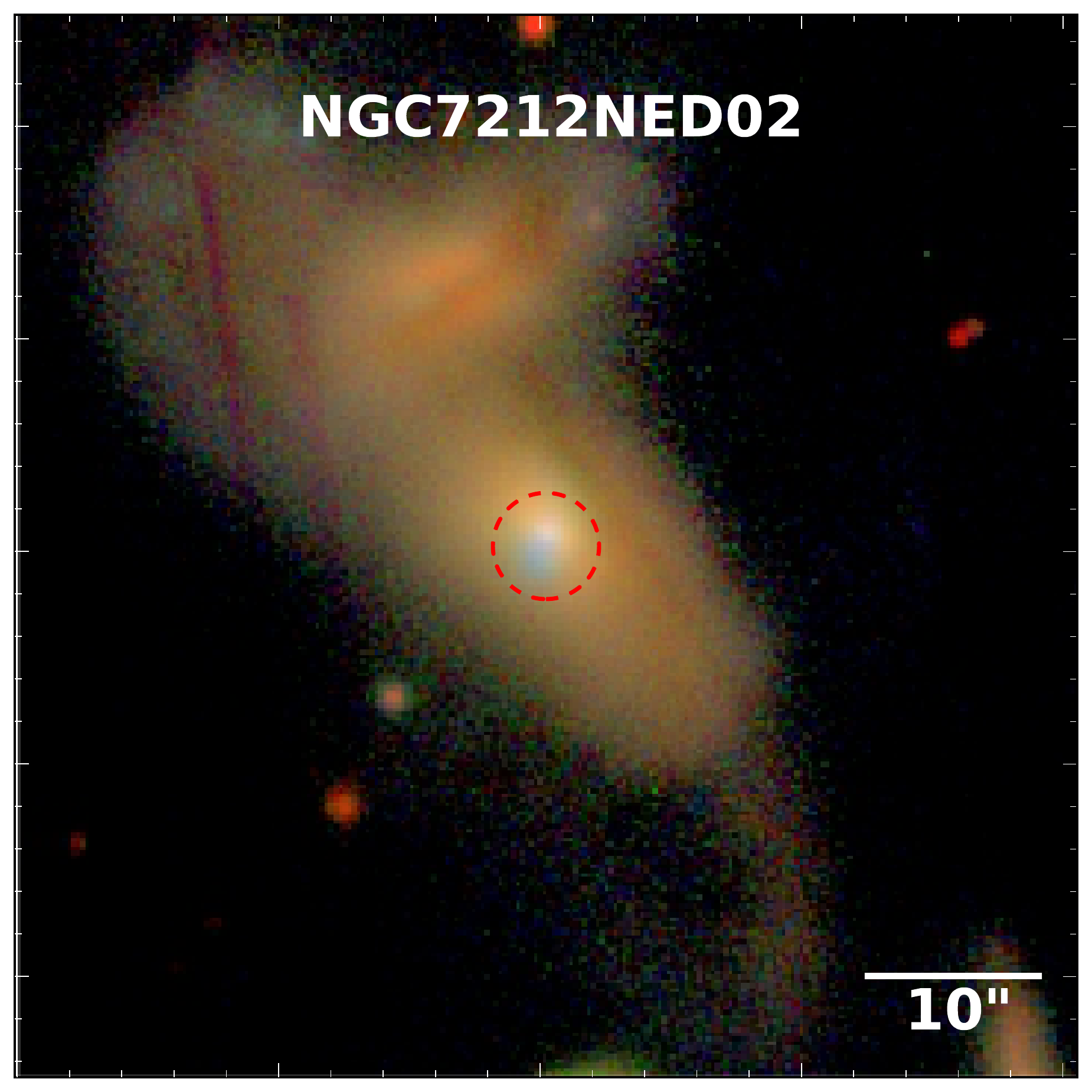}
\includegraphics[width=6.0cm]{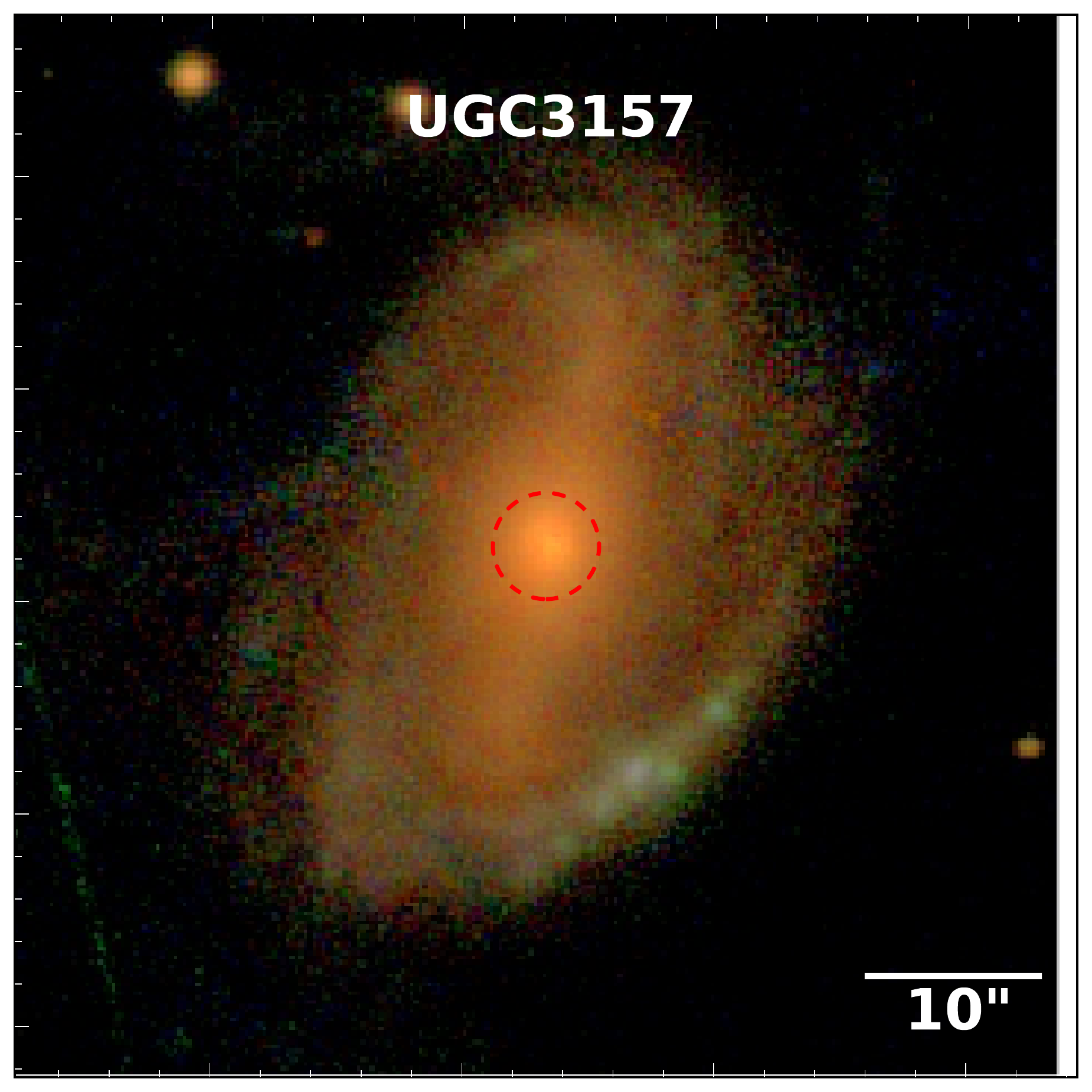}
\caption{Tricolor optical images in $gri$ displayed with an arcsinh scale for nine sources selected based on \swiftbat SC.  Images are 1$\arcmin$ on a side except for NGC 3079 which is 8$\arcmin$ on a side.  A red dashed circle indicates the BAT-detected counterpart based on soft X-ray data from \swiftxrtsh.  For NGC 3588 NED02 and NGC 7212 NED02, \chandra data confirms that the majority of the hard X-ray emission is coming from the galaxy nucleus and not the merging counterpart ($>$95\% at 2-8~keV).   The high fraction of sources in close ($<$10 kpc) mergers  (22\%, 2/9) and/or highly edge-on ($b/a<$0.22)  galaxies (22\%, 2/9) suggests a possible connection of high levels of obscuration to the final merger stage and host galaxy inclination.}
\label{optical_images}
\end{figure*}

We investigate whether our sources have unique host morphologies or accretion rates compared to the rest of the nearby BAT AGN. Tricolor {\it gri} filter images for the nine sources selected based on \swiftbat SC are shown in Figure \ref{optical_images}.  We find that 22\% (2/9) of the sample are in close mergers ($<10$ kpc).  In both sources, faint tidal tails and radial velocity differences of less than 500 km s$^{-1}$ (from NED) between the sample galaxy and its possible companion suggest an ongoing major merger rather than a chance association.  NGC 3588 NED02 has a separation of 4.2 kpc (8.1$\arcsec$).  NGC 7212 NED02 is in a galaxy triple with a separation of 9.8 kpc (18.3$\arcsec$) from NGC 7212 NED03 and a separation of 22 kpc (41$\arcsec$) from NGC 7212 NED01.  This fraction is higher than typically found for BAT AGN; \citet[][]{Koss:2010:L125} found 8\% (11/144) of BAT AGN are in close mergers ($<10$ kpc), though consistent within Poisson errors.  With such a small sample size this difference is not significant based on a Fisher exact test.  \\

To investigate this further, we use the \citet{Koss:2010:L125} study, which had high quality imaging and optical spectroscopy to study the companion separation. We find that the other two \nustarsh-observed AGN in close mergers, NGC 6240 and 2MASX J00253292+6821442, are also in the SC Compton-thick regime (\scnustar=0.49$\pm$0.01 and \scnustar=0.56$\pm$0.05; NGC 6240 and 2MASX J00253292+6821442).    The likelihood of finding all four sources being Compton-thick is $<1\%$ based on a Fisher test.\\

Another interesting morphological feature is that 22\% (2/9) of the galaxies from the program (NGC 3079 and 2MFGC 02280) are nearly edge-on.  \citet{Koss:2011:57} derived the axis ratio ($b/a$) from the major and minor axes derived from isophotal $r$-band photometry for both of these galaxies (NGC 3079; $b/a=0.15$ and 2MFGC 02280; $b/a=0.21$).  By comparison, 6\% of BAT AGN have $b/a<0.22$ \citep{Koss:2011:57}.  The frequency of edge-on galaxies selected using SC vs. the other BAT AGN is not statistically significant because of the small sample size based on a Fisher test, (11\% chance), implying larger samples are needed.\\    

We compare the black-hole mass and Eddington ratio of our sources to the other nearby BAT-detected AGN.  We use the velocity dispersion measurements for measurements of black-hole mass (from Koss et al., in prep) and the median and median absolute deviation (MAD) to compare the populations because of the spread over several orders of magnitude.     We find that the typical black-hole mass of our sample is a factor of four smaller than typical BAT-detected AGN ($M_{\mathrm{BH}}=(1.3\pm0.4)\times10^7$\Msun\ vs.  $M_{\mathrm{BH}}=(5.1\pm0.4)\times10^7$\Msun) where the error refers to the MAD 1$\sigma$ error.  \\

\begin{figure}
\plotone{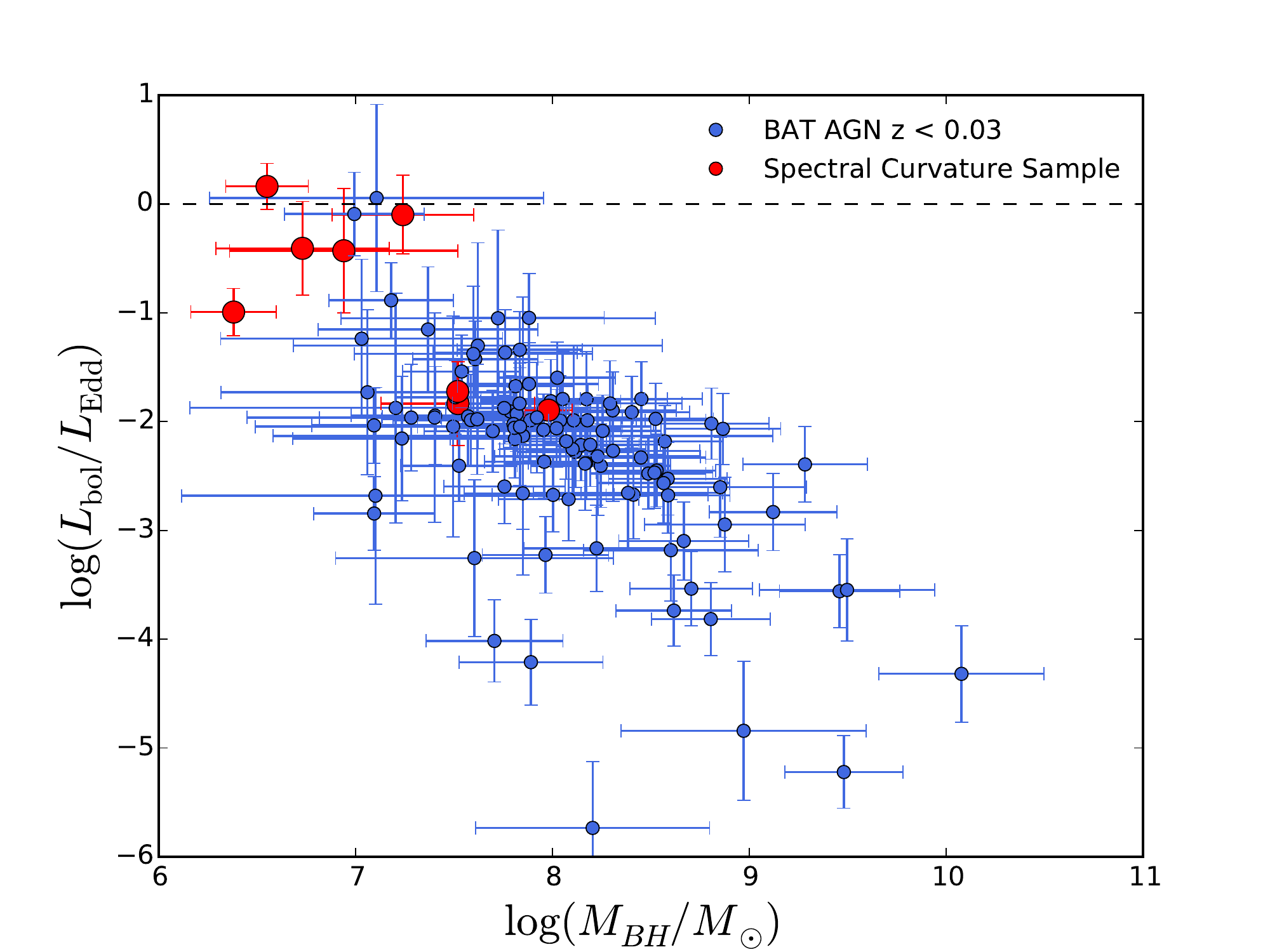}
\caption{Eddington ratio of our heavily obscured AGN compared to other nearby ($z<0.03$) BAT-detected AGN.  We find that these sources typically are in the upper left quadrant of the sample with higher accretion rates and smaller black holes than typical BAT-detected AGN. }
\label{edd_ratio}
\end{figure}

We also estimated the bolometric luminosity $L_{bol}$ from the X-ray luminosity (\Lbat) using the bolometric corrections from \citet{Vasudevan:2009:1124}.  Including the absorption corrected 14-195~keV emission based on the \nustar spectral fitting of our sources, the typical Eddington ratio of our sources is about a factor of four larger ($\lambda_{\mathrm{Edd}}=0.068\pm0.023$ compared to $\lambda_{\mathrm{Edd}}=0.016\pm0.004$)  where the error refers to the 1$\sigma$ error in the MAD (Figure \ref{edd_ratio}).  A Kolmogorov--Smirnov (K-S) test indicates a $<$1\% chance that the distribution of Eddington ratios for SC-selected BAT AGN are from the same parent distribution as the other BAT AGN.  This indicates that the SC-selected BAT AGN have, on average, higher accretion rates.


\clearpage
\section{Summary}

We define a new Spectral Curvature measure of Compton-thickness using weighted averages of different energy bands in the low sensitivity \swiftbat survey.  We then select nine AGN for \nustar follow-up to study their possible Compton-thick nature.

\begin{enumerate}
\renewcommand{\theenumi}{(\roman{enumi})}
\renewcommand{\labelenumi}{\theenumi}

\item We find that all nine targeted sources are consistent with Compton-thick AGN in the majority of indicators (e.g., \feka EW$>$1~keV, $\Gamma<$1, X-ray spectra fitting, mid-IR indicators) confirming the effectiveness of the SC method to identify new Compton thick AGN. Using \nustar spectroscopy, the majority of the nine targets are consistent with Compton-thick AGN using  \mytorus models (78\%,7/9)  and the remaining two are nearly Compton-thick (\nh$\simeq5-8\times10^{23}$ \nhunit).  The observed 2-10~keV emission compared to the 6 $\mu$m emission is also consistent with a Compton-thick AGN for most sources (8/9, 89\%; $F^{\rm obs}_{\mathrm{2-10\ keV}}$/$F^{\rm pred\  6\mu m}_{\rm 2-10\ keV}>20$).

\item Our results suggest the $>10$~keV emission may be the only way to identify this population of Compon-thick AGN other than through detailed SED fitting. We find only two sources show evidence of an excess in the Balmer decrement corrected \oiii vs. X-ray luminosity consistent with a Compton-thick AGN  ($F_{\rm [O\ {\scriptscriptstyle III}]}/F^{\rm obs}_{\mathrm{2-10\ keV}}>1$).  As expected for lower luminosity AGN, we find that most sources (6/9, 67\%) would not be identified as AGN using \wise colors, though detailed SED fitting with the mid-IR would identify most sources.  

\item We find the \scbat and \scnustar measurements to be consistent on average (\scbat=$0.29\pm0.07$ vs. \scnustar=$0.27\pm0.03$). This suggests this measure can be used with other satellites with $>10$~keV coverage such as \astroh, or high-redshift AGN ($z>3$) observed with \chandra or \athena where the bands are shifted to cover rest frame $10-30$~keV.

\item We find that the SC measure is much more effective at selecting Compton-thick AGN than band ratios (8-24/3-8~keV) finding 10/10 well known Compton-thick AGN compared to only 2/10 using band ratios. 


\item We find that these heavily obscured AGN have smaller black holes ($\langle M_{\mathrm{BH}} \rangle=(1.3\pm0.36)\times10^7$\Msun\ vs.  $\langle M_{\mathrm{BH}} \rangle=(5.1\pm0.39)\times10^7$\Msun) and higher accretion rates than other BAT-detected AGN ($\langle \lambda_{\mathrm{Edd}}\rangle=0.068\pm0.023$ compared to $\langle \lambda_{\mathrm{Edd}} \rangle=0.016\pm0.004$).

\end{enumerate}

	We find that the four \nustarsh-observed sources in very close mergers ( $<$10 kpc) are all found to be Compton-thick, suggesting a physically plausible link between increased gas supply and obscuration which might be natural in the early stages of a merger \citep[e.g.,][]{Sanders:1988:L35,Hopkins:2005:L71}.  Based on simulations, the timescale within 10 kpc for major mergers is relatively short, on the order of 100 to 200 Myr \citep{VanWassenhove:2012:L7}, so finding even a small number of galaxies may be significant.  Another interesting morphological feature is that 2/9 program sources are in extremely edge-on galaxies ($b/a<0.25$) suggesting that galaxy wide extinction may be important for some sources. This compares to only 6\% of BAT AGN.  The likelihood of this occurring by chance is 11\%, implying larger samples are needed.\\
		

	
	Based on the robustness of SC in identifying Compton-thick AGN, we measure the fraction of Compton-thick nearby BAT AGN ($z<0.03$) as $\approx22\%$ (\scbat=22\%, \scnustar=21\%).  The Compton-thick number density is $2.3\pm0.3\times10^{-6}$~Mpc$^{-3}$ above \Lsoft$>10^{43}$~\ergps.  This is a conservative estimate since \swiftbat likely misses reflection-dominated AGN. This number is significantly higher than previous work with \swiftbat which reported only a handful of Compton-thick AGN corresponding to fractions of a few percent \citep[e.g.,][]{Tueller:2008:113, Winter:2009:1322, Burlon:2011:58,Vasudevan:2013:111}.  This 22\% fraction is in line with estimates of the intrinsic Compton-thick fraction in X-ray background population synthesis models \citep[5-52\% of obscured AGN, for review see][]{Ueda:2014:104}. \\\
	
	These Compton-thick AGN show high Eddington ratios consistent with other well-known Compton-thick AGN in the BAT sample already observed with \nustar \citep[e.g., Circinus, $\lambda_{\mathrm{Edd}}=0.2$; NGC 4945, $\lambda_{\mathrm{Edd}}=$0.1--0.3; NGC 1068, $\lambda_{\mathrm{Edd}}=$0.5--0.8 ---] []{Arevalo:2014:81,Puccetti:2014:26,Bauer:2014:670} and also in recent results from the XMM-COSMOS survey \citep{Lanzuisi:2015:A137}.   This suggests that the sum of black hole growth in Compton-thick AGN (Eddington ratio times population percentage) may be nearly as much as the rest of the population of mildly obscured AGN and unobscured AGN.  A highly obscured (\nh $>10^{24}$ \nhunit), high-Eddington population ($\lambda_{\mathrm{Edd}}>0.1$) like these AGN could be important for resolving discrepancies based on considerations of the Soltan argument \citep[e.g.,][]{Brandt:2015:1,Comastri:2015:L10}.    Additionally, the high Eddington ratio with relatively weak \oiii to X-ray ratio despite being Compton-thick, are consistent with the ``Eigenvector 1" relationships \citep[e.g.,][]{Boroson:1992:109}.   In further studies, we will use much larger samples of BAT-detected AGN with measured black-hole masses and accretion rates to study which populations have most of the black hole growth (Koss et al., in prep).

\section{Acknowledgments}
We acknowledge financial support from: Ambizione fellowship grant PZ00P2\textunderscore154799/1 (M.K.), the Swiss National Science Foundation (NSF) grant PP00P2 138979/1 (M.K. and K.S.), the Center of Excellence in Astrophysics and Associated Technologies (PFB 06), by the FONDECYT regular grant 1120061 and by the CONICYT Anillo project ACT1101 (E.T.), NASA Headquarters under the NASA Earth and Space Science Fellowship Program, grant NNX14AQ07H  (M.B.), NSF award AST
1008067 (D.B.), Caltech NuSTAR subcontract 44A-1092750 and NASA ADP grant NNX10AC99G (W.N.B.), and the ASI/INAF grant I/037/12/0Ð 011/13 and the Caltech Kingsley visitor program (A.C.).  M.K. also acknowledges support for this work was provided by the National Aeronautics and Space Administration through \chandra Award Number AR3-14010X issued by the \chandra X-ray Observatory Center, which is operated by the Smithsonian Astrophysical Observatory for and on behalf of the National Aeronautics Space Administration under contract NAS8-03060.   This work was supported under NASA Contract No.\ NNG08FD60C, and made use of data from the \nustar mission, a project led by the California Institute of Technology, managed by the Jet Propulsion Laboratory, and funded by the National Aeronautics and Space Administration. We thank the \nustar Operations, Software and Calibration teams for support with the execution and analysis of these observations. This research has made use of the \nustar Data Analysis Software (NuSTARDAS) jointly developed by the ASI Science Data Center (ASDC, Italy) and the California Institute of Technology (USA).  This research made use of the XRT Data Analysis Software (XRTDAS), archival data, software and on-line services provided by the ASDC.  This work made use of data supplied by the UK Swift Science Data Centre at the University of Leicester.  The scientific results reported in this article are based on data obtained from the \chandra Data Archive (Obs ID=4078, 4868, 12290, 13895).  Based on observations obtained with XMM-Newton (Obs ID=0110930201, 0147760101, 0200430201), an ESA science mission with instruments and contributions directly funded by ESA Member States and NASA.
{\it Facilities:}  \facility{NuSTAR}, \facility{Swift}, \facility{XMM}, \facility{SDSS}, \facility{CXO}
\bibliographystyle{/Applications/astronat/apj/apj}

\end{document}